\documentclass[useAMS,usenatbib,onecolumn]{mnras}
\usepackage{graphicx}
\usepackage{times}
\usepackage{amssymb}
\usepackage{amsmath}
\usepackage{multirow}
\usepackage{subfig}
\usepackage{hyphenat}
\usepackage{hyperref}
\usepackage{sansmath}
\usepackage{aas_macros}
\usepackage{tabularx}
\usepackage{multicol}
\newcommand{\hompc}{\,h\,{\rm Mpc}^{-1}}
\newcommand{\mpcoh}{\,h^{-1}\,{\rm Mpc}}

\newcommand{\mpcohV}{\,h^{-3}\,{\rm Mpc^3}}
\newcommand{\kmsmpcohV}{\,{\rm km^{2}\,s^{-2}}\,h^{-3}\,{\rm Mpc^3}}

\newcommand{\bk}{\boldsymbol{k}}
\newcommand{\bq}{\boldsymbol{q}}
\newcommand{\bx}{\boldsymbol{x}}
\newcommand{\br}{\boldsymbol{r}}
\newcommand{\bv}{\boldsymbol{v}}

\begin{document}

\title[Redshift-Space Momentum Power Spectrum I] 
{The Redshift-Space Momentum Power Spectrum I: Optimal Estimation From Peculiar Velocity Surveys}

\author[C. Howlett et. al.]{\parbox{\textwidth}{
Cullan Howlett\thanks{Email: c.howlett@uq.edu.au}$^{1,2,3}$
}
  \vspace*{4pt} \\ 
$^{1}$ International Centre for Radio Astronomy Research, The University of Western Australia, Crawley, WA 6009, Australia. \\
$^{2}$ ARC Centre of Excellence for All-sky Astrophysics (CAASTRO). \\
$^{3}$ School of Mathematics and Physics, University of Queensland, Brisbane, QLD 4072, Australia. \\
}

\pagerange{\pageref{firstpage}--\pageref{lastpage}} \pubyear{2019}
\maketitle
\label{firstpage}

\begin{abstract}
Low redshift surveys of galaxy peculiar velocities provide a wealth of cosmological information. We revisit the idea of extracting this information by directly measuring the redshift-space momentum power spectrum from such surveys. We provide a comprehensive theoretical and practical framework for estimating and fitting this from data, analogous to well understood techniques used to measure the galaxy density power spectrum from redshift surveys. We formally derive a new estimator, which includes the effects of shot noise and survey geometry; we evaluate the variance of the estimator in the Gaussian regime; we compute the optimal weights for the estimator; we demonstrate that the measurements are Gaussian distributed, allowing for easy extraction of cosmological parameters; and we explore the effects of peculiar velocity measurement errors. We finish with a proof-of-concept using realistic mock galaxy catalogues, which demonstrates that we can measure and fit both the redshift-space galaxy density \textit{and momentum} power spectra from peculiar velocity surveys and that including the latter substantially improves our constraints on the growth rate of structure. We also provide theoretical descriptions for modelling the non-linear redshift-space density and momentum power spectrum multipoles, and forecasting the constraints on cosmological parameters using the Fisher information contained in these measurements for arbitrary weights. These may be useful for measurements of the galaxy density power spectrum even in the absence of peculiar velocities.
\end{abstract}


\section{Introduction}
The nature of dark matter and dark energy is one of the greatest mysteries in cosmology. Modification to our current theory of gravity, General Relativity (GR; \citealt{Einstein1916}) offers a way to solve this. Large galaxy surveys, through the use of Redshift Space Distortion (RSD; \citealt{Kaiser1984}), give us a way to test different theories of gravity on scales dominated by dark matter and dark energy. RSD induces an anisotropy in the clustering of galaxies proportional to the rate at which structure grows, commonly parameterised on linear scales by the parameter $f$. GR provides stringent predictions on this `growth rate of structure', namely that it is scale-independent and evolves with redshift as $f(z) \approx \Omega_{m}(z)^{0.55}$.

Many galaxy surveys have been used to place constraints on the growth rate of structure (e.g., \citealt{Blake2011,Beutler2012,delaTorre2013,Howlett2015a,Alam2017,GilMarin2018}, to name a few), and these will tighten significantly with the advent of next generation surveys such as DESI \citep{Levi2013}, 4MOST \citep{deJong2012} and Euclid \citep{Laureijs2011}. In addition, there exists a long history of works (e.g., \citealt{Strauss1989,Kaiser1991,Nusser1994,Hudson1995,Park2000,Silberman2001,Pike2005,Carrick2015} or see \cite{Strauss1995} for an excellent and comprehensive review) which demonstrate that combining peculiar velocity (PV) surveys with galaxy redshifts offers a way to test our cosmological model. More recent studies \citep{Burkey2004,Koda2014,Adams2017,Howlett2017a,Howlett2017b} have emphasised that growth rate measurements from future redshift surveys will be fundamentally limited by cosmic variance - the volume of the universe, and hence the number of independent `modes' with which we can measure cosmology, is too small - but have also repeatedly demonstrated that the addition of PV data is one of the most promising routes to overcome this fundamental limitation. The use of PV data acts in the same way as the multi-tracer method \citep{McDonald2009b}, where one tracer has a known \textit{large scale} bias of one \citep{Desjacques2010,Jennings2015,Chen2018}. Multi-tracer analyses allow us to improve our constraints on the growth rate by including information from the \textit{relative} clustering amplitudes of multiple tracers of the same underlying density field \citep{Abramo2013}. In the case of PV surveys, this is the density field traced separately by the distribution of luminous matter, and the motions of galaxies induced through gravity.

Realising the potential of PV surveys for constraining gravity requires development of techniques to measure the correlations in the velocity field and between the density and velocity fields. \cite{Kolatt1997} estimated the power spectrum by first reconstructing the density field on a grid from a set of PV measurements. Many studies \citep{Zaroubi1997,Macaulay2012,Johnson2014,Howlett2017c,Huterer2017} have also successfully used a likelihood based method to estimate the velocity power spectrum from existing datasets. This was extended in \cite{Adams2017} to include measurements of the galaxy density power spectrum, and the cross-power spectrum between the density and velocity fields. This latter approach, in theory, fully captures the information in the two fields. However, it requires a computation of the integral of the power spectrum for each individual pair of galaxies, and each choice of model. Next generation datasets such as the Taipan Galaxy Survey \citep{DaCunha2017}, WALLABY survey \citep{Koribalski2012}, or supernovae \citep{Howlett2017b,Kim2019} will contain up to an order of magnitude as many PV measurements as all current surveys combined \citep{Tully2016}, and require more complex non-linear modelling. This may make the above approaches prohibitively expensive. An alternative is measuring the velocity correlation function \citep{Gorski1989,Groth1989,Juszkiewicz2000} or the density-velocity cross-correlation \citep{Nusser2017} directly, but recent results \citep{Wang2018} suggest the presence of non-Gaussian posteriors in such measurements makes extracting cosmological information difficult.

In this work, we present an alternative; estimation of the momentum (or mass-weighted velocity) power spectrum from a real survey. Much like the galaxy density power spectrum and redshift surveys, this presents an efficient compression of the information contained in a PV survey. This is not the first work to use this; measurements of the momentum power spectrum are common when analysing simulations \citep{Scoccimarro2004,Pueblas2009,Jennings2011,Chen2018} although often the underlying goal is analysis of the non-linear velocity field, which cannot be easily extracted from such a measurement. This has also been applied to data. Two key references that present this as a viable method for extracting cosmology from PV surveys are \cite{Park2000} and \cite{Park2006}. However, to the best of the author's knowledge this has not been revisited over the last decade, whilst techniques to measure cosmology from RSD have improved tremendously. Recently, PV surveys have also seen a resurgence, with significant improvements in data quality and volume that will only continue over the coming years. As such, we feel it is time to revisit this, and we completely re-interpret the formalism in the framework of modern galaxy surveys, bringing this technique up to the same standard as is used for modern RSD measurements. As we will show, this approach overcomes some of the short comings with the above methods, although does come with its own challenges (in particular estimation of the covariance matrix, as with the galaxy density power spectrum, is theoretically difficult and computationally demanding).

With this work, we aim to present the momentum power spectrum as a  viable method to extract the full cosmological information from PV surveys and provide a complete description for how this can be measured in current or future surveys. This paper has the following structure: In Sections~\ref{sec:momentum} and~\ref{sec:estimator} we introduce the momentum field and present a new estimator for the redshift-space momentum power spectrum based on the well-known \citealt{Feldman1994} (FKP) and \cite{Yamamoto2006} estimators for the galaxy density power spectrum. We then measure this in simulated halo catalogues and identify the similarities and differences between the two power spectra; i.e., the similar normalisations and window function convolutions. In Sections~\ref{sec:covariance} and~\ref{sec:weights}, we derive the covariance of our estimator in the Gaussian regime, compare this to the same halo simulations and derive the optimal weights for our estimator. Throughout these three Sections there is, by design, considerable analogy between the momentum and galaxy density power spectra. In Section~\ref{sec:mocks}, we then turn to realistic mock PV catalogues to demonstrate how our estimator works in reality, including several quirks of PV surveys that do not affect galaxy redshift surveys such as PV measurement errors. In particular, in Section~\ref{sec:proof} we bring everything together and demonstrate that we can use combined measurements of the galaxy density power spectrum and momentum power spectrum multipoles to improve our constraints on the growth rate of structure. We also compare our results to forecasts for the same mock data to show that we are recovering the expected amount of information on the growth rate. We conclude in Section~\ref{sec:conclusion}. Finally, Appendices~\ref{sec:appendix} and~\ref{sec:appendixB} provide theoretical descriptions of how to compute the non-linear redshift-space density and momentum power spectrum, and forecast the constraints on cosmological parameters using the Fisher information contained in the multipoles of the galaxy density and velocity/momentum power spectrum for arbitrary weights. These may be useful for measurements of the galaxy density power spectrum even in the absence of peculiar velocities.

This is just the first in a planned series of works using the momentum power spectrum. In a follow up paper in this series (Qin et. al, in preparation; Paper II) we apply this to the combined 2MTF (\citealt{Hong2014}; Hong et al., in preparation) and 6dFGSv \citep{Campbell2014} data and show that we can recover accurate and robust constraints on the growth rate of structure at low redshift. We also plan to investigate how to extract the additional information in the cross-power spectrum of the density and momentum fields (which as shown in Section~\ref{sec:proof} is expected to improve the growth rate constraints by at least a further $\sim 10\%$), and to investigate potential constraints on alternative models of gravity that could be achieved with our method. 

\section{The Momentum Field} \label{sec:momentum}
We start by defining the momentum field, following \cite{Park2000}, as the velocity field traced by discrete galaxies. Mathematically, we write the momentum field, $\boldsymbol{\rho}$ as
\begin{equation}
\boldsymbol{\rho}(\br) = (1+\delta_{g}(\br))\bv(\br),
\end{equation}
where $\delta_{g}(\br)$ and $\bv(\br)$ are the galaxy overdensity and the velocity field, respectively, at some location $\br$. In a real survey, we are only able to measure the radial peculiar velocities of galaxies $u(\br)$, which means that we are also only able to construct the radial momentum field $\rho_{||}(\br)$. Additionally, both the measured density and velocity fields contain non-linear contributions and are potentially biased tracers of the true underlying density and velocity fields (although there is good reason to believe the latter is unbiased on large scales, \citealt{Desjacques2010,Jennings2015,Chen2018}). Nonetheless, the measured radial momentum field can still be used as a tracer of cosmology.

As with any field, we can consider measuring the radial momentum field on a grid and taking the Fourier transform. Given the nature of the radial momentum field, we can simply do this by averaging the peculiar velocities of galaxies in cells using some interpolation scheme, which can be corrected for after the Fourier transform has been taken. Note that, unlike when trying to estimate the velocity field directly, the momentum field does not suffer from `zero-valued' regions, i.e., parts of the gridded space where the velocity field is artificially set to zero due to the absence of tracers. In the absence of any galaxies due to a survey selection function, the momentum field is both measured \textit{and expected} to be zero (because the overdensity is expected to be zero given our knowledge of the survey selection function), and so we are able to obtain results without complex corrections or gridding schemes (such as those used in \citealt{Jennings2011,Jennings2012} or \citealt{Koda2014}). However, we do have to modify our expectations for what we have measured; we have not measured the velocity field on the grid, and so we are not able to measure the velocity power spectrum directly.

We can demonstrate what we \textit{have} measured by correlating the radial momentum field in Fourier space. We can write the radial momentum power spectrum as 
\begin{align}
(2\pi)^{3}\delta^{D}(\bk-\bk')P^{p}(\bk) &= \langle (1+\delta_{g}(\bk))u(\bk)(1+\delta_{g}(\bk'))u(\bk') \rangle \notag \\
&= \langle u(\bk)u(\bk') \rangle + \langle u(\bk)\delta_{g}(\bk')u(\bk') \rangle + \langle \delta_{g}(\bk)u(\bk)u(\bk') \rangle + \langle \delta_{g}(\bk)u(\bk)\delta_{g}(\bk')u(\bk')\rangle,
\label{eq:mom}
\end{align}
which highlights the fact that the power spectrum of the radial momentum field is \textit{related} to the velocity power spectrum (it is the first term in the above equation), but is not fully equivalent. On linear scales, the velocity power spectrum dominates due to the $1/k^{2}$ scaling of the velocity field relative to the density field. However, on non-linear scales higher order terms arise from the convolution of the density and velocity fields at the same location. Although at first glance these appear to be three- and four-point functions, in practice they still only contain two distinct locations, and so can be computed relatively simply using perturbation theory (\citealt{Vlah2012,Vlah2013,Okumura2014}, see also Appendix~\ref{sec:appendix}). The equivalence of the velocity and momentum power spectra on linear scales makes this measurement an excellent tracer of gravity, whilst the density field dependence on small scales can be used to partially break the degeneracy between galaxy bias and the growth rate of structure seen in measurements of the galaxy density power spectrum.   

Applying this to a real survey introduces some complexity. Firstly, in order to assign the velocities to a grid, we need an estimate of their position. This can be done one of three ways: using the measured redshifts and assuming a cosmological model to put the galaxies at their \textit{redshift-space} distance; using the redshifts and peculiar velocities to estimate the true distance to each galaxy; or using their positions inferred from reconstructions of the density field using linear theory. In this work, we advocate the former. This is in contrast to the work of \cite{Park2000} and \cite{Park2006}, but as we will show, the use of redshift-space positions (which effectively replaces $\delta_{g}$ is Eq.~\ref{eq:mom} with the redshift space overdensity $\delta^{s}_{g}$ and can also be modelled with perturbation theory) only changes the measured momentum power spectrum on small, noise-dominated scales. Using instead the true distances (from the peculiar velocities or via reconstruction) could introduce considerable measurement or systematic error which may be harder to theoretically model.

Secondly, to apply this to a real survey, we need to formally derive an estimator that accounts for the discrete sampling of the fields and the survey geometry. This will be the focus of the next Section. It is also worth noting that extracting constraints from these measurements does require accurate knowledge of the survey geometry and selection function. However, ensuring a well-defined selection function for future peculiar velocity surveys will be extremely advantageous for a number of reasons beyond the requirements of the method developed herein.

\section{Estimating the Redshift Space Power Spectrum} \label{sec:estimator}
Galaxies typically fall towards dense regions of the Universe and have 3-dimensional velocities, however the redshift-space and real-space positions only differ due to the line-of-sight velocity. The net result of this on large scales is that an otherwise spherical distribution of galaxies will appear squashed or extended along the line-of-sight; the transverse positions are unaffected, whilst the line-of-sight positions are moved closer to the overdensities influencing the large scale motion of the galaxy. In terms of the clustering of galaxies, whereas in an isotropic and homogeneous universe we might expect the correlation function or power spectrum to depend only on the separation between galaxies, the use of redshift-space positions introduces a measurable difference along and transverse to the line-of-sight. The strength of this anisotropy is proportional to the typical velocity of objects along of the line-of-sight and so can be used to determine the growth rate of structure.

\begin{figure}
\centering
\includegraphics[width=0.30\textwidth]{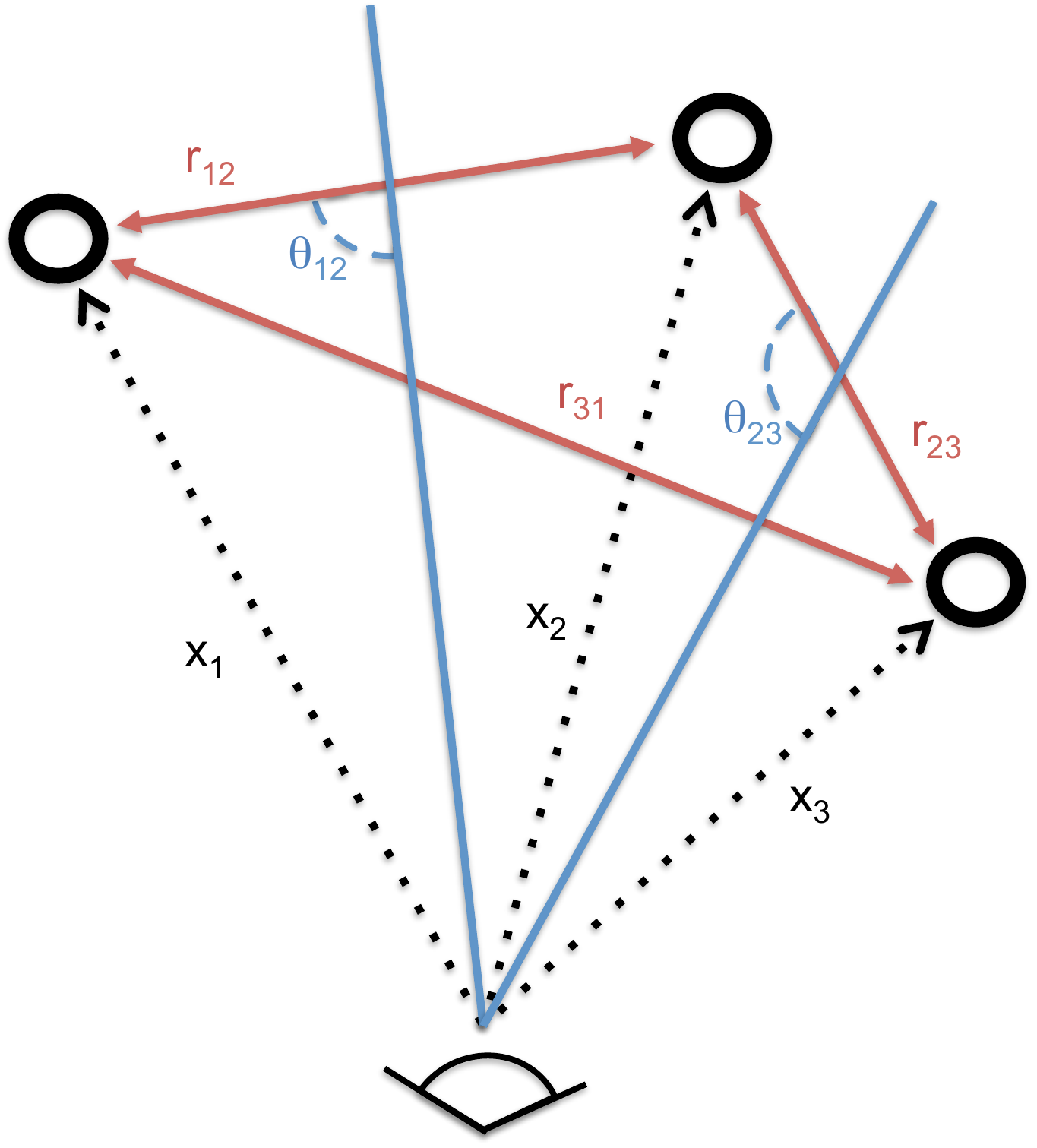}
  \caption{A diagram describing how the coordinates and angles between the observer and the redshift-space positions of galaxies are usually defined when computing the anisotropic clustering. The vectors $\bx_{1}$, $\bx_{2}$ and $\bx_{3}$ are the vectors of the line of sight to each galaxy. $\br_{ij}=\bx_{i}-\bx_{j}$ denotes the separation between objects. The angle between the line-of-sight and the separation vector is then typically defined by assuming that each pair of objects is `locally parallel', such that $\mu_{ij}=\mathrm{cos}(\theta_{ij})=1/2(\bx_{i}+\bx_{j}) \cdot \br_{ij}$ and varies across the sky. If the separation between pairs of objects is sufficiently small, or the objects are sufficiently far away, we can treat all lines-of-sight $\bx_{i}$ and angles $\theta_{ij}$ as equal. This is commonly called the ``distant-observer" or ``global plane-parallel" approximation.}
  \label{fig:rsddiagram}
\end{figure}

Although the spherical symmetry of two-point clustering measurements is broken in redshift-space, they do retain symmetry about the line-of-sight. Hence, a common technique is to decompose the redshift-space correlation function $\xi(\br)$ or power spectrum $P(\bk)$ into multipoles using the Legendre polynomials $L(\mu)$ e.g., 
\begin{equation}
\xi(\br) = \xi(r,\mu) = \sum_{\ell}\xi_{\ell}(r)L_{\ell}(\mu),
\label{eq:pkmulti}
\end{equation}
where $r$ is the separation and $\mu$ is the cosine of the angle to the line-of-sight. Fig.~\ref{fig:rsddiagram} shows a diagrammatic representation of this. As the angle $\mu$ arises from the bisection of the vector between pairs of objects and the line-of-sight it will vary across a survey. For sufficiently distant galaxies or small survey areas all possible values of $\mu$ are approximately equal and we only need to consider a single fixed line-of-sight. This is commonly referred to as the ``distant-observer'' or ``global plane-parallel'' approximation. Given the wide-area and low redshift of most peculiar velocity surveys we do not wish to make such an approximation here \textit{a priori}, although it will be used in some cases later.

Allowing for a varying line-of-sight, measuring the anisotropic correlation function can be achieved by counting the number of pairs of galaxies as both a function of their separation and $\mu$. The power spectrum on the other hand cannot be calculated so easily because the varying line-of-sight does not align with a Cartesian grid, which in turn makes Fourier transforms difficult. One alternative is to decompose the redshift-space fields into spherical harmonics \citep{Heavens1995}, but a more popular method is to define a different $\mu$ for different subsamples of objects \citep{Cole1994}. This technique was extended to its logical conclusion by defining a separate line-of-sight for each pair of galaxies \citep{Yamamoto2006} such that each pair is considered `locally parallel', but the line-of-sight can still vary across the survey. As such this assumption is typically called the ``local plane-parallel'' approximation \citep{Samushia2012,Beutler2014,Yoo2015}. It was recently demonstrated \citep{Bianchi2015,Scoccimarro2015} that the monopole, quadrupole and hexadecapole ($l=0$,$2$,$4$ moments) of the galaxy power spectrum can be computed with only 1+6+15 Fast Fourier transformations respectively under this approximation (although there is a subtly different convention used to calculate the line-of-sight, see \citealt{Beutler2019}) and this technique has been quickly adopted into the standard practice for estimating the redshift-space clustering of galaxies from large galaxy redshift surveys (e.g., \citealt{Blake2011,Oka2014,Beutler2014,GilMarin2016,Beutler2017,GilMarin2018}). In this work, we will demonstrate that a similar technique can be developed for measuring the redshift-space clustering of the momentum field. 

It is worth noting here that neither the \cite{Yamamoto2006} estimator, nor it's implementation using Fast Fourier Transformations \citep{Bianchi2015,Scoccimarro2015}, are immune from `wide-angle effects' on large scales, which if left unaccounted may bias constraints obtained from anistropic power spectrum measurements. There is a long history of works attempting to account for these effects (see i.e., \cite{Hamilton1996,Zaroubi1996} for some early examples, or \cite{Castorina2018,Beutler2019} and references therein for more recent studies). We do not consider such effects in this work and did not find  any associated bias in the results when the methods developed in this paper were applied to current data in Paper II. However, given the low redshift, large sky area and increased precision of upcoming and planned peculiar velocity surveys similar techniques should be explored to ensure we can obtain unbiased results in the future.

\subsection{Momentum Power Spectrum}

In this section, we present a new estimator for the multipoles of the redshift-space line-of-sight \textit{momentum} power spectrum (which from now on will simply be called the momentum power spectrum), as measured from a PV survey with arbitrary number density and weighting. This formalism effectively combines the methods of \cite{Park2006}, \cite{Yamamoto2006} and \cite{Blake2018} resulting in an estimator that looks and behaves similar to the multipoles of the galaxy \textit{density} power spectrum.
 
We begin by defining the weighted radial momentum field
\begin{equation}
F^{p}(\br) = \frac{w(\br)n_{g}(\br)u(\br)}{A},
\label{eq:momfield}
\end{equation}
where $n_{g}(\br)$ and $u(\br)$ are the number of galaxies and line-of-sight peculiar velocity at location $\br$ respectively. $w(\br)$ and $A$ are some convenient weights and normalisation that we will define later\footnote{For comparison, the weighted density field used to define the galaxy power spectrum estimator in \cite{Feldman1994,Yamamoto2006} is
\begin{equation}
F^{\delta}(\br) = w(\br)[n_{g}(\br)-\alpha n_{s}(\br)]/A,
\label{eq:momfield}
\end{equation}
where a synthetic random catalogue containing $1/\alpha$ times as many objects as there are galaxies is used to calculate $n_{s}$. The weights for these two fields may not be the same.}.

If we then take the expectation value of the correlation of the weighted momentum field at different locations we obtain
\begin{equation}
\langle F^{p}(\br)F^{p}(\br') \rangle = \frac{1}{A^{2}}w(\br)w(\br')\langle n_{g}(\br)n_{g}(\br')u(\br)u(\br') \rangle.
\end{equation}
Using the classic trick of dividing the survey volume V into infinitesimally small cells with occupancy $0$ or $1$, \cite{Park2006} showed that we can write
\begin{equation}
\langle n_{g}(\br)n_{g}(\br')u(\br)u(\br') \rangle = \bar{n}(\br)\bar{n}(\br')\xi^{p}(\br-\br') + \bar{n}(\br)\langle v^{2}(\br) \rangle\delta^{D}(\br-\br'),
\label{eq:park}
\end{equation}
such that
\begin{equation}
\langle F^{p}(\br)F^{p}(\br') \rangle = \frac{1}{A^{2}}[w(\br)w(\br')\bar{n}(\br)\bar{n}(\br')\xi^{p}(\br-\br') \\
									+ w^{2}(\br)\bar{n}(\br)\langle v^{2}(\br) \rangle\delta^{D}(\br-\br')],
\label{eq:corrmomex}
\end{equation}
and where $\bar{n}(\br)$ is the expected number of galaxies at $\br$, $\xi^{p}(\br-\br')$ is the radial momentum correlation function, and $\langle v^{2}(\br) \rangle$ is the value of this at zero-lag. Switching now to Fourier space, we define the multipole moments of the momentum power spectrum in the usual manner
\begin{equation}
P^{p}(\bk) = \sum_{\ell}P^{p}_{\ell}(k)L_{\ell}(\mu),
\label{eq:pkmulti}
\end{equation}
and, under the `local plane-parallel' approximation (such that $L_{\ell}(\hat{\bk}\cdot\hat{\br})=L_{\ell}(\hat{\bk}\cdot\hat{\br}')$), write the momentum version of the \cite{Yamamoto2006} estimator
\begin{equation}
|F^{p}_{\ell}(k)|^{2} = \frac{2\ell + 1}{V}\int \frac{d\Omega_{k}}{4\pi}\int d^{3}r \int d^{3}r' F^{p}(\br)F^{p}(\br') L_{\ell}(\hat{\bk}\cdot\hat{\br}') e^{i\bk\cdot(\br-\br')}, \\
\label{eq:Fp}
\end{equation}
where $L_{\ell}(\hat{\bk}\cdot\hat{\br}')$ are the Legendre polynomials, and we average over shells in $\bk$-space as represented by the integral over $d\Omega_{k}$. The volume factor, $1/V$ arises due to our Fourier transformation convention. Taking the expectation value of $|F^{p}_{\ell}(\bk)|^{2}$, and making use of Eq.~\ref{eq:corrmomex}
\begin{align}
\langle |F^{p}_{\ell}(k)|^{2} \rangle &= \frac{2\ell + 1}{A^{2}V}\int \frac{d\Omega_{k}}{4\pi}\int d^{3}r \int d^{3}r' w(\br)w(\br')\langle n_{g}(\br)n_{g}(\br')u(\br)u(\br') \rangle L_{\ell}(\hat{\bk}\cdot\hat{\br}')e^{i\bk\cdot(\br-\br')} \\
&=\frac{2\ell + 1}{A^{2}}\int \frac{d\Omega_{k}}{4\pi}\biggl[\frac{1}{V} \int d^{3}r \int d^{3}r' w(\br)w(\br')\bar{n}(\br)\bar{n}(\br')\int \frac{d^{3}k'}{(2\pi)^{3}} \,P^{p}(\bk',\br')L_{\ell}(\hat{\bk}\cdot\hat{\br}')e^{i(\bk-\bk')\cdot(\br-\br')} + N^{p}_{\ell}(\bk)\biggl],
\end{align}
where we have substituted the momentum correlation function for its Fourier counterpart, the momentum power spectrum $P^{p}(\bk,\br)$, and compressed the $\bk$-dependent shot-noise term into
\begin{equation}
N^{p}_{\ell}(\bk) = \int d^{3}r\,w^{2}(\br)\bar n(\br)\langle v^{2}(\br)\rangle L_{\ell}(\hat{\bk}\cdot\hat{\br}).
\label{eq:shot-noise}
\end{equation}
From here, we follow the method of \cite{Blake2018} (Eqs. 13 and 14 therein) and substitute the multipole expansion in Eq.~\ref{eq:pkmulti} into the above expression
\begin{equation}
\langle |F^{p}_{\ell}(k)|^{2} \rangle = \frac{(2\ell + 1)}{A^{2}}\int \frac{d\Omega_{k}}{4\pi}\biggl[\sum_{\ell'}\int \frac{d^{3}k'}{(2\pi)^{3}}P^{p}_{\ell'}(k')G(\bk-\bk')S_{\ell,\ell'}(\bk,\bk') + N^{p}_{\ell}(\bk)\biggl],
\label{eq:convterms1}
\end{equation}
where
\begin{align}
G(\bk-\bk') &= \int d^{3}r\,w(\br)\bar{n}(\br)e^{i(\bk-\bk')\cdot \br}, \\
S^{*}_{\ell,\ell'}(\bk,\bk') &= \frac{1}{V}\int d^{3}r\,w(\br)\bar{n}(\br)L_{\ell}(\hat{\bk}\cdot\hat{\br})L_{\ell'}(\hat{\bk'}\cdot\hat{\br})e^{-i(\bk-\bk')\cdot \br}.
\label{eq:convterms}
\end{align}
Finally, in the absence of any window function, the weights and number density must be constant as a function of $\br$. In this case $G(\bk-\bk')=\bar{n}wV\delta^{D}(\bk-\bk')$ and the orthogonality condition of the Legendre polynomials means 
$S^{*}_{\ell,\ell'}(\bk,\bk') = \bar{n}w\delta^{D}(\ell-\ell')\delta^{D}(\bk-\bk')/(2\ell+1)$ where $\delta^{D}$ is the Dirac delta. Hence, if we set the normalisation constant to the same value used when estimating the galaxy density power spectrum $A^{2}=\int d^{3}r\,w^{2}(\br)\bar{n}^{2}(\br)$, then in the absence of any window function $\langle |F^{p}_{\ell}(k)|^{2} \rangle = P^{p}_{\ell}(k) + N^{p}_{\ell}(k)$ and we can formally express our estimator for the multipoles of the redshift-space momentum power spectrum in the familiar form\footnote{Again, for comparison, the same derivation for the galaxy density field in \cite{Yamamoto2006} results in 
\begin{equation}
\widehat{P^{\delta}_{\ell}}(k) = |F^{\delta}_{\ell}(k)|^{2} - N^{\delta}_{\ell}(k),
\end{equation}
where $N^{\delta}_{\ell}(k) = \int d^{3}r\,w^{2}(\br)\bar n(\br) L_{\ell}(\hat{\bk}\cdot\hat{\br})$.}
\begin{equation}
\widehat{P^{p}_{\ell}}(k) = |F^{p}_{\ell}(k)|^{2} - N^{p}_{\ell}(k).
\end{equation}

This completes the formal definition of our estimator for the multipoles of the momentum power spectrum. In practice, when estimating this from a dataset we use the Fourier-based method of \cite{Bianchi2015} (see also \citealt{Scoccimarro2015}), replacing the weighted density field $F^{\delta}(\br)$ used in their work with $F^{p}(\br)$ from Eq.~\ref{eq:momfield}. For the shot-noise terms, we approximate the momentum correlation function at zero-lag as a sum over $N$ measurements
\begin{equation}
\langle v^{2}(\br)\rangle=\frac{1}{N}\sum_{N} u^{2}(\br).
\end{equation}
Looking at the shot-noise term in Eq.~\ref{eq:shot-noise}, the above derivation demonstrates that the measured momentum power spectrum depends on both the use of discrete tracers of the velocity field, and the variance in the velocity field itself. This can be seen more clearly if we take the case for $\ell=0$ without any window function $N^{p}_{0} = \langle v^{2} \rangle/\bar{n}$, which can be compared to the well-known limit for the shot-noise on the density power spectrum in this regime $N^{\delta}_{0} = 1/\bar{n}$. The variance in the velocity field contains contributions from the intrinsic variance and the measurement errors, and so remains even for a field where the peculiar velocities are perfectly known. In the absence of any errors on the PV measurements the expectation of the velocity field is independent of position, and so we can use all the measurements to evaluate the shot noise term. In practice, the presence of errors on the PV measurements adds a radial dependence to $\langle v^{2}(\br)\rangle$.  A suitable value for this can be obtained by fitting the errors on the data as a function of redshift or distance (i.e., as in \citealt{Howlett2017c}), or by averaging over measurements in radial bins. We will further examine the contribution from measurement errors, and the fact that the mean velocity itself may not be zero, in Section~\ref{sec:mocks}.

The obvious similarities between the estimator for the multipoles of the density and momentum power spectra highlight its potential as a tool for extracting cosmological information from current and future PV surveys:
\begin{itemize}
\item{Codes to estimate the galaxy density power spectrum can be trivially modified to also measure the momentum power spectrum, allowing for efficient measurement of both of these from a survey.}
\item{In a real survey, the measurements of the momentum power spectrum are affected by the survey window function through the convolution with the $G$ and $S_{\ell,\ell}$ terms in Eqs.~\ref{eq:convterms1}-\ref{eq:convterms}. However, direct comparison between these equations and the convolution between the galaxy density power spectrum and the window function in \cite{Feldman1994}, \cite{Yamamoto2003} or \cite{Blake2018}, shows that the convolution of the the momentum power spectrum with the survey window function is the same. Hence, it can be evaluated using the same random catalogue that mimics the survey geometry.}
\item{The covariance matrix of the multipoles of the momentum power spectrum estimated from a PV survey can be estimated both theoretically (under some approximations, see Section~\ref{sec:covariance}) or using an ensemble of realistic simulations of the survey, again in an identical manner to that used for measurements of the galaxy density power spectrum.}
\item{Given all of the above, and a model for the redshift-space momentum power spectrum (e.g., Appendix~\ref{sec:appendix}) extracting cosmological constraints from the momentum power spectrum can be performed using the same methods as used for the galaxy density power spectrum, e.g., using a Monte-Carlo Markov Chain approach to compute the posterior probabilities.}
\end{itemize}  

We demonstrate our estimator of the momentum power spectrum using simulated halos in two different mass bins at redshift $z=0$. The simulation consists of $2560^{3}$ particles in a box of length $1800\mpcoh$ and halo catalogues were generated using the implementation of the 3D Friend-of-Friends algorithm in the \textsc{VELOCIraptor} code \citep{Elahi2011}. As a first step towards creating conditions representative of real observations, we calculate the momentum power spectrum by placing an observer in the centre of the simulation box and computing line-of-sight velocities for each halo. These are then assigned to a $512^{3}$ grid using nearest neighbour interpolation. The corresponding Nyquist frequency is $k_{Ny}=0.89\hompc$, much larger than the scales we are interested in. We compare the momentum power spectra in both real- and redshift-space, where for the latter we use the line-of-sight velocity to perturb the positions of the halos.

\begin{figure}
\centering
\includegraphics[width=0.8\textwidth]{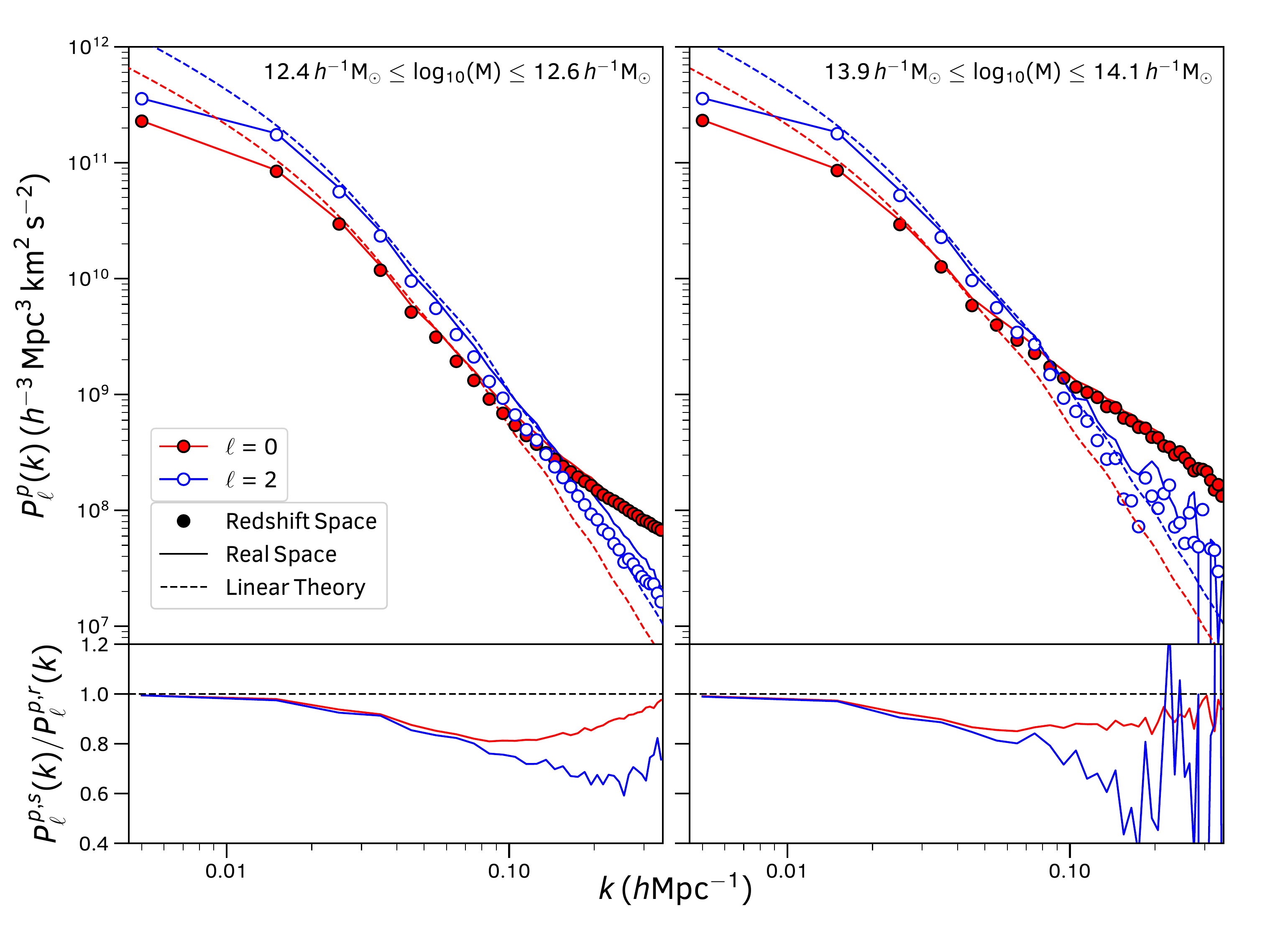}
  \caption{Measurements of the monopole and quadrupole of the momentum power spectrum in real-space and redshift-space for simulated halos in two different mass bins. The points and solid lines show the measurements, whilst the dashed lines show the prediction from linear theory. The lower panel shows the ratio between the redshift- and real-space spectra, with a dashed line to guide the eye.}
  \label{fig:pkmomhalos}
\end{figure}

Our measurements are shown in Fig.~\ref{fig:pkmomhalos} and highlight some interesting features of the momentum power spectrum:
\begin{itemize}
\item{Unlike the galaxy density power spectrum, the real-space momentum power spectrum measured using line-of-sight velocities is \textit{not} isotropic. The use of peculiar velocities adds a direction dependence to the momentum power spectrum even in real-space.}
\item{Using the redshift-space positions of the halos does not change the momentum power spectrum on linear scales. The lower panels of Fig.~\ref{fig:pkmomhalos} show the ratio of the redshift- and real-space spectra. The only effect of using the redshifts to infer the positions of objects when computing the momentum power spectrum is a damping on non-linear scales. The equivalence of the real and redshift-space momentum power spectrum on linear scales is demonstrated theoretically in \cite{Vlah2013} and \cite{Okumura2014} and arises due to the fact that the velocity power spectrum contribution to the momentum power spectrum does not change if we use redshift-space positions.}
\item{In the linear regime, all odd multipoles and multipoles beyond the quadrupole are expected to vanish. This can be demonstrated using the fact that the large scale radial momentum power spectrum is equivalent to the radial velocity power spectrum \citep{Park2000}, which in turn means
\begin{equation}
P^{p}_{L}(k,\mu) \approx \mu^{2}f^{2}(aH)^{2}k^{-2}P_{L}(k)
\end{equation}
where $\mu$ is cosine of the angle to the line-of-sight, $f$ is the growth rate of structure, $a$ is the scale factor, $H$ is the Hubble parameter and $P^{p}_{L}(k)$, $P_{L}(k)$ are the linear momentum and matter power spectra. Multiplying by the relevant Legendre polynomials and integrating over the line-of-sight gives
\begin{equation}
P^{p}_{L,0}(k) = \frac{f^{2}(aH)^{2}}{3k^{2}}P_{L}(k), \qquad P^{p}_{L,2}(k) = \frac{2f^{2}(aH)^{2}}{3k^{2}}P_{L}(k), \qquad P^{p}_{L,\ell}(k) = 0\,\mathrm{for}\,\ell\ne0,2.
\end{equation}
We plot the linear theory predictions for the monopole and quadrupole of the momentum power spectrum in Fig.~\ref{fig:pkmomhalos}. The momentum power spectrum is a pure tracer of the growth rate of structure on large scales, independent of galaxy bias. This property makes it competitive as a probe of the growth rate, even given the typically small numbers of galaxies with direct PV measurements and the intrinsic error in each velocity measurement.
}
\item{On non-linear scales the momentum power spectrum becomes dominated by the power spectrum of the convolution between density and velocity. That is, the power spectrum arising from correlations between $\delta(\br)v(\br)$ and $\delta(\br')v(\br')$. This causes an increase in the momentum power spectrum above that expected from the velocity power spectrum and proportional to the halo/galaxy bias. This can be seen in Fig.~\ref{fig:pkmomhalos} as the excess in power for $k \gtrsim 0.1\hompc$, which is more evident for larger mass halos. As shown in Section~\ref{sec:conclusion} and Appendix~\ref{sec:appendix} this can be well modelled using perturbation theory.}
\end{itemize}

The final feature of note in Fig.~\ref{fig:pkmomhalos} is the departure from linear theory seen on large scales. This arises due to the window function imposed on the simulation when placing an observer. As with the galaxy density power spectrum, this window function convolves the observed power, increasing the correlation between modes and reducing the large scale power. Given the relatively shallow depth of PV surveys, we expect this to cause a significant reduction in the measure power spectra from real data and will need to be well modelled. Fortunately, the effect of a window function on the momentum power spectrum has already been derived in Eqs.~\ref{eq:convterms1}-\ref{eq:convterms}. In Section~\ref{sec:conclusion} we include this convolution using the same (matrix-multiplication) method as is commonly used for the galaxy density power spectrum.

\section{Covariance matrix of the momentum power spectrum} \label{sec:covariance}
Given the above estimator for the momentum power spectrum we can compute the covariance matrix of the momentum power spectrum as
\begin{equation}
C^{P}_{\ell\ell'}(k,k') = \langle |F^{p}_{\ell}(\bk)|^{2}|F^{p}_{\ell'}(\bk')|^{2} \rangle  - \langle |F^{p}_{\ell}(\bk)|^{2} \rangle \langle |F^{p}_{\ell'}(\bk')|^{2} \rangle.
\end{equation}
Substituting Eqs.~\ref{eq:momfield} and~\ref{eq:Fp} into this expression, we find
\begin{multline}
C^{P}_{\ell\ell'}(k,k') = \frac{(2\ell + 1)(2\ell' + 1)}{A^{4}V^{2}}\int \frac{d\Omega_{k}}{4\pi}\int \frac{d\Omega_{k'}}{4\pi} \int d^{3}r_{1} \int d^{3}r_{2}\int d^{3}r_{3} \int d^{3}r_{4} w(\br_{1})w(\br_{2})w(\br_{3})w(\br_{4}) L_{\ell}(\hat{\bk}\cdot\hat{\br_{2}})L_{\ell'}(\hat{\bk'}\cdot\hat{\br_{4}})\\ 
[\langle n_{g}(\br_{1})n_{g}(\br_{2})n_{g}(\br_{3})n_{g}(\br_{4})u(\br_{1})u(\br_{2})u(\br_{3})u(\br_{4}) \rangle -  \langle n_{g}(\br_{1})n_{g}(\br_{2})u(\br_{1})u(\br_{2}) \rangle \langle n_{g}(\br_{3})n_{g}(\br_{4})u(\br_{3})u(\br_{4}) \rangle]e^{i\bk\cdot(\br_{1}-\br_{2})}e^{i\bk\cdot(\br_{3}-\br_{4})}.
\label{eq:covp1}
\end{multline}
Although this expression is daunting, we can make some progress by rewriting $n_{g}(\br_{i})u(\br_{i}) = \bar{n}(\br_{i})(1 + \delta_{g}(\br_{i}))u(\br_{i}) = \bar{n}(\br_{i})\rho(\br_{i})$ (where $\delta_{g}(\br)$ is the galaxy overdensity) and assuming that the radial momentum field $\rho(\br_{i})$ is Gaussian. Although this assumption is certainly not true (even if the density and velocity fields are both Gaussian, their product would not be), on large scales the radial momentum field is simply proportional to the radial velocity. The large scale radial velocity \textit{is} approximately Gaussian in nature\footnote{Note, this is not to be confused with the pairwise velocity between objects, which is non-Gaussian \citep{Scoccimarro2004}, or the distribution of velocity amplitudes, which has a Maxwell-Boltzmann distribution \citep{ColesLucchin}} and so this gives credence to the assumption that the radial momentum field is too.

In this case, we can then use Wick's Theorem for a Gaussian random field to write
\begin{equation}
\langle n_{g}(\br_{1})n_{g}(\br_{2})n_{g}(\br_{3})n_{g}(\br_{4})u(\br_{1})u(\br_{2})u(\br_{3})u(\br_{4}) \rangle = \langle n_{g}(\br_{1})n_{g}(\br_{2})u(\br_{1})u(\br_{2})\rangle \langle n_{g}(\br_{3})n_{g}(\br_{4})u(\br_{3})u(\br_{4})\rangle + \mathrm{2cyc.},
\end{equation}
and substituting this into Eq.~\ref{eq:covp1},
\begin{equation}
C^{P}_{\ell\ell'}(k,k') = \frac{(2\ell + 1)(2\ell' + 1)}{A^{4}V^{2}}\int \frac{d\Omega_{k}}{4\pi}\int \frac{d\Omega_{k'}}{4\pi}\biggl[Q_{\ell\ell'}(\bk,\bk')Q^{*}_{00}(\bk,\bk') + Q_{0\ell'}(\bk,\bk')Q^{*}_{\ell0}(\bk,\bk')\biggl],
\label{eq:pkmomcov}
\end{equation}
where
\begin{align}
Q_{\ell\ell'}(\bk,\bk') &= \int d^{3}r \int d^{3}r' w(\br)w(\br')\langle n_{g}(\br)n_{g}(\br')u(\br)u(\br') \rangle L_{\ell}(\hat{\bk}\cdot\hat{\br})L_{\ell'}(\hat{\bk}'\cdot\hat{\br'})e^{i(\bk\cdot \br-\bk'\cdot \br')} \\
&= \int d^{3}r \int d^{3}r' w(\br)w(\br')\bar{n}(\br)\bar{n}(\br')\int\frac{d^{3}k''}{(2\pi)^{3}} \,P^{p}(\bk'',\br')L_{\ell}(\hat{\bk}\cdot\hat{\br})L_{\ell'}(\hat{\bk}'\cdot\hat{\br'})e^{i(\bk\cdot \br-\bk'\cdot \br'+\bk''\cdot(\br-\br'))} + M^{p}_{\ell\ell'}(\bk,\bk'),
\end{align}
and
\begin{equation}
M^{p}_{\ell\ell'}(\bk,\bk') = \int d^{3}\br\,w^{2}(\br)\bar{n}(\br)\langle v^{2}(\br)\rangle L_{\ell}(\hat{\bk}\cdot\hat{\br})L_{\ell'}(\hat{\bk}'\cdot\hat{\br'})e^{i(\bk-\bk')\cdot \br}.
\label{eq:mval}
\end{equation}

Finally, we substitute in Eq.~\ref{eq:park} and follow the steps in Eqs.~(33) and (34) of \cite{Blake2018}, neglecting the convolution with $P(\bk'',\br')$ which allows us to take this outside the integral over $\bk''$, and defining an effective momentum power spectrum
\begin{equation}
P^{p}_{\mathrm{eff}}(\bk,\bk',\br) = \frac{1}{2}\biggl[P^{p}(\bk,\br)+P^{p}(\bk',\br)\biggl] = \frac{1}{2}\sum_{\ell''}\biggl[P^{p}_{\ell''}(k)L_{\ell''}(\hat{\bk}\cdot\hat{\br}) + P^{p}_{\ell''}(k')L_{\ell''}(\hat{\bk}'\cdot\hat{\br})\biggl].
\end{equation}
Following these steps,
\begin{equation}
Q_{\ell\ell'}(\bk,\bk') = \int d^{3}r\,w^{2}(\br)\bar{n}^{2}(\br)P_{\mathrm{eff}}^{p}(\bk,\bk',\br)L_{\ell}(\hat{\bk}\cdot\hat{\br})L_{\ell'}(\hat{\bk}'\cdot\hat{\br})e^{i(\bk-\bk')\cdot \br} + M^{p}_{\ell\ell'}(\bk,\bk').
\label{eq:qval}
\end{equation}

From here, the covariance matrix of the momentum power spectrum in the Gaussian limit has the same form as the galaxy density power spectrum in \cite{Blake2018}. Hence, this can be computed for a given survey using the same decomposition into spherical harmonics, and a model for the momentum power spectrum and associated shot noise. Given that the theory and procedure is the same as that in \cite{Blake2018} it will not be repeated here. It is interesting however, to consider the case without a survey window function and constant variance in the velocity field (such that $\langle v^{2}(\br) \rangle$ is also independent of radius). In this limit, the weights, galaxy number density and velocity field variance must be independent of position and can be moved outside of the integral in Eqs~\ref{eq:mval} and~\ref{eq:qval}. Following some simplification, reverting to the distant-observer approximation (as opposed to the local plane-parallel approximation used to derive the previous expressions), and rewriting the equation in terms of $\mu$ instead of $\hat{\bk}\cdot\hat{\br}$, we find that the covariance matrix of the momentum power spectrum can be expressed
\begin{equation}
C^{p}_{\ell\ell'}(k,k') = \frac{(2\pi)^{3}}{V_{k}V}(2\ell+1)(2\ell'+1)\delta^{D}(k-k')\int_{-1}^{1}d\mu\,L_{\ell}(\mu)L_{\ell'}(\mu)\biggl[P^{p}(k,\mu)+\frac{\langle v^{2} \rangle}{\bar{n}}\biggl]^{2}, 
\label{eq:covgauss}
\end{equation}
i.e., in the same form as the galaxy density power spectrum covariance matrix \citep{Yamamoto2006,Taruya2010,Grieb2016}, where $V_{k}=4\pi k^{2}\Delta k$ and $\Delta k$ is the bin width.

We can demonstrate the validity of our analytic formulae by comparing to cubic simulations. We generated $256$ approximate dark matter simulations using the \textsc{l-picola} simulation code \citep{Howlett2015b} and the same simulation parameters as in Section~\ref{sec:estimator}. Halos are also identified in these simulations using the same procedure. From these simulations we computed the redshift space density and momentum power spectrum for halos in the mass range $12.4h^{-1}\mathrm{M_{\odot}} < M < 12.6h^{-1}\mathrm{M_{\odot}}$ using the z-axis of the simulation as the line-of-sight. The full covariance matrix for $P^{p}_{0}(k)$ and $P^{p}_{2}(k)$ as well as the galaxy density power spectrum multipoles was then computed from all 256 measurements. The use of a fixed line-of-sight for the whole simulation differentiates this from the results shown in Figure~\ref{fig:pkmomhalos}, and was chosen to remove the effect of the window function, which is expected to change the diagonal elements of the covariance matrix of both the density and momentum power spectrum away from the Gaussian prediction (see \citealt{Howlett2017d} for a detailed discussion of how a window function affects the galaxy density power spectrum covariance matrix). 

\begin{figure}
\centering
\subfloat{\includegraphics[width=0.5\textwidth]{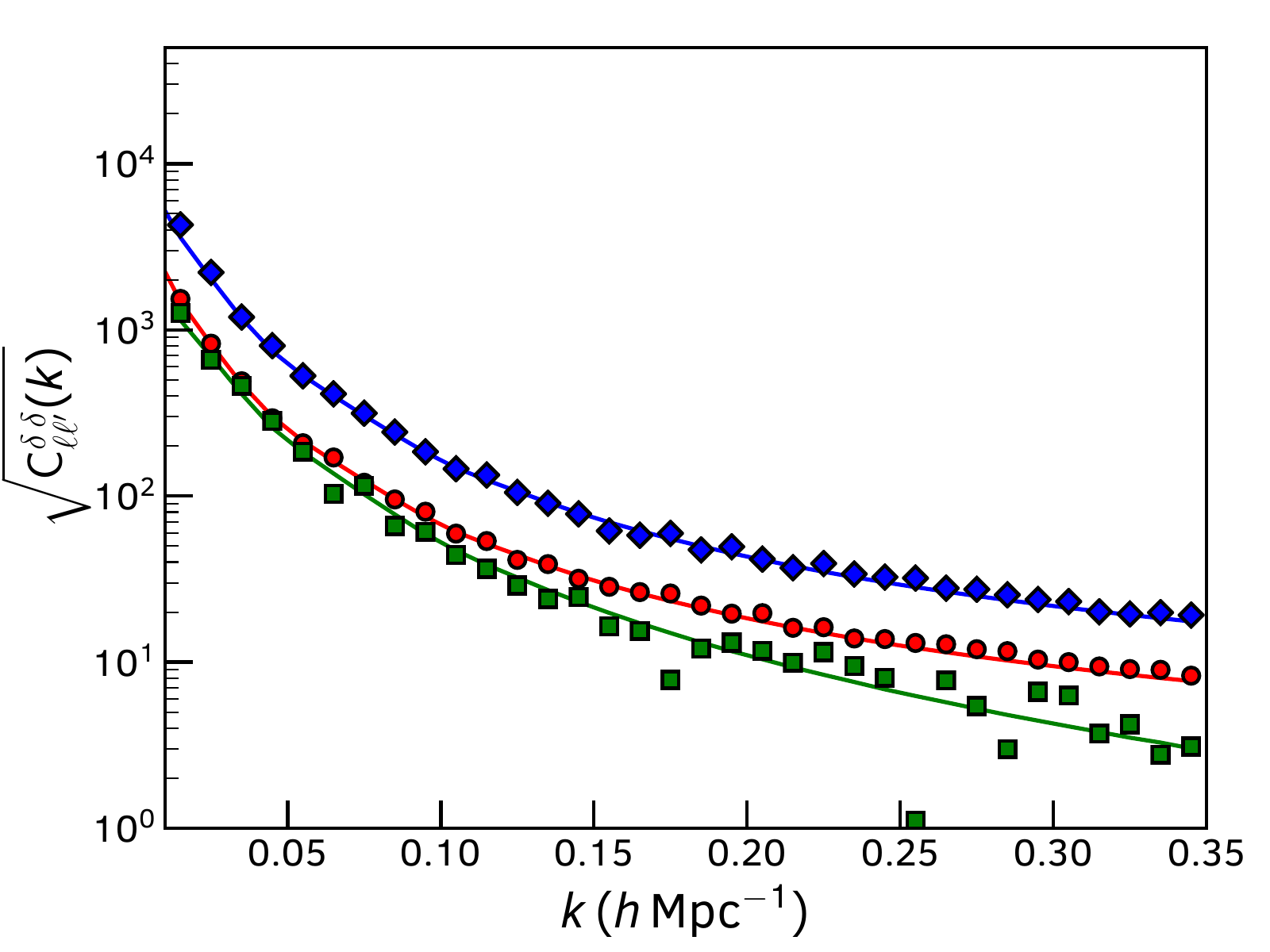}}
\subfloat{\includegraphics[width=0.5\textwidth]{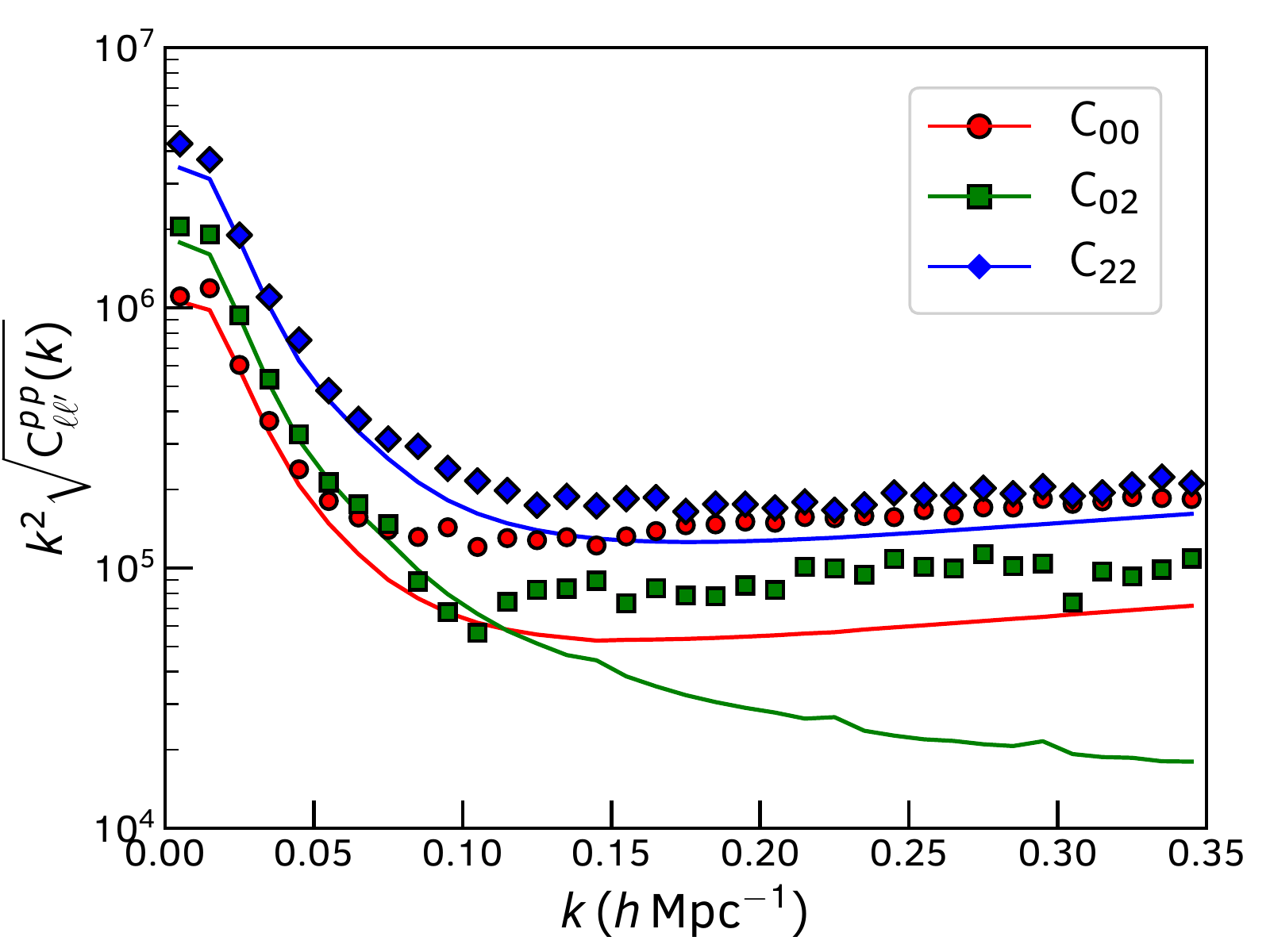}}
\caption{Measurements (points) of the standard deviation of the multipoles of the density (left) and momentum (right) power spectra along with the Gaussian prediction (lines). The variance of the momentum power spectrum multipoles is multiplied by $k^{2}$ to reduce the dynamic range. Different colours/symbols represent the variance of the monopole and quadrupole moments (denoted $C_{00}$ and $C_{22}$ respectively) and the cross-covariance between the monopole and quadrupole, denoted $C_{02}$.}
  \label{fig:pkcov}
\end{figure}

We calculated the Gaussian prediction using a model for the anisotropic power spectrum (both density and momentum) based on the averaged  measurements of the multipoles from the simulations, $P(k,\mu)=P_{0}(k)+L_{2}(\mu)P_{2}(k)$. This was then used as input to Eq.~\ref{eq:covgauss}, and the similar equation for the galaxy density power spectrum multipoles. Although the galaxy density power spectrum should contain higher order multipoles, even in the linear regime, these measurements are typically noise dominated and are not expected to contribute significantly to the Gaussian covariance matrix. Fig.~\ref{fig:pkcov} shows the measured diagonal elements of the covariance matrices of the density and momentum power spectra, as well as the cross-covariance between the monopole and quadrupole alongside these Gaussian predictions.

In all cases we see good agreement between the measurements and theory on the largest scales where the density and momentum fields are approximately Gaussian. However, the variance is underestimated as we go to smaller scales, where non-Gaussian terms enter the covariance matrices and the disagreement is much more severe for the momentum power spectrum. This is due to the contribution to the momentum power spectrum from the correlation between the density and velocity fields, which even in the case of Gaussian random fields would produce a non-Gaussian signature.

\begin{figure}
\centering
\subfloat{\includegraphics[width=0.5\textwidth]{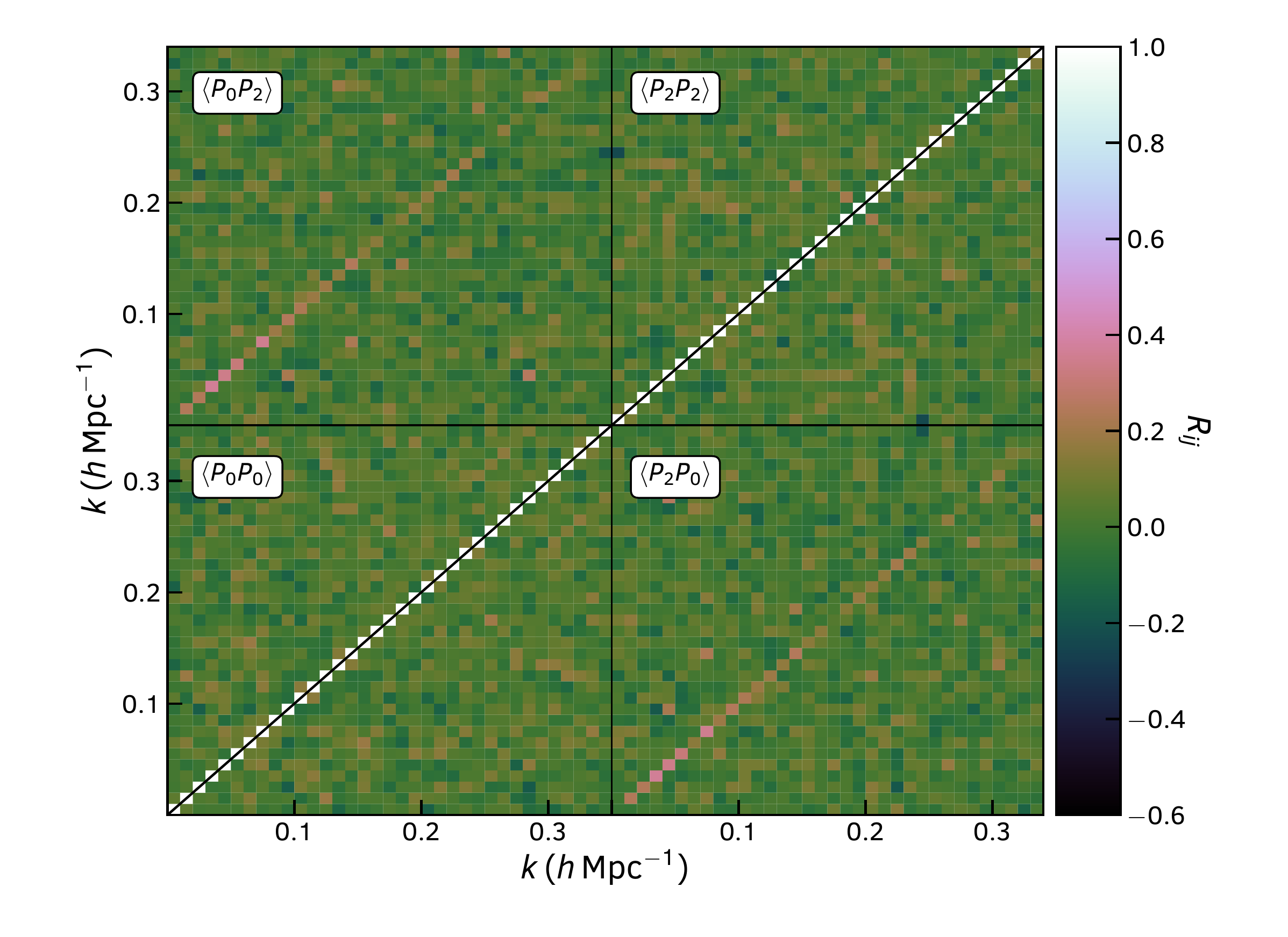}}
\subfloat{\includegraphics[width=0.5\textwidth]{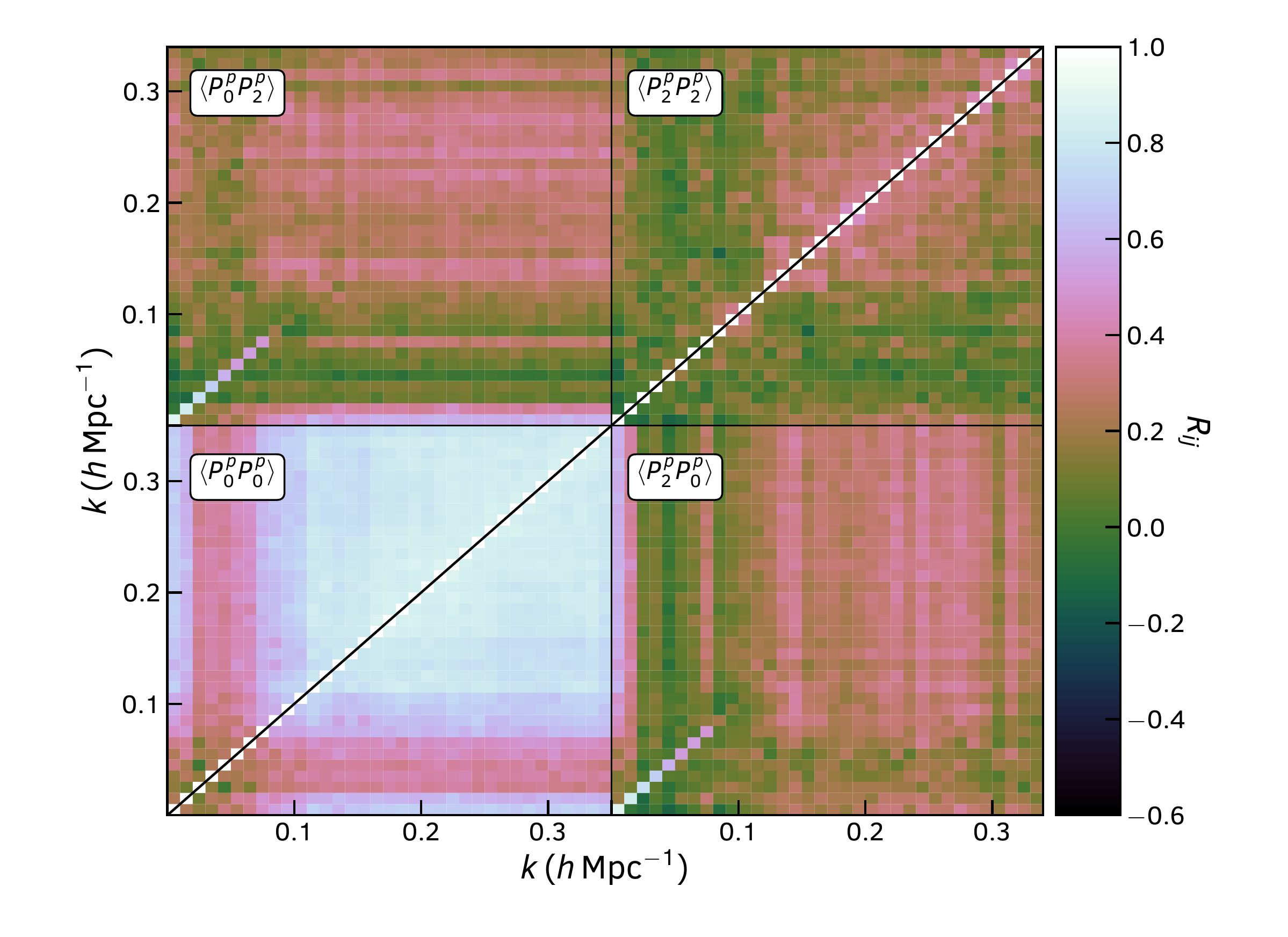}}
\caption{\textbf{\textit{Left:}} Measurements of the correlation matrix for the density power spectrum multipoles. In the lower left/top right quadrants we show the monopole and quadrupole correlations respectively. The top-left and lower-right then show the cross-correlation between the monopole and quadrupole. \textbf{\textit{Right:}} The same for the momentum power spectrum multipoles. Comparison of the left and right plots highlights the significant cross-correlation between bins for the momentum power spectrum compared to the density power spectrum.}
  \label{fig:pkcov2}
\end{figure}

Using the same simulations, in Fig.~\ref{fig:pkcov2}, we highlight portions of the correlation matrix $R_{ij}=C_{ij}/\sqrt{C_{ii}C_{jj}}$ for the density and momentum multipoles. We can see that the the cubic simulations show negligible off-diagonal covariance for the galaxy density power spectrum, as to be expected for halos in the absence of a window function and given the agreement between the diagonal elements and the Gaussian prediction. However, the momentum power spectrum shows significant correlation between modes on quasi-linear scales driven by the correlated nature of the density and velocity fields. This is particularly striking for the monopole, but reduced for the quadrupole moments. The cross-covariance between the two multipoles is also much more significant away from $k_{i}=k_{j}$ for the momentum power spectra compared to the galaxy density power spectrum. 

\begin{figure}
\centering
\includegraphics[width=0.48\textwidth, trim={0pt 0pt 0pt 20pt}]{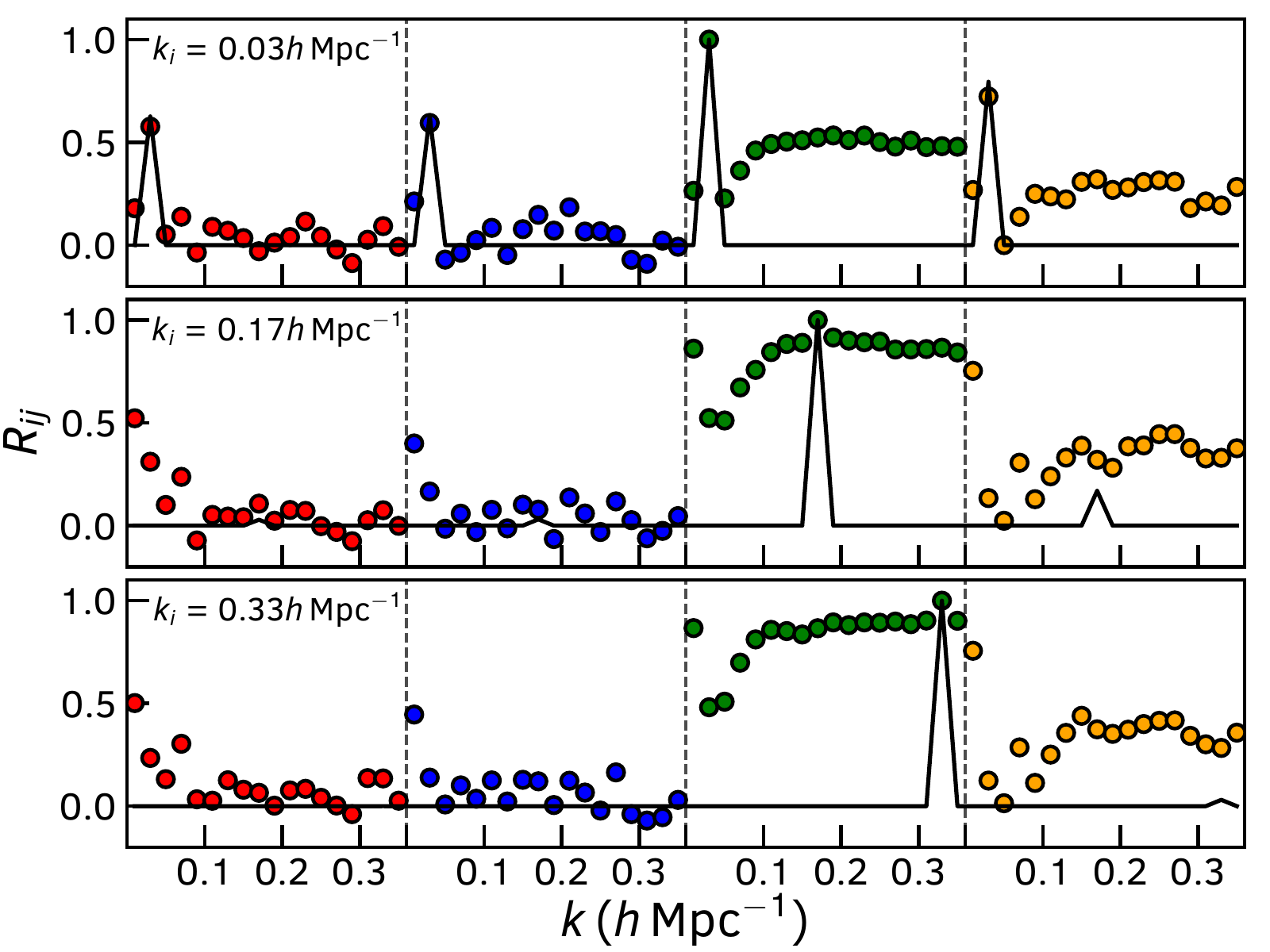}
\caption{Slices through the measured correlation matrix for the monopole moment of the momentum power spectrum compared to all multipoles of the density and momentum power spectrum at $k=[0.03,0.17,0.35] h^{-1}\mathrm{Mpc}$. Different parts of the correlation matrix are separated by dashed vertical lines, and from left to right correspond to the correlation between $P_{0}P^{p}_{0}$, $P_{2}P^{p}_{0}$, $P^{p}_{0}P^{p}_{0}$ and $P^{p}_{2}P^{p}_{0}$. The black solid line shows the Gaussian prediction, including for the cross-covariance between the density and momentum power spectra detailed in Section~\ref{sec:crosscovariance}. Whilst the Gaussian prediction is relatively accurate for the cross-covariance, there is significant correlation between the multipoles of the momentum power spectrum that is non-Gaussian in origin and arises from the convolution between density and velocity fields.}
  \label{fig:pkcov3}
\end{figure}

In Fig.~\ref{fig:pkcov3} we plot slices through the joint correlation matrix of the density and momentum power spectra alongside the Gaussian prediction of Eq.~\ref{eq:covgauss}. From this we see that the high degree of cross-covariance between bins for the momentum power spectrum is a result of non-Gaussian contributions to the covariance matrix of the momentum power spectra. However, we also find that the density and momentum power spectra are largely uncorrelated except when looking at measurements in the same $k$-bin. Given that the momentum field contains information from the density field, this is perhaps unexpected. This will be explored further in the next section, where we also demonstrate how we can compute a Gaussian prediction \textit{cross}-covariance matrix of the density and momentum power spectra.

\subsection{Cross-covariance between density and momentum power spectra} \label{sec:crosscovariance}

In order to combine measurements of the density and momentum power spectrum we need to account for potential cross-covariance between these two sets of measurements. Given a set of mock catalogues, these can be computed by brute force. However, we can also compute these in the Gaussian limit using the same method as for the covariance of the momentum power spectrum.

Starting from the weighted density \citep{Feldman1994} and momentum fields, we express the cross-covariance between density and momentum power spectra as
\begin{align}
C^{p\delta}_{\ell\ell'}(k,k') &= \langle |F^{p}_{\ell}(\bk)|^{2}|F_{\ell'}(\bk')|^{2} \rangle  - \langle |F^{p}_{\ell}(\bk)|^{2} \rangle \langle |F_{\ell'}(\bk')|^{2} \rangle \\
&= \frac{(2\ell + 1)(2\ell' + 1)}{A^{4}V^{2}}\int \frac{d\Omega_{k}}{4\pi}\int \frac{d\Omega_{k'}}{4\pi} \int d^{3}r_{1} \int d^{3}r_{2}\int d^{3}r_{3} \int d^{3}r_{4} w(\br_{1})w(\br_{2})w(\br_{3})w(\br_{4}) L_{\ell}(\hat{\bk}\cdot\hat{\br_{2}})L_{\ell'}(\hat{\bk'}\cdot\hat{\br_{4}}) \notag \\
&[\langle n_{g}(\br_{1})n_{g}(\br_{2})n_{g}(\br_{3})n_{g}(\br_{4})u(\br_{1})u(\br_{2}) \rangle -  \alpha\langle n_{g}(\br_{1})n_{g}(\br_{2})n_{g}(\br_{3})n_{r}(\br_{4})u(\br_{1})u(\br_{2}) \rangle \notag \\
&- \alpha\langle n_{g}(\br_{1})n_{g}(\br_{2})n_{r}(\br_{3})n_{g}(\br_{4})u(\br_{1})u(\br_{2})\rangle + \alpha^{2}\langle n_{g}(\br_{1})n_{g}(\br_{2})n_{r}(\br_{3})n_{r}(\br_{4})u(\br_{1})u(\br_{2}) \rangle \notag  \\
&- \langle n_{g}(\br_{1})n_{g}(\br_{2})u(\br_{1})u(\br_{2}) \rangle\langle (n_{g}(\br_{3})-\alpha n_{r}(\br_{3}))(n_{g}(\br_{4})-\alpha n_{r}(\br_{4}))\rangle] e^{i\bk\cdot(\br_{1}-\br_{2})}e^{i\bk\cdot(\br_{3}-\br_{4})}.
\label{eq:crosscovp1}
\end{align}
The large number of terms compared to the previous derivation arises due to the use of a random field in the calculation of the weighted density field which is not used for the weighted velocity field. Although this seems tedious to evaluate, we can again make headway using Wick's theorem, .e.g., 
\begin{multline}
\langle n_{g}(\br_{1})n_{g}(\br_{2})n_{g}(\br_{3})n_{g}(\br_{4})u(\br_{1})u(\br_{2}) \rangle = \langle n_{g}(\br_{1})n_{g}(\br_{2})u(\br_{1})u(\br_{2})\rangle \langle n_{g}(\br_{3})n_{g}(\br_{4})\rangle \\
+ \langle n_{g}(\br_{1})u(\br_{1})n_{g}(\br_{3})\rangle \langle n_{g}(\br_{2})u(\br_{2})n_{g}(\br_{4})\rangle + \langle n_{g}(\br_{1})u(\br_{1})n_{g}(\br_{4})\rangle \langle n_{g}(\br_{2})u(\br_{2})n_{g}(\br_{3})\rangle
\end{multline}
and similarly for the other terms. Substituting these into Eq.~\ref{eq:crosscovp1} results in a large cancellation of terms and a number of terms involving the correlation between the momentum field and random field at two different locations (which vanishes), such that all that remains is
\begin{multline}
C^{p\delta}_{\ell\ell'}(k,k') = \frac{(2\ell + 1)(2\ell' + 1)}{A^{4}V^{2}}\int \frac{d\Omega_{k}}{4\pi}\int \frac{d\Omega_{k'}}{4\pi} \int d^{3}r_{1} \int d^{3}r_{2}\int d^{3}r_{3} \int d^{3}r_{4} w(\br_{1})w(\br_{2})w(\br_{3})w(\br_{4}) L_{\ell}(\hat{\bk}\cdot\hat{\br_{2}})L_{\ell'}(\hat{\bk'}\cdot\hat{\br_{4}}) \\
[\langle n_{g}(\br_{1})u(\br_{1})n_{g}(\br_{3})\rangle\langle n_{g}(\br_{2})u(\br_{2})n_{g}(\br_{4})\rangle + \langle n_{g}(\br_{1})u(\br_{1})n_{g}(\br_{4})\rangle\langle n_{g}(\br_{2})u(\br_{2})n_{g}(\br_{3})\rangle]e^{i\bk\cdot(\br_{1}-\br_{2})}e^{i\bk\cdot(\br_{3}-\br_{4})}.
\end{multline}
Finally, we introduce the cross-correlation function between the density and momentum \textit{fields} $\xi^{p\delta}(\br-\br')$, which has Fourier counterpart $P^{p\delta}(\bk,\br')$, using
\begin{equation}
\langle n_{g}(\br)u(\br)n_{g}(\br') \rangle = \bar{n}(\br)\bar{n}(\br')\xi^{p\delta}(\br-\br'),
\end{equation}
where we have neglected shot-noise in the cross-correlation function. Then, following the same steps as in Section~\ref{sec:covariance}, we finally arrive at
\begin{equation}
C^{p\delta}_{\ell\ell'}(k,k') = \frac{(2\ell + 1)(2\ell' + 1)}{A^{4}V^{2}}\int \frac{d\Omega_{k}}{4\pi}\int \frac{d\Omega_{k'}}{4\pi}\biggl[Q^{p\delta}_{\ell\ell'}(\bk,\bk')Q^{p\delta,*}_{00}(\bk,\bk') + Q^{p\delta}_{0\ell'}(\bk,\bk')Q^{p\delta,*}_{\ell0}(\bk,\bk')\biggl],
\end{equation}
where
\begin{equation}
Q^{p\delta}_{\ell\ell'}(\bk,\bk') = \int d^{3}r\,w^{2}(\br)\bar{n}^{2}(\br)P^{p\delta}_{\mathrm{eff}}(\bk,\bk',\br)L_{\ell}(\hat{\bk}\cdot\hat{\br})L_{\ell'}(\hat{\bk}'\cdot\hat{\br})e^{i(\bk-\bk')\cdot \br}.
\end{equation}
In the limit of no window function and following the same procedure used to derive Eq.~\ref{eq:covgauss}, this reduces to 
\begin{equation}
C^{p\delta}_{\ell\ell'}(k,k') = \frac{(2\pi)^{3}}{V_{k}V}(2\ell+1)(2\ell'+1)\delta^{D}(k-k')\int_{-1}^{1}d\mu\,L_{\ell}(\mu)L_{\ell'}(\mu)[P^{p\delta}(k,\mu)]^{2},
\label{eq:crosscovgauss}
\end{equation}
i.e., the cross-\textit{covariance} between the density and momentum power spectra is a function of the cross \textit{power spectrum} between the two fields. This mimics the cross-covariance between two different measurements of the galaxy density power spectrum from different galaxy samples often considered in a multi-tracer analysis (i.e., \citealt{Smith2009,White2009}).

\begin{figure}
\centering
\subfloat{\includegraphics[width=0.5\textwidth]{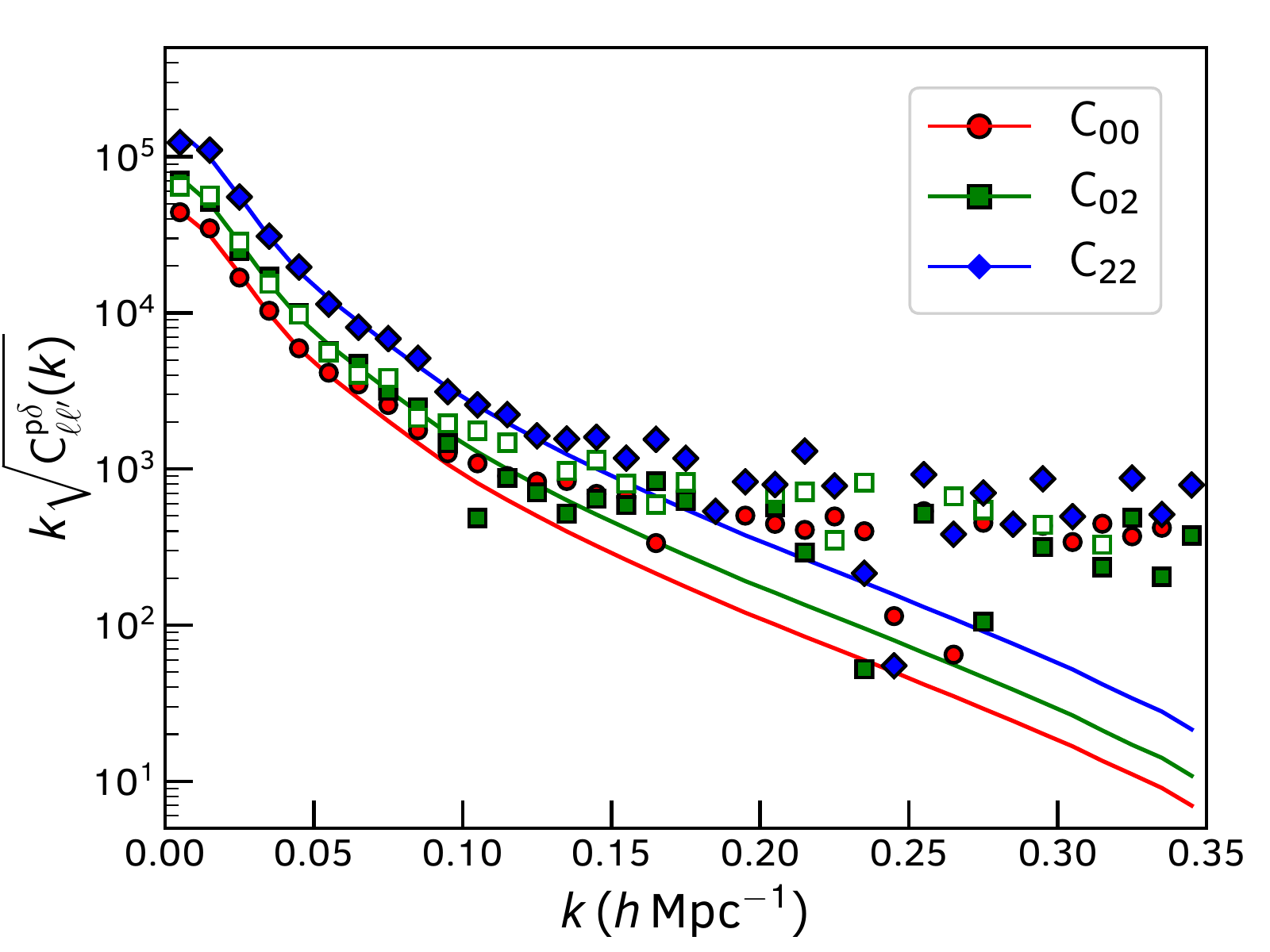}}
\subfloat{\includegraphics[width=0.5\textwidth]{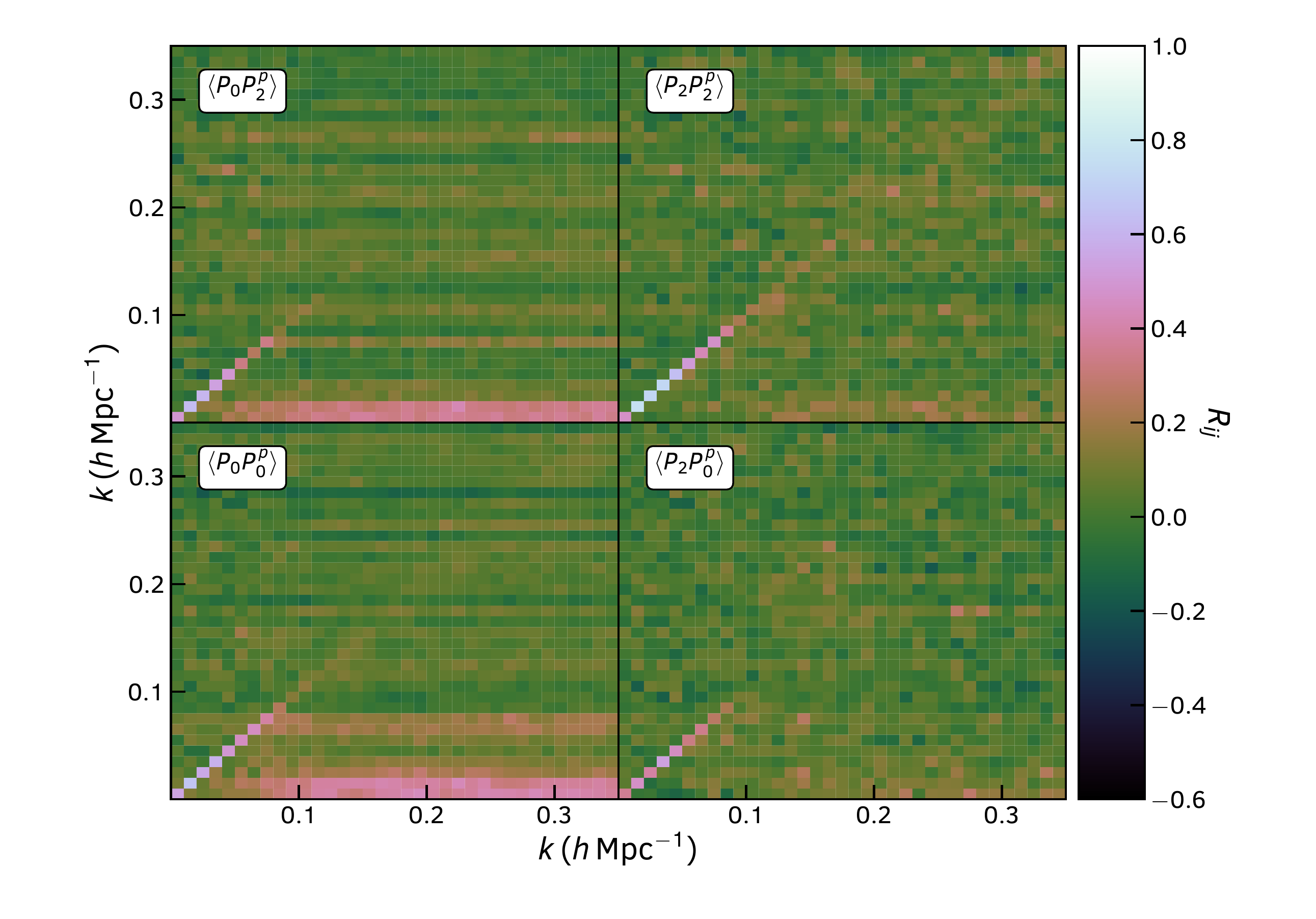}}
\caption{\textbf{\textit{Left:}} Measurements (points) of the square root of the cross-covariance between the density and momentum power spectra along with the Gaussian prediction (lines). The measurements and theory are multiplied by $k$ to reduce the dynamic range. Different colours/symbols represent the cross-covariance of the monopole and quadrupole moments (denoted $C_{00}$ and $C_{22}$ respectively) and the cross-covariance between the monopole of the momentum power spectrum and the quadrupole of the galaxy density power spectrum, denoted $C_{02}$. The open symbols show the reverse of this; the covariance between the momentum field monopole and the density field quadrupole, which in the Gaussian limit is equivalent. This is corroborated by the measurements. \textbf{\textit{Right:}} Measurements of the cross-correlation matrix for the density and momentum multipoles. Each quadrant shows a different portion of the full cross-correlation matrix between the four multipoles, and is labelled accordingly.}
  \label{fig:pkcrosscov}
\end{figure}

As before, we can use simulations to test this theory. We use the same halos as Section~\ref{sec:covariance} and measure the multipoles of the cross-power spectrum using the same procedure before computing the analytic cross \textit{covariance}. The results are shown in Fig.~\ref{fig:pkcrosscov}, where we plot the diagonal elements of the cross-covariance  of the density and momentum power spectra alongside the correlation matrix. Again, the theoretical result agrees well with the measurements on large-scales, and then underestimates the cross-covariance when non-Gaussian terms become important. This transition occurs on smaller scales than for the momentum power spectrum covariance matrix. In Gaussian theory, the cross-covariance between the monopole and quadrupole of the density and momentum fields respectively should be the same as the cross-covariance between the monopole and quadrupole of the momentum and density fields respectively, i.e., Eq.~\ref{eq:crosscovgauss} is invariant under the exchange of $\ell$ and $\ell'$. The measurements from the simulations agree with this prediction within noise.  In Fig.~\ref{fig:pkcrosscov}, we also show the cross-correlation matrix between the multipoles of the density and momentum fields. The full set of measurements is only correlated on large scales when $k_{i}=k_{j}$; on small scales, the large intrinsic variance in the velocity field (on the order of $150^{2}-300^{2}\mathrm{km^{2}\,s^{-2}}$; \citealt{Turnbull2012}) decorrelates the density and momentum power spectra. An alternative way of viewing this is that the $\langle v^{2}(\br)\rangle$ component of the shot-noise adds such a large component to the variance of the non-linear momentum power spectrum that the cross correlation coefficient, which is proportional to the inverse of this variance, becomes negligible.

As a final note, it is worth stating that we do not explore further the cross-power spectrum between the density and momentum fields in this work, using it only to test the covariance matrix of the auto-correlation of the density and momentum fields. Our forecasts for the mock catalogues used in Section~\ref{sec:proof} suggest this can improve the growth rate constraints by $\sim10\%$ or more over the already considerable improvement obtained from including the momentum power spectrum multipoles in addition to the galaxy density power spectrum multipoles. As such this is an interesting measurement in its own right and an important step toward fully realising the information in the combined velocity and density fields measured from PV surveys. However, measuring this in the presence of observational effects is significantly more involved than in the cubic simulations used in this sections and requires in-depth study. As such we elect to study this in future work and focus the remainder of this paper on the auto-power spectrum of the momentum field. It is important to realise that we do include the cross-\textit{covariance} between the density and momentum power spectra in Section~\ref{sec:proof}; it is only the measurement of the cross-\textit{spectrum} and the resulting cross-covariance between this measurement and the density and momentum power spectra that we defer to later work.

\section{Optimal weights} \label{sec:weights}
In the above derivation of our estimator for the momentum power spectrum and its covariance matrix we have purposely included a free weight parameter $w(\br)$. Following the standard set by \cite{Feldman1994} and \cite{Yamamoto2003,Yamamoto2006}, we can compute the value for this weight that minimises the variance in our estimation of the momentum power spectrum multipoles. To make an analytic solution for this weight tractable, we start by applying the distant observer approximation to Eq.~\ref{eq:pkmomcov} such that we can combine the $Q_{\ell,\ell}$ terms into a single integral over $\br$,
\begin{equation}
\langle \Delta P^{p}_{\ell}(k)^{2} \rangle = C^{P}_{\ell\ell}(k,k) = 2(2\ell + 1)^{2}\int \frac{d\Omega_{k}}{4\pi} \frac{\int d^{3}r\,w^{4}(\br)\bar{n}^{4}(\br) \biggl[P^{p}(\bk,\br) + \frac{\langle v^{2}(\br) \rangle}{\bar{n}(\br)}\biggl]^{2}L^{2}_{\ell}(\bk\cdot\br)}{\biggl[\int d^{3}r\,w^{2}(\br)\bar{n}^{2}(\br)\biggl]^{2}}
\label{eq:covweight}
\end{equation}
The optimal weights are then found by requiring this variance to be stationary with respect to a small change in weights $w(\br) \rightarrow w_{0}(\br)+\delta w(\br)$, .i.e, $\delta \langle \Delta P^{p}_{\ell}(k)^{2} \rangle/\delta w(\br) = 0$. This can be solved as in \cite{Feldman1994,Yamamoto2003,Yamamoto2006} using functional derivatives (see also \citealt{Smith2015} for a much more in depth discussion of how this is accomplished). Alternatively, we can save considerable time by direct comparison with these previous results, finding 
\begin{equation}
w(\br) = \frac{1}{\langle v^{2}(\br) \rangle+\bar{n}(\br)P^{p}(\bk,\br)},
\label{eq:weights}
\end{equation}
as the weight factor that minimises the variance\footnote{We did verify that performing the full derivation using functional derivatives gives the same solution!}. As with the FKP weights, this weighting scheme balances contributions from shot-noise, which in this case depends on both the number density of galaxies and the variance in the velocity field, and the power spectrum itself. In practice, we set the weights by fixing $\langle v^{2}(\br) \rangle$ based on the variance in the measured PVs in the data and choosing a constant value $P^{p}(\bk,\br)=P^{P}_{FKP}$, as is done for the usual FKP weights. We explore different choices for $P^{p}_{FKP}$ in Section~\ref{sec:mocks}, where we test our estimator on realistic mocks with varying number density.

The above weight factor minimises the variance of the estimator in the Gaussian limit. This is not necessarily the optimum choice of weights when including non-Gaussian contributions (which as shown in Section~\ref{sec:covariance} are significant for the momentum power spectrum), nor when optimising cosmological parameter constraints. In particular, one of the main uses of the momentum power spectrum will be constraining the growth rate of structure in combination with galaxy density power spectrum measurements. In this case, optimal weights could be obtained numerically following the method of \cite{Pearson2016b}. In Section~\ref{sec:proof} we use such a method to identify the optimum choices for the constants $P_{FKP}$ (used for the density power spectrum measurement) and $P^{p}_{FKP}$ (used for the momentum power spectrum measurement) by maximising the Fisher information (or equivalently minimising the ideal error) on $f\sigma_{8}$.

\section{Tests on realistic mocks} \label{sec:mocks}
In the remainder of this work, we focus on applying the new estimator for the momentum power spectrum to a set of realistic simulated peculiar velocity surveys (mocks) that have a sky coverage and redshift distribution representative of what could be achieved with next generation surveys. For our test case we consider a sample of Type IA supernovae measured with LSST \citep{Ivezic2008} matching the predictions of \cite{Howlett2017b,Kim2019}. We also incorporate errors on the PVs for each mock galaxy and look at the effects of a non-zero Bulk Flow on the measured power spectra.

The mocks were generated using the same approximate \textsc{l-picola} halo catalogues as in Section~\ref{sec:covariance}. Galaxies were added into these halo catalogues using a hybrid Halo Occupation Distribution (HOD, e.g., \citealt{Zheng2007}) and Subhalo Abundance Matching (SHAM, e.g., \citealt{Conroy2006}) approach. Full details of the construction of the mock catalogues is detailed in a companion paper (Qin et. al., in preparation, herein Paper II), where we use the same technique to produce mock galaxy catalogues as part of the application of this work to real data.

In brief, whilst the approximate nature of the \textsc{l-picola} simulations allows us to reproduce halos reasonably well (see \cite{Howlett2015a} for a comparison with fully non-linear N-Body simulations) it makes them unsuitable for identification of subhalos. Instead, motivated by measurements of the subhalo mass function \citep{Springel2008,Giocoli2008,Han2016,Elahi2018}, we add subhalos to halos by drawing from a power law. Subhalos are placed within the halos assuming an NFW profile; drawing from the inverse CDF for the positions \citep{Robotham2018} and using the virial theorem to then determine the satellite velocity as a function of position \citep{Lagos2018}. This prescription contains two free parameters, which are tuned to the galaxy density power spectrum in the data. For the mock catalogues in this paper, we use the parameters listed under the ``2MTF" sample in Paper II, which is chosen to roughly account for the idea that Type IA supernovae are preferentially found in star-forming galaxies \citep{Smith2012,Andersen2018}, which are also those used in the construction of the 2MTF sample.

Given a catalogue now containing central and satellite galaxies, we assign luminosities to these following \cite{Howlett2017c} and \cite{Qin2018}, drawing from the K-band luminosity function of \cite{Kochanek2001}. These are assigned to the halos and subhalos by rank ordering against their mass. We artificially introduce scatter into this matching to mimic the observed scatter between halo mass and luminosity by drawing the `matching' log luminosity from a Gaussian distribution where the standard deviation is an additional free parameter, also tuned to the data (i.e., in the same spirit as \citealt{Avila2018}). Note that this `matching' log luminosity is only used to assign the luminosities to the galaxies, and not for each galaxy's actual luminosity. 

Next, we place 8 maximally separated observers within each simulation, compute apparent magnitudes for the galaxies and apply a magnitude cut of $K\leq 14.75$, a redshift cut of $z\leq 0.2$ and keep only objects with declination $-80.0<\delta<15.0$, which roughly matches what could be observed with LSST. This returns a total number of objects $(\sim150,000)$ and a  redshift distribution similar to that proposed in \cite{Kim2019}. Finally, errors are assigned to the velocities of each galaxy using $0.08H_{0}r(z)$ where $r(z)$ is the comoving distance to the galaxy \citep{Howlett2017a,Howlett2017b}.

Optimal measurements of the momentum power spectra require an estimate of $\langle v^{2}(\br) \rangle$. As mentioned in Section~\ref{sec:estimator}, for real PV measurements this will be dominated by the measurement error on the individual galaxies, which typically increases as a function of distance from the observer. As such, we can obtain an estimate for this by fitting the errors in the data as a function of redshift or distance. For the mocks used herein we have assigned errors using the distance itself and so we simply adopt $\langle v^{2}(\br) \rangle = (0.08H_{0}r(z))^{2} + 300^{2} \mathrm{km^{2}\,s^{-2}}$, where the additional $300^{2} \mathrm{km^{2}\,s^{-2}}$ models the contribution from the intrinsic variance in the velocity field. This value is used for calculating the optimal weights and the window function of the data. For computing the shot-noise we use a sum over all galaxies (and their velocities) as is done for the shot-noise in the galaxy density power spectrum. 

\subsection{Effects of PV measurement errors}

\begin{figure}
\centering
\includegraphics[width=0.5\textwidth,trim=0mm 0mm 0mm 0mm]{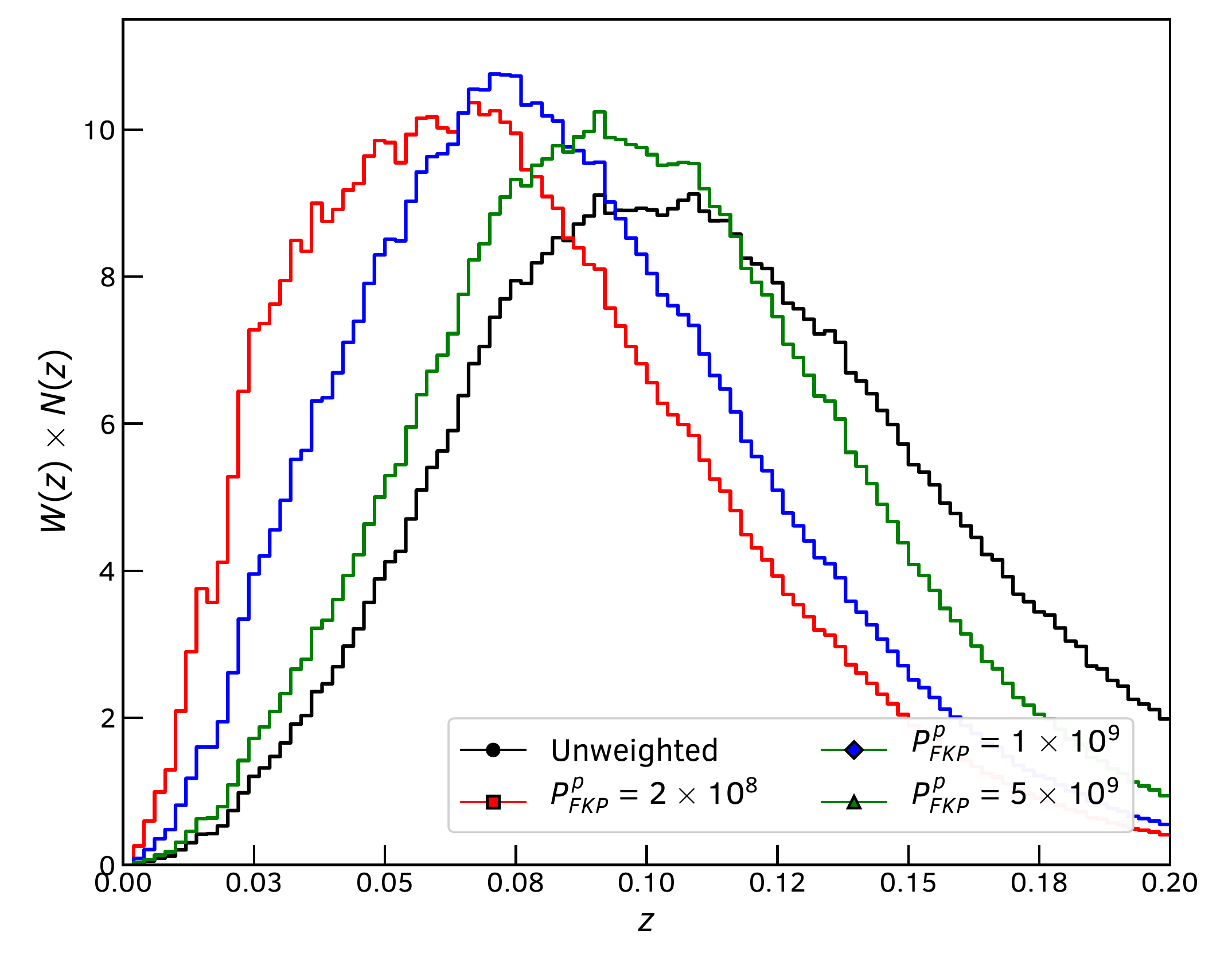}
\caption{The normalised, weighted number of mock galaxies in our realistic mocks as a function of redshift for different weights (unweighted and $P^{\,p}_{FKP}=[2\times10^{8},1\times10^{9},5\times10^{9}]\kmsmpcohV$). Different weights change the balance between upweighting higher redshift data to increase the cosmic volume (and hence decrease the power spectrum variance) and upweighting closer objects where the velocity errors are smaller.}
  \label{fig:mocktest}
\end{figure}

\begin{figure}
\centering
\subfloat{\includegraphics[width=0.5\textwidth,trim=0mm 0mm 0mm 0mm]{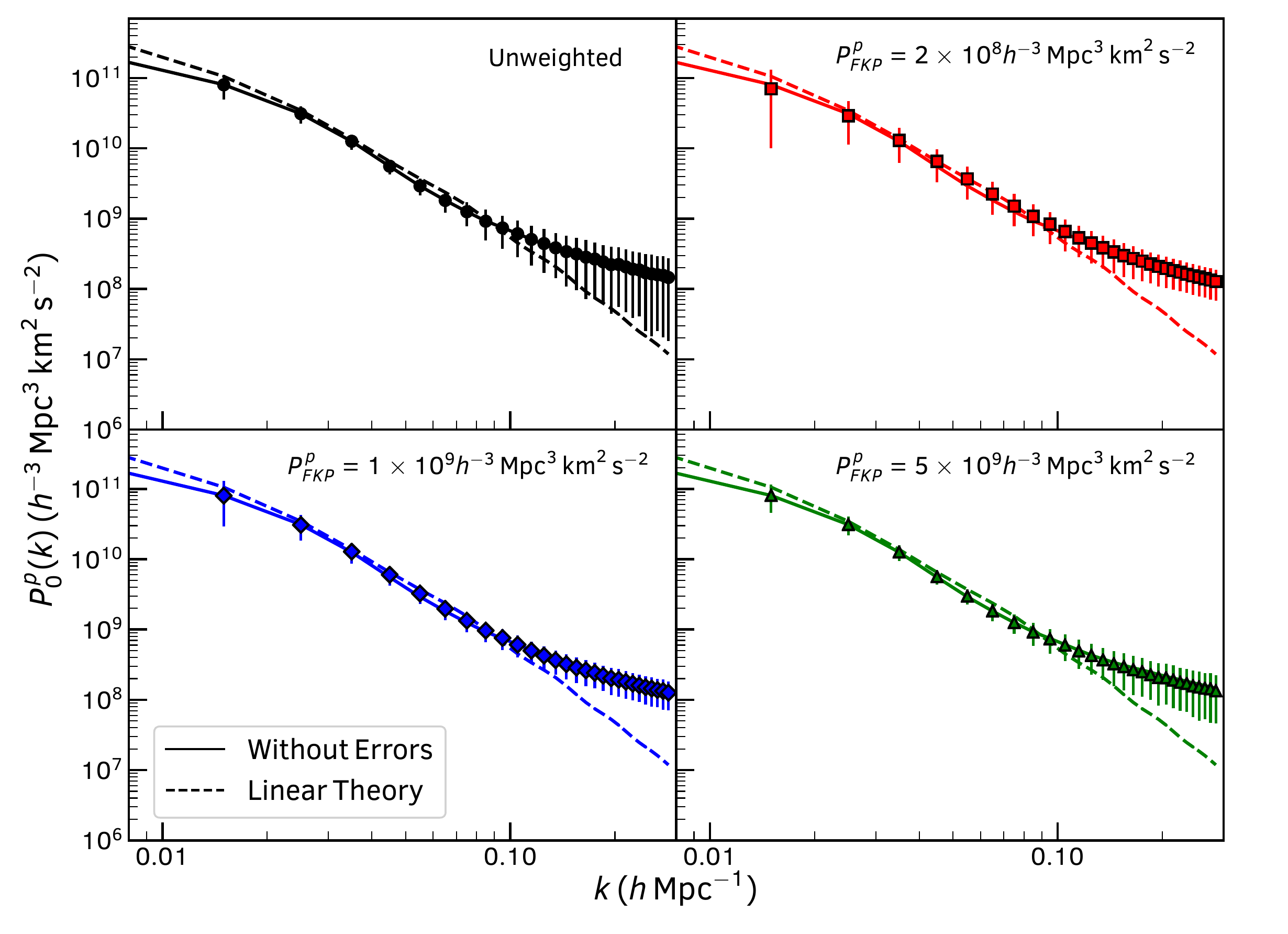}}
\subfloat{\includegraphics[width=0.5\textwidth,trim=0mm 0mm 0mm 0mm]{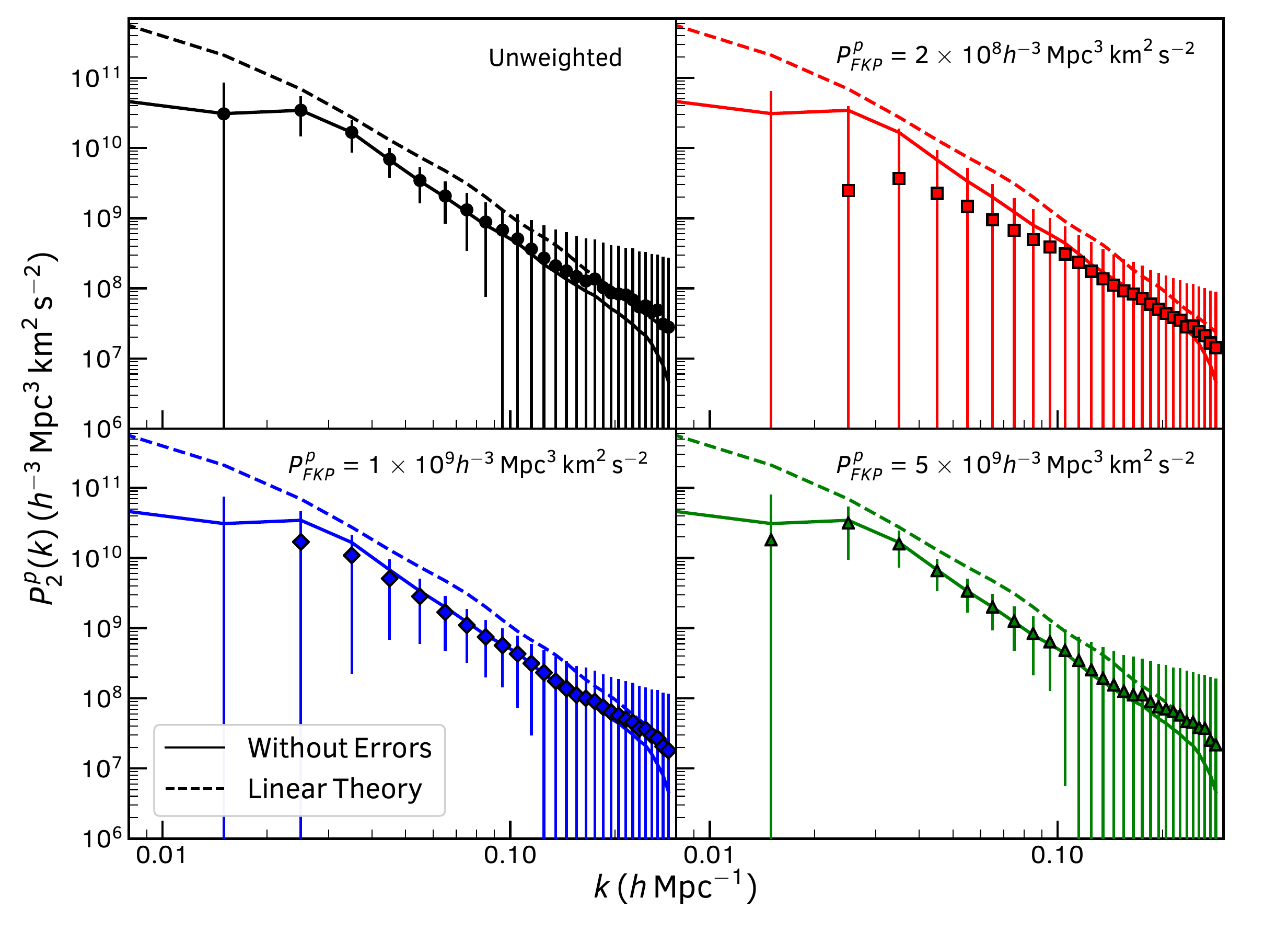}}
\caption{\textit{\textbf{Left:}} Measurements of the monopole of the momentum power spectrum averaged over all mocks, with errors calculated from the variance in the mocks. The points in each panel show a different choice for the weights: unweighted and $P^{\,p}_{FKP}=[2\times10^{8},1\times10^{9},5\times10^{9}]\kmsmpcohV$. The solid lines show the same measurements without any errors on the PVs. The dashed line is the linear theory model. Both of these are unaffected by the choice of weights. \textit{\textbf{Right:}} The same for the quadrupole of the momentum power spectrum. Different weighting schemes result in consistent average measurements but change the relative errors on different scales. The disagreement between the quadrupole measurements and linear theory is a result of non-linear redshift-space distortions and the survey window function (see Section~\ref{sec:proof}).}
  \label{fig:mocktest2}
\end{figure}

\begin{figure}
\centering
\subfloat{\includegraphics[width=0.33\textwidth,trim=0mm 0mm 0mm 0mm,clip]{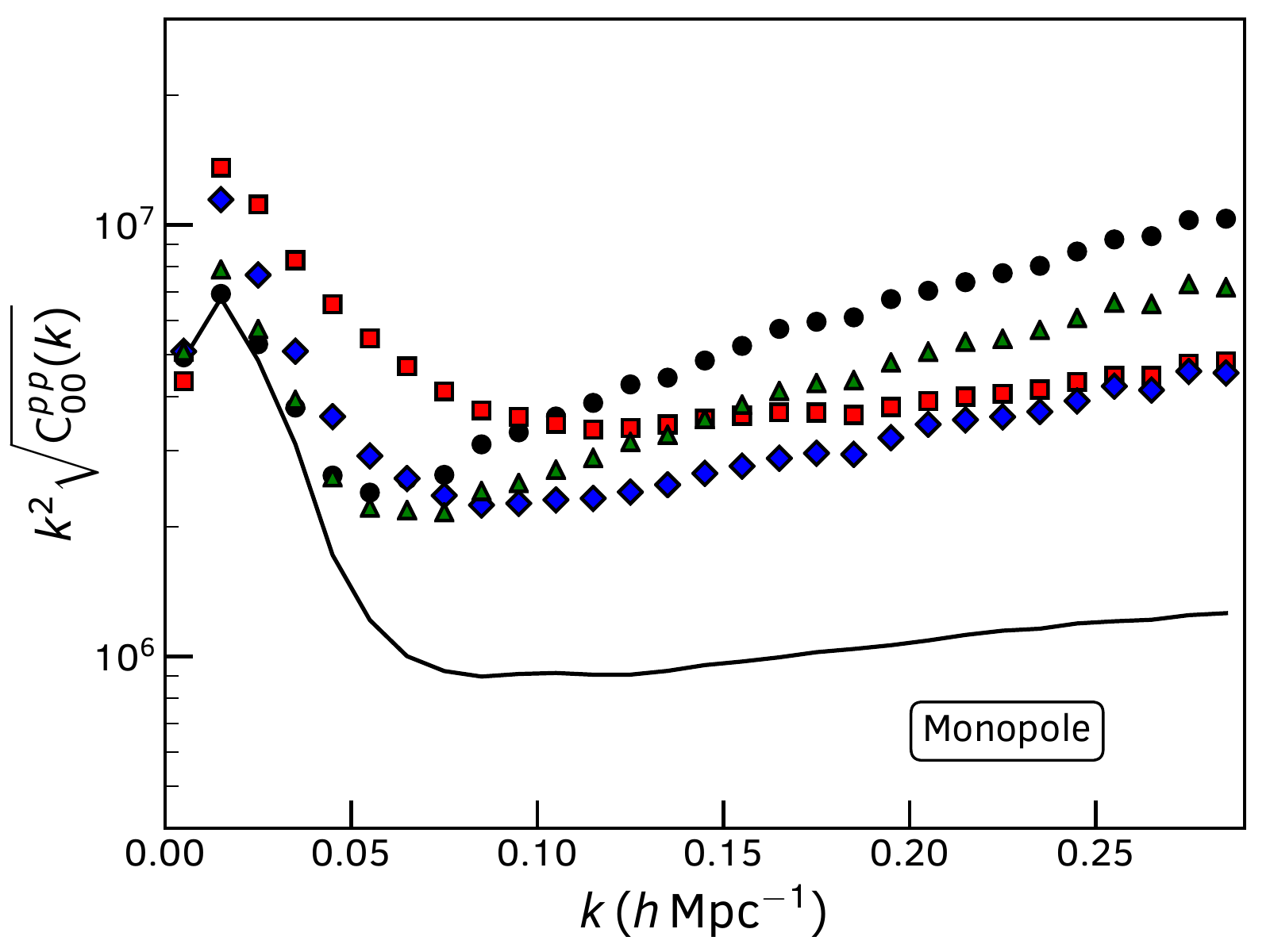}}
\subfloat{\includegraphics[width=0.33\textwidth,trim=0mm 0mm 0mm 0mm,clip]{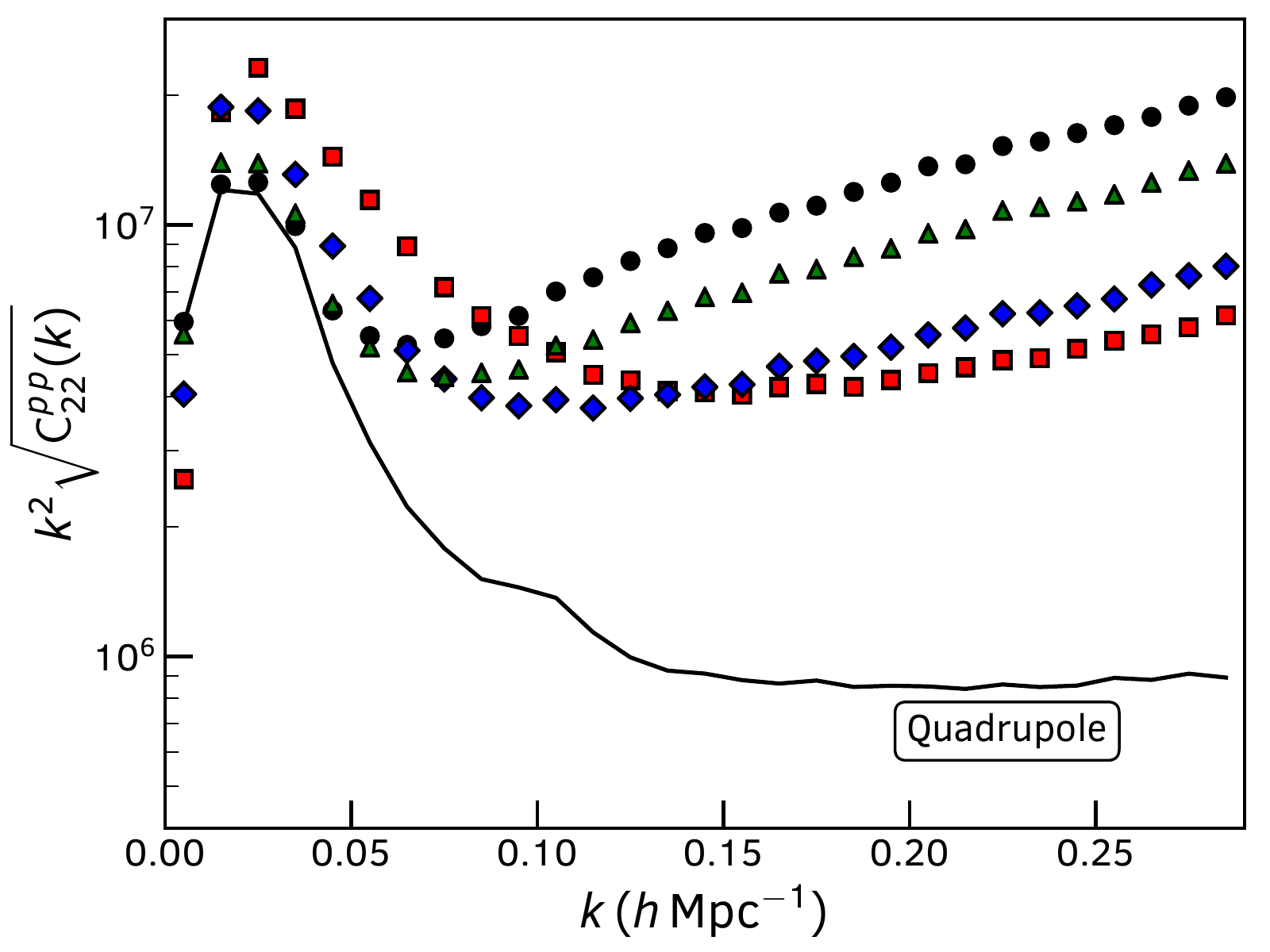}}
\subfloat{\includegraphics[width=0.33\textwidth,trim=0mm 0mm 0mm 0mm,clip]{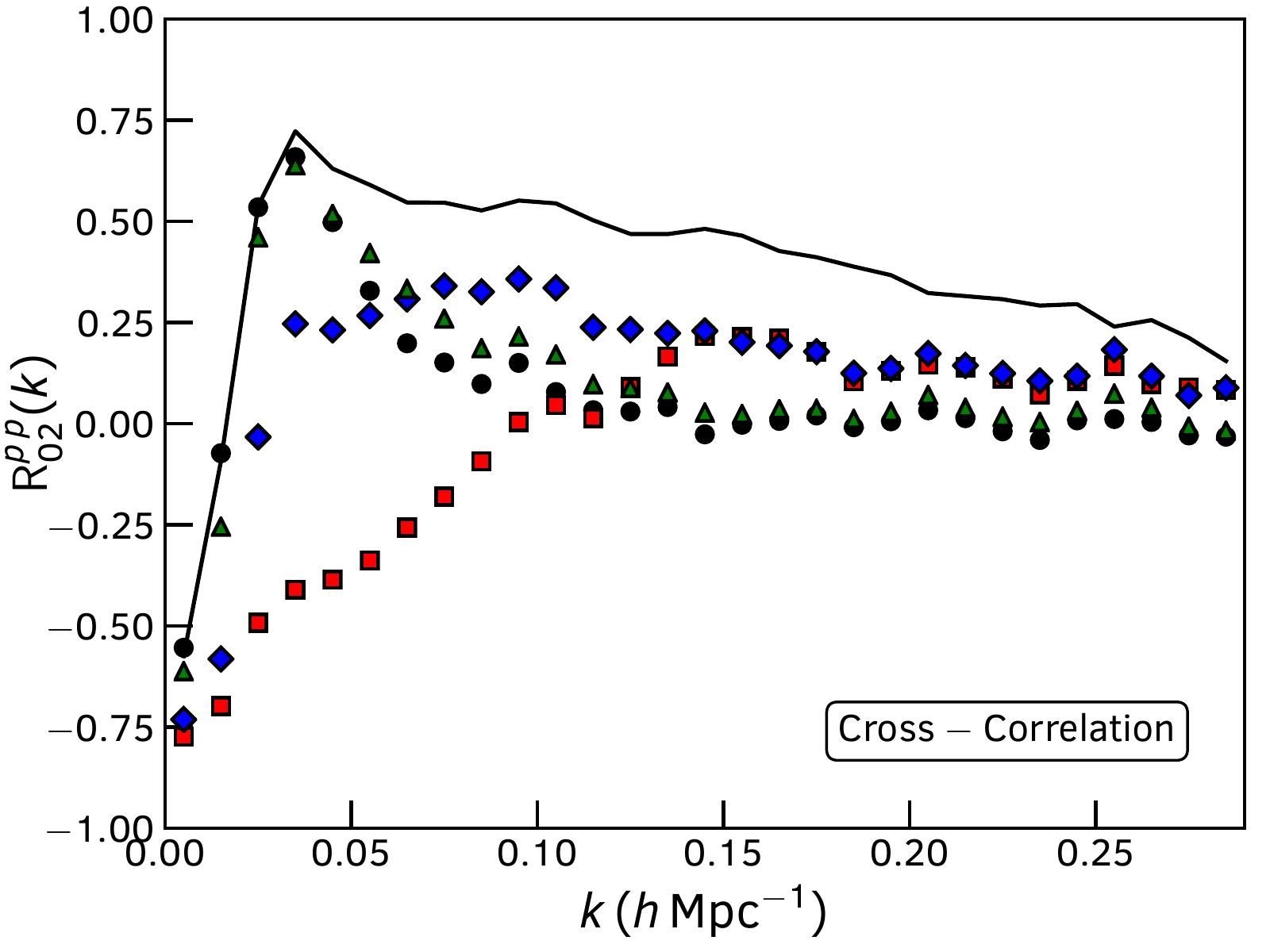}}
 \caption{\textit{\textbf{Left:}} The errors (diagonal elements of the covariance matrix) on the monopole of the momentum power spectrum for four different weighting schemes (unweighted and $P^{\,p}_{FKP}=[2\times10^{8},1\times10^{9},5\times10^{9}]\kmsmpcohV$) are plotted as points. The diagonal elements of the covariance matrix for the momentum power spectrum without any PV measurement errors are shown as a solid line. \textit{\textbf{Middle:}} The same for the quadrupole moment of the momentum power spectrum. \textit{\textbf{Right:}} The (diagonal) cross-correlation between the monopole and quadrupole of the momentum power spectrum for different weighting schemes. In all cases we see that different choices of weighting change the relative errors on different scales for the momentum power spectra, which indicates a particular choice should be made depending on the scales we wish to recover most accurately.}
\label{fig:mocktest3}
\end{figure}

\begin{figure}
\centering
\includegraphics[width=0.5\textwidth,trim=0mm 0mm 0mm 0mm]{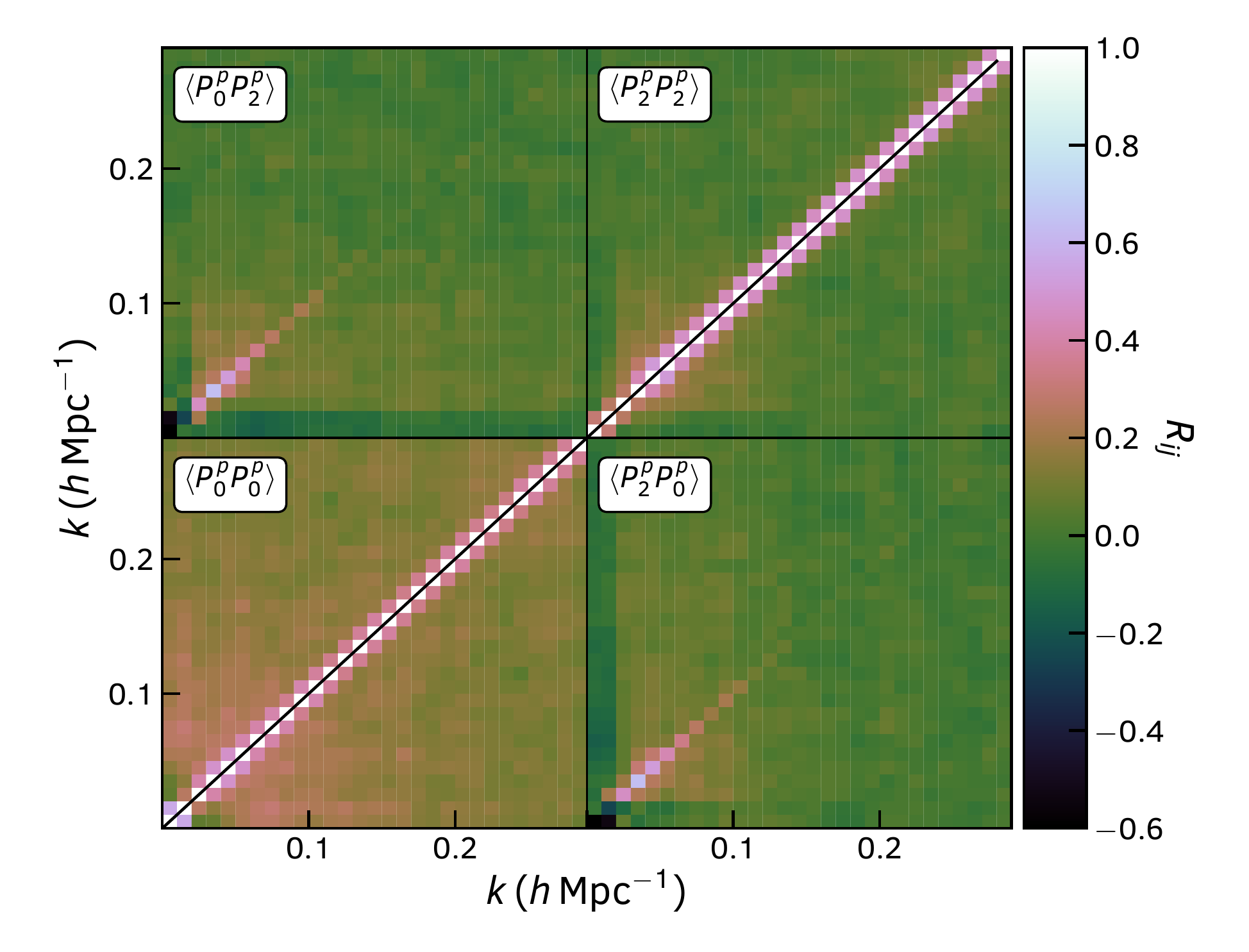}
\caption{The correlation matrix for the monopole and quadrupole of the momentum power spectrum in the presence of errors on the peculiar velocities and with weights calculated using $P^{\,p}_{FKP}=5\times10^{9}\kmsmpcohV$. Lower-left and upper-right quadrants show the correlation for the monopole and quadrupole respectively. The two other quadrants show the cross-correlation between the monopole and quadrupole. Measurement errors on the peculiar velocities increase the diagonal covariance of the monopole and quadrupole of the momentum power spectrum by up to an order of magnitude while keeping the off-diagonal terms largely unchanged. This results in a much more diagonal correlation structure between different $k$-bins compared to the case without PV measurement errors (which can be seen in the right-panel of Fig.~\ref{fig:pkcov2}).}
  \label{fig:mocktest4}
\end{figure}

We begin by looking at the effect of PV measurement errors on the measurements of the momentum power spectrum. For our mocks we find we are able to produce robust measurements of both the monopole and quadrupole, although this may not be true for current or future surveys depending on the cosmological volume, shot noise or velocity errors. We do not explore higher order multipoles as these are expected to be zero from linear theory and were found to be completely noise-dominated in our mocks. Figures~\ref{fig:mocktest}-\ref{fig:mocktest4} show a series of measurements without the inclusion of measurement errors and also when including errors and four different weighting schemes; unweighted, and $P^{\,p}_{FKP}=[2\times10^{8},1\times10^{9},5\times10^{9}]\kmsmpcohV$. 

Turning first to Fig.~\ref{fig:mocktest} we see that different choices for $P^{\,p}_{FKP}$ change the effective depth of the measurement, in the same way as the standard FKP weights. However, the FKP weights typically (relatively) upweight more distant regions of the survey where the number density of objects is lower, increasing the effective cosmological volume and decreasing the variance in the power spectrum. Instead, the optimal weights for the momentum power spectrum have to also account for the fact that more distant objects have larger errors, and hence larger $\langle v^{2}(\br) \rangle$. This means that decreasing the variance is not as simple as upweighting more distant objects.

Next, looking at the monopole and quadrupole power spectra for the different cases (Fig~\ref{fig:mocktest2}), we see that the inclusion of PV errors does not significantly change the mean power spectrum measured from the mocks. For all choices of weights the measurements with errors are almost indistinguishable from those without for the monopole. There are some differences for the quadrupole depending on the choice of $P^{p}_{FKP}$, however on large scales we attribute this to the fact that the distribution of quadrupole measurements can become non-Gaussian especially for the case of $P^{p}_{FKP}=2\times 10^{8}\kmsmpcohV$ given the emphasis this places on lower redshift measurements (a point which will be explored later and is a discussed at length in Paper II when measurements are made from real data). Overall, the measurements match what we would expect; little change in the power spectrum in the case of Gaussian distributed peculiar velocity errors, which, whilst included in our mocks by design, are also obtainable for modern PV surveys\footnote{Modern PV surveys such as 2MTF \citep{Hong2014} or 6dFGSv \citep{Campbell2014,Springob2014} typically contain measurements of the log-distance ratio, which is (for the most part) Gaussian distributed. Whilst the velocities are then generally log-normally distributed, we can obtain Gaussian velocities if we make use of the \cite{Watkins2015} estimator.}. A subtle point of note is the change in the amplitude of the non-linear momentum power spectrum for different weightings, which arises due to different weights changing the effective galaxy bias of the sample, which in turn contributes to the portion of the momentum power spectrum arising from the convolution between the density and velocity fields.

Whilst we believe that the presence of Gaussian distributed PV errors should not bias the power spectrum measurements, what they do do is significantly increase the variance of the power spectrum, acting as additional shot-noise. The introduction of PV errors also adds random non-linear noise to the momentum power spectrum, significantly reducing the correlation between the monopole and quadrupole and between different scales. Although not highlighted here, the cross-correlation between the momentum and galaxy density power spectra is also reduced. These effects can be seen in Figs~\ref{fig:mocktest3} and~\ref{fig:mocktest4}, where regardless of the weighting scheme the errors on the monopole and quadrupole are increased by up to an order of magnitude due to the PV errors whilst the cross-correlation coefficient between the monopole and quadrupole at fixed scale is reduced. The reduction in the correlation between different scales in the monopole and quadrupole can be best seen by comparing Fig.~\ref{fig:mocktest4} with the right-hand panel of Fig,~\ref{fig:pkcov2}. It's worth noting that whilst the correlation is reduced between Figs~\ref{fig:mocktest4} and~\ref{fig:pkcov2}, the overall amplitude of the off-diagonal covariance is not; the window function actually increases the off-diagonal covariance in the mocks compared to the cubic simulations used in previous sections, but the inclusion of uncorrelated PV errors then adds so substantially to the diagonal covariance that the net effect is a large reduction in \textit{correlation}.    

Finally, although the effect of PV errors matches our expectations, different choices for $P^{\,p}_{FKP}$ in the presence of these errors change the relative variance of linear and non-linear scales and the cross-correlations in a complex way. The overall take-away from this is that whilst we have derived the optimal weights for the momentum power spectrum, the fact that it is common to choose a fixed $P^{p}(k)=P^{\,p}_{FKP}$ in evaluating these weights means some choices for $P^{\,p}_{FKP}$ will be better than others depending on how the momentum power spectrum is to be used. In particular, for measuring the growth rate of structure it may be more beneficial to minimize the errors on linear scales, whilst sacrificing the constraining power on small scales where galaxy bias becomes a nuisance parameter. We explore this further in Section~\ref{sec:proof} and Appendix~\ref{sec:appendixB} where we identify the choice of $P^{\,p}_{FKP}$ that maximises the information on $f\sigma_{8}$ obtainable from our mocks. 

\subsection{Effects of non-zero Bulk Flow} \label{sec:bulkflow}

Another feature of real PV surveys is the presence of a non-zero Bulk Flow. Physically, this arises due to our own motion within the cosmic web, and is equivalent in our notation to a non-zero mean velocity, $\langle v(\br) \rangle \neq 0$. A \textit{measured} Bulk Flow (even if the true Bulk Flow is zero) can also arise due to non-uniform survey geometry and sampling of the velocity field \citep{Andersen2016}. It is interesting to ask what effect this has on the measured momentum power spectrum. To test this, we measure the momentum power spectrum monopole in our mock catalogues for a single weighting scheme ($P^{\,p}_{FKP}=2.4\times10^{9}\kmsmpcohV$) after removing the observers Bulk Flow from each mock galaxy. Although modern, sophisticated methods for this exist \citep{Nusser2011,Peery2018,Qin2018}, we do this simply by first estimating the 3-dimensional Bulk Flow $\boldsymbol{B}$ using the standard Maximum Likelihood estimator \cite{Kaiser1987} (which is valid given how our measured PVs are largely Gaussian) and then subtracting this component from the radial velocity of each galaxy using $u(\br) \rightarrow u(\br) - \boldsymbol{B}\cdot\hat{\br}$. 

\begin{figure}
\centering
\subfloat{\includegraphics[width=0.33\textwidth,trim=0mm 0mm 0mm 0mm,clip]{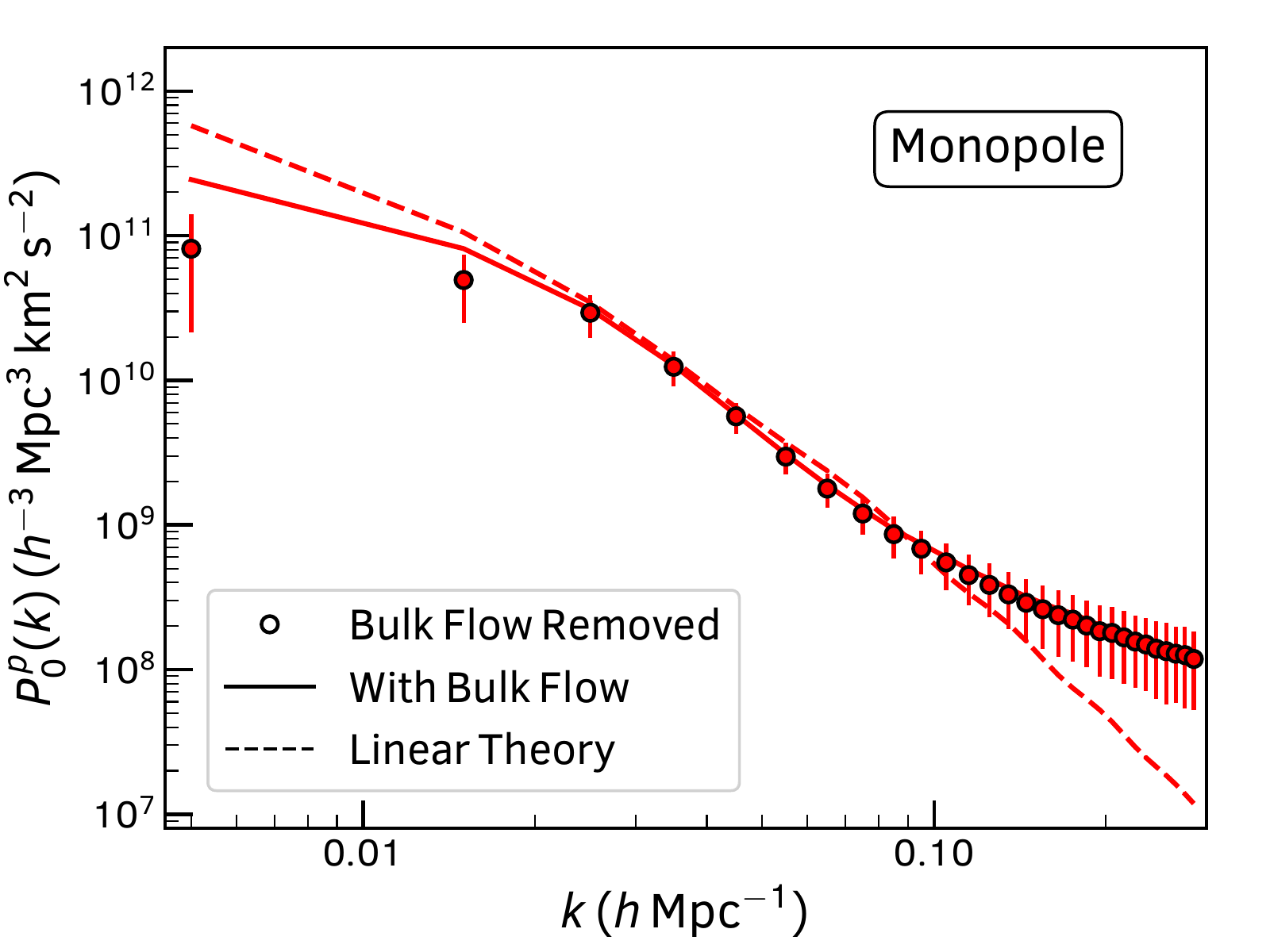}}
\subfloat{\includegraphics[width=0.33\textwidth,trim=0mm 0mm 0mm 0mm,clip]{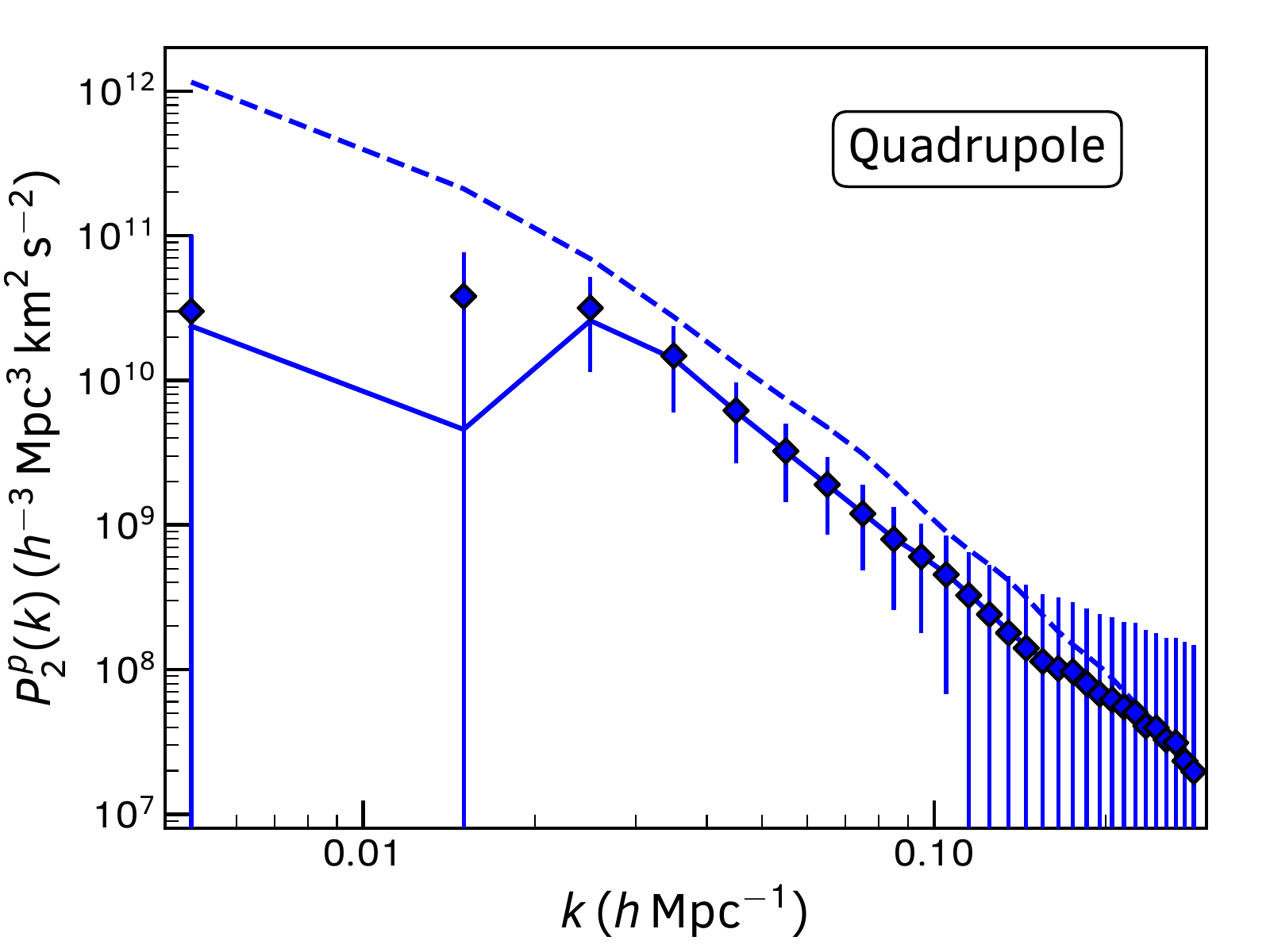}}
\subfloat{\includegraphics[width=0.33\textwidth,trim=0mm 0mm 0mm 0mm,clip]{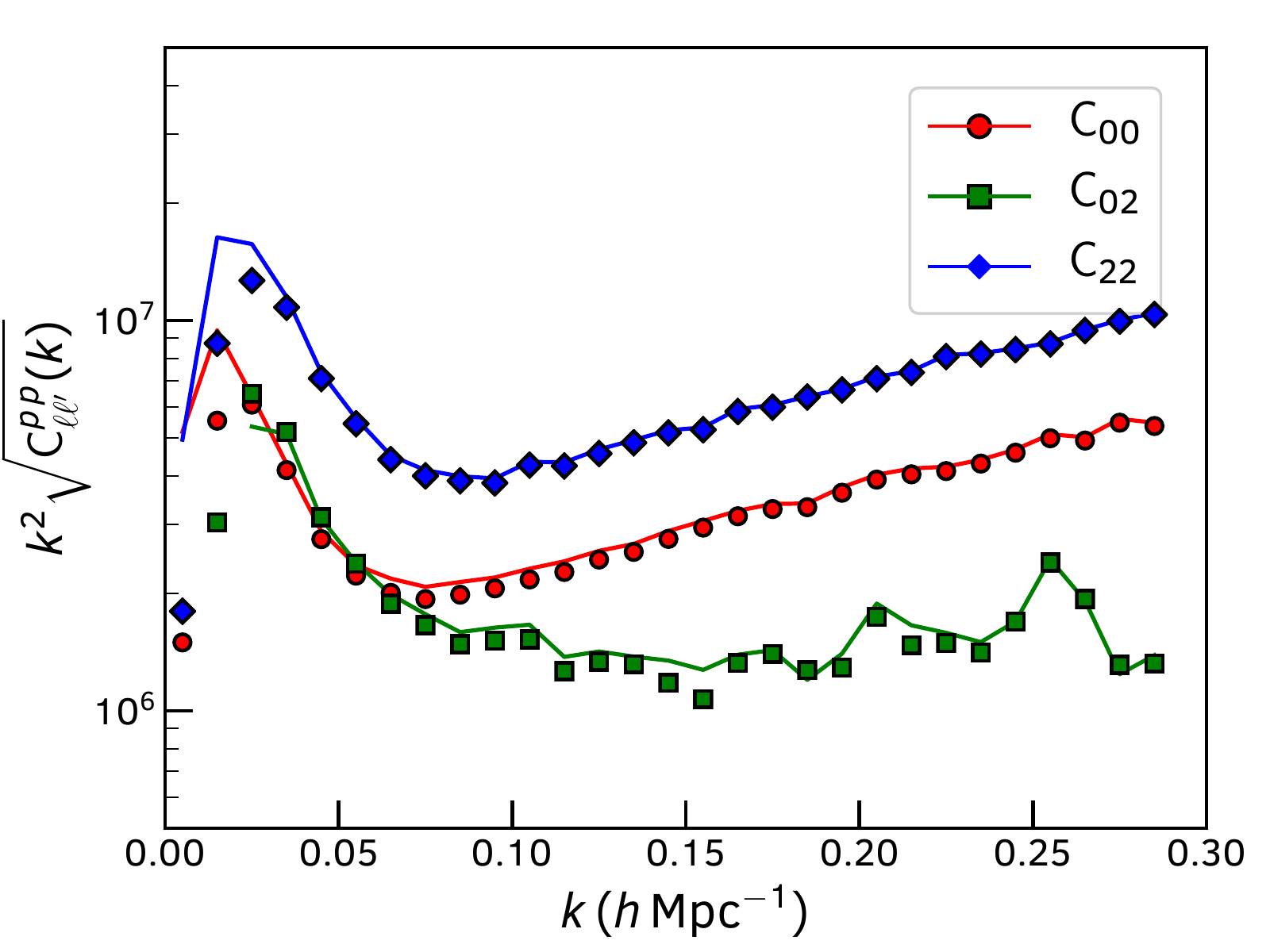}}
\caption{Measurements of the momentum power spectrum monopole and quadrupole from mock galaxy catalogues with and without removing the observer's Bulk Flow. In all plots, points show measurements after removing the Bulk Flow whilst solid lines show the measurements when this is kept in. \textit{\textbf{Left:}} Measurements of the monopole of the momentum power spectrum. The dashed line is the predicted power spectrum from linear theory.  \textit{\textbf{Middle:}} Measurements of the quadrupole of the momentum power spectrum. The dashed line is the predicted power spectrum from linear theory. \textit{\textbf{Right:}} The diagonal elements of the covariance matrix for the monopole and quadrupole (denoted $C_{00}$ and $C_{22}$ respectively) of the momentum power spectrum and the diagonal elements of the cross-covariance between these (denoted $C_{02}$). The only notable effect of removing the Bulk Flow is a reduction in the monopole of the power spectrum on large scales and a reduction in the variance and cross-covariance of the monopole and quadrupole on similar scales.}
  \label{fig:mockBF}
\end{figure}

The results for this test are shown in Figure~\ref{fig:mockBF}. We find that removing the Bulk Flow reduces the large scale monopole of the momentum power spectrum and slightly increases the quadrupole (although the latter remains unchanged within the errors). The variance and covariance of both of these is reduced on large ($k<0.05\hompc$) scales. In subtracting the Bulk Flow, we are artificially setting $\langle v(\br)\rangle = 0$, which forces the spherically averaged power spectrum to go to zero as $k=0$. This is related to the well-known integral constraint in the galaxy density power spectrum arising from the assumption that the average overdensity over the observed survey volume is zero (which we are also assuming here too). The variance of the power spectrum is reduced as we are removing a portion of the cosmic variance associated the observer's own motion, which even in $\Lambda$CDM can have a significant range of values. Although not shown, we also found that the presence of a Bulk Flow has negligible effect on the correlation between different scales in the density/momentum power spectrum multipoles.

When fitting the mocks in Section~\ref{sec:proof} we tested the fits with and without removing the Bulk Flow, finding that the errors on the growth rate were marginally improved after its removal, consistent with the reduction in the variance of the momentum power spectrum. The best-fit value of the growth rate did not change. During the fitting we included the standard correction for the integral constraint (i.e., following \citealt{Ross2013, Beutler2014} for the galaxy density power spectrum), but do not explicitly include further correction for artificially setting the mean velocity to zero. Our mocks advocate that there is some small benefit in removing the bulk flow. In their original work, \cite{Park2006} removed the bulk flow and our mocks seem to validate their approach. However, our results could be strongly influenced by the fact that the mocks used here extend over quite a large cosmological volume and we only fit for $k\ge 0.02\hompc$; our mock Bulk Flows are likely smaller than in current datasets with lower redshift limits or that cover smaller sky areas and their effects do not extend much past our lower $k$-limit. Without further theoretical work towards understanding the effect that removing the Bulk Flow has on the momentum power spectrum and its relation to the well-known integral constraint, we conclude that this should be tested on a case-by-case basis.

In addition to the Bulk Flow, peculiar velocity surveys often suffer from an unknown `zero-point' and the monopole of the velocity field is artificially set to zero. \cite{Howlett2017a} demonstrate that the impact of a constant addition to the velocity field would be to add as shot-noise to the velocity power spectrum. The same would be true for the momentum power spectrum. However, even for the optimistic sample we consider in this work, the contribution to the shot-noise from a zero-point offset would remain subdominant compared to the shot-noise arising from the distance errors. Hence we believe that any systematic error associated with the monopole of the velocity field will be small compared to the typical statistical error on the momentum power spectrum.

\subsection{Gaussianity of power spectrum measurements}

\begin{figure}
\centering
\subfloat{\includegraphics[width=0.5\textwidth,trim=0mm 0mm 0mm 0mm]{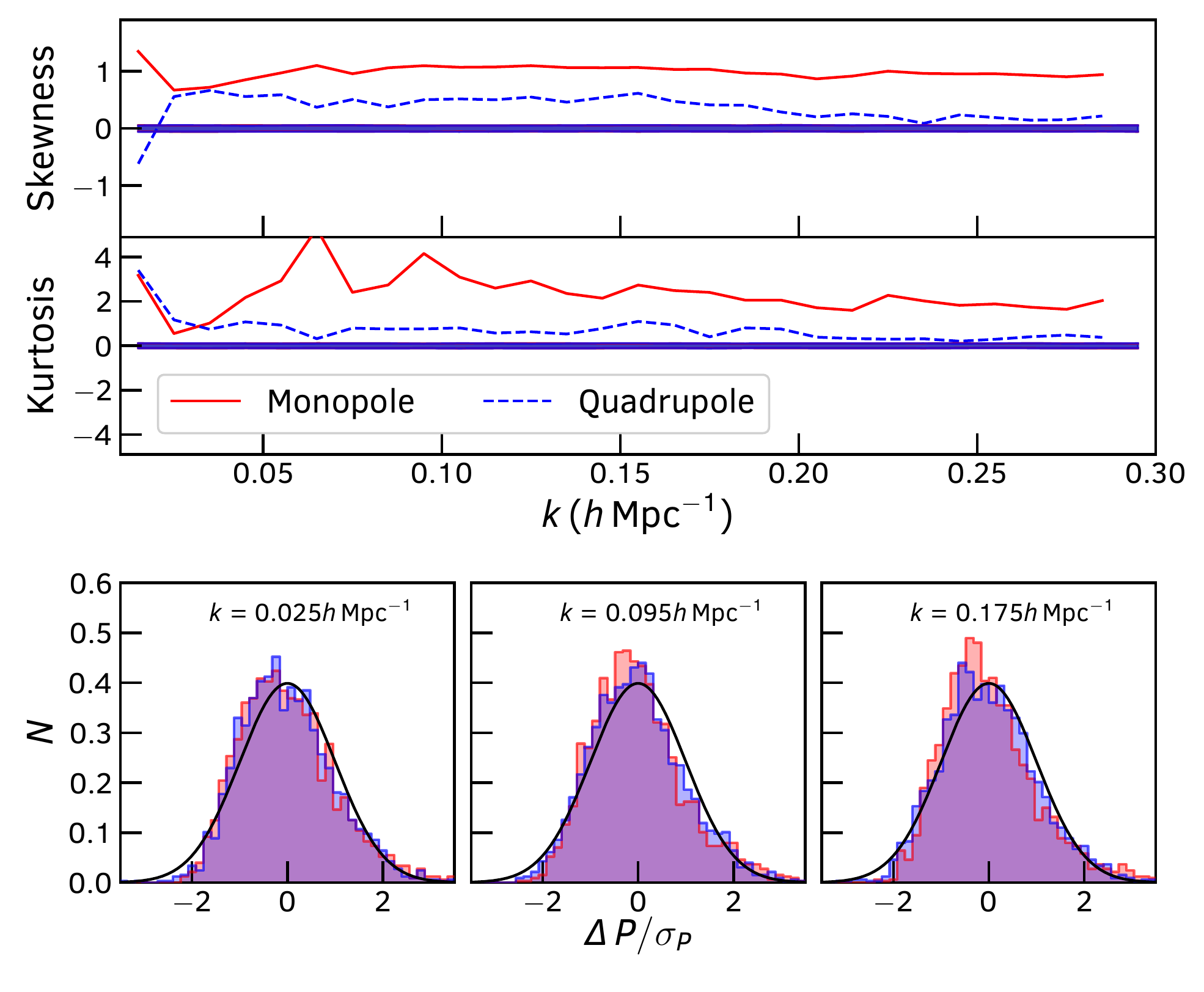}}
\subfloat{\includegraphics[width=0.5\textwidth,trim=0mm 0mm 0mm 0mm]{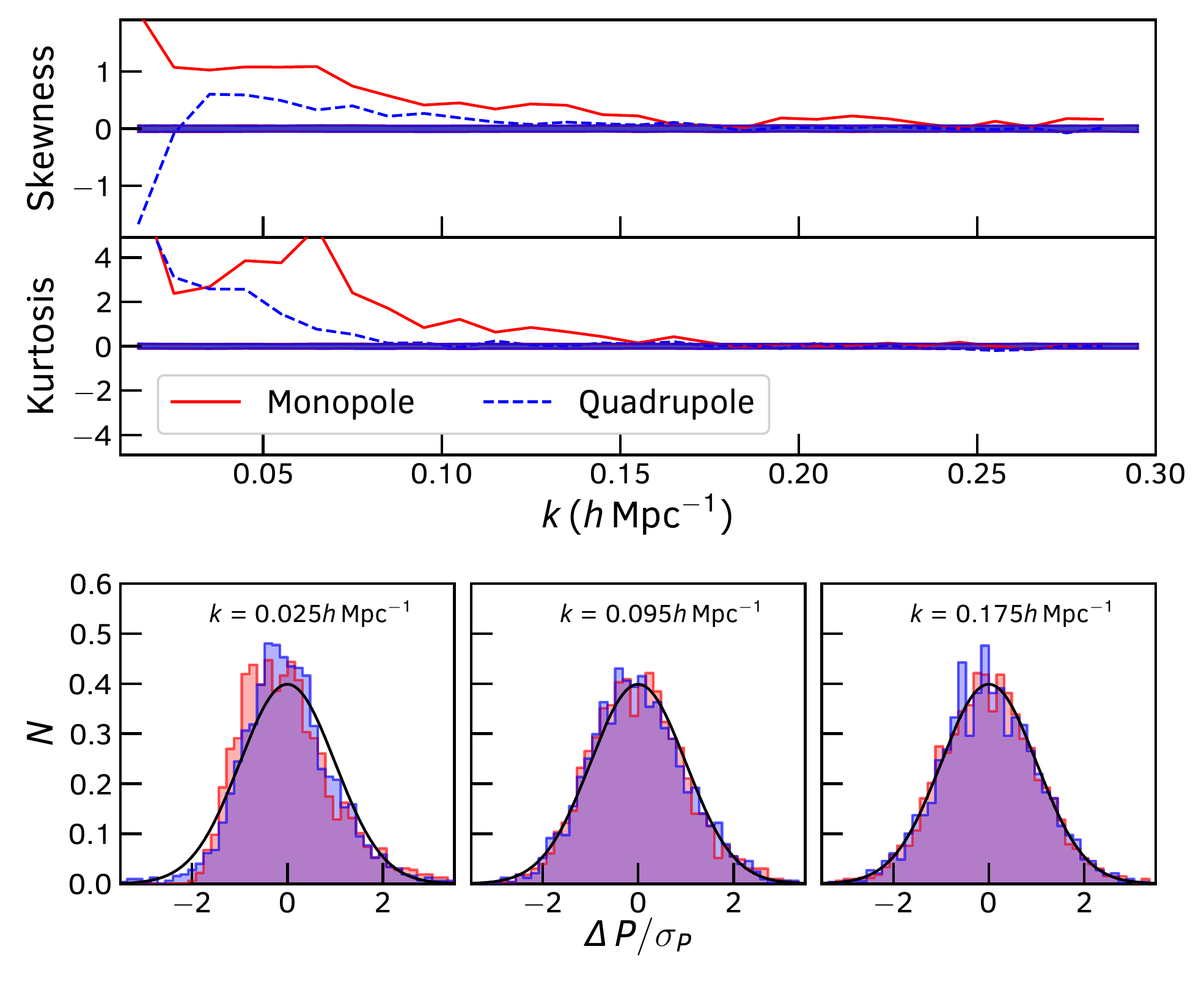}}\\
\caption{The distribution of measured momentum power spectra with respect to the mean as a function of $k$ without (left panels) and with (right panels) PV errors. In all cases red lines and histograms correspond to the monopole, whilst blue lines/histograms are the quadrupole measurements. In the top two panels of each side we show the skewness and kurtosis as a function of scale (lines) against the expectation for true Gaussian distributions with a number of measurements equal to the number of mocks (bands). In the lower panels we show individual normalised distributions for different $k$ bins alongside a Gaussian PDF (solid line) to guide the eye. For the measurements without PV errors (left panels) the distributions are approximately log-normal, with high levels of skewness/kurtosis. Once measurement errors are included (right panels) the measurements become more Gaussian although some residual non-Gaussianity remains on large scales where there are fewer modes and the central-limit theorem does not hold. In the tests performed in this work, this was not found to bias the growth rate constraints. A method to account for this is also presented in Paper II.}
  \label{fig:mockgauss}
\end{figure}

One final thing that we wish to test before applying this work to real data is the Gaussianity of the distribution of momentum power spectrum measurements. This affects how well we can describe the data using only the measurements themselves and an estimate of their covariance matrix; and how easily we are able to recover cosmological constraints from these. We use our mock catalogues as a test of the Gaussianity of the moments of the momentum power spectrum with and without PV errors by evaluating the difference between the distribution of the $P^{\,p}_{\ell}(k)$ values and the mean at fixed $k$. Our results are shown in Fig.~\ref{fig:mockgauss}, where we compute the skewness and kurtosis of this distribution compared to a true Gaussian with the same number of measurements\footnote{More precisely, we draw a number of `fake' momentum power spectra from a multivariate Gaussian distribution equal to the number of mocks, with mean and covariance equal to the measured mean and covariance. We then compute the skewness and kurtosis. We repeat this 1000 times, obtaining 1000 values for the skewness and kurtosis in each $k$-bin. We then plot the $16^{th}$ and $84^{th}$ percentiles of the skewness and kurtosis values.}. We also plot the distribution for three individual $k$ bins.

For the measurements without errors, we find the monopole and quadrupoole of the momentum power spectrum to be closer to log-normally distributed rather than Gaussian (similar results are found for the galaxy density power spectrum e.g., \citealt{Ross2015}) resulting in considerable skewness and kurtosis. If we instead perform the experiment above on the logarithm of the monopole, the distribution becomes close to Gaussian; there is some residual skewness and kurtosis, but this is not particularly evident in the individual distributions and we do not believe it will have much impact on parameter constraints if the logarithm is fit. However, such a transformation cannot be naively applied to the quadrupole due to the presence of negative $P^{\,p}_{2}(k)$ values. Once PV measurement errors are included, the distribution of monopole and quadrupole power spectrum measurements becomes much more Gaussian given that the PV errors are also Gaussian (seen by comparing the left and right panels of Fig~\ref{fig:mockgauss}). On large scales the distribution still contains significant skewness and kurtosis which, again, we cannot get round using the logarithm of the measurements due to the presence of negative $P^{\,p}(k)$ values in both the monopole and quadrupole. 

Fortunately, this is not unmanageable. When fitting the mocks in Section~\ref{sec:proof} we find that this does not overly bias the growth rate recovered from fitting the momentum power spectrum. In Paper II we also found no bias in the growth rate constraints due to non-Gaussianity in measurements from the 6dFGSv survey. For the 2MTF data used in Paper II we did find that the small volume of the survey leads to a high degree of non-Gaussianity which biases the growth rate measurements\footnote{As highlighted in \cite{Kalus2016}, for a field whose real and imaginary parts are close to individually Gaussian distributed (as might be expected for our Universe) the power spectrum is actually Rayleigh distributed. The distribution of measured power spectra only approaches Gaussian if there are sufficient modes that the central limit theorem holds. For the small volume of the 2MTF survey, the number of modes is sufficiently low that the central-limit theorem does not hold even for quite non-linear scales.}. However, in this same work, we developed and verified that performing a simple transformation to the measurements from the mocks and data prior to fitting is enough to fully account for this. Hence even though we do not apply them to our mocks, the techniques therein could be applied to the momentum power spectrum measurements from any dataset. We conclude that, given an estimated covariance matrix, we can easily extract cosmological information from measurements of the momentum power spectrum from real data. The presence of Gaussian PV errors (even small ones used here) is enough to ensure that whilst the intrinsic measurement in roughly log-normal, the measurement from data is Gaussian enough for robust constraints to be obtained.

\section{Final proof of concept} \label{sec:proof}
As a final proof of concept that measurements of the momentum power spectrum can be used to access the cosmological information in the velocity field from current and next-generation surveys, we bring everything discussed so far in this work together and measure the growth rate of structure from our mock PV surveys introduced in Section~\ref{sec:mocks}.

We take measurements of the monopole, quadrupole and hexadecapole ($\ell=0$,$2$ and $4$ moments) of the galaxy density field of the mock galaxies using an FKP weight with constant $P_{FKP}=8000\mpcohV$ and measurements of the monopole and quadrupole of the momentum power spectrum using weights with constant $P^{\,p}_{FKP}=2.4\times 10^{9}\kmsmpcohV$. These weights were chosen to maximise the information on the growth rate based on the procedure given in Appendix~\ref{sec:appendixB}, where we show how to estimate the Fisher information on the growth rate contained in different multipoles of the density and momentum power spectrum using different weights. The same calculation is used to produce forecasts for the constraints on the growth rate for a scenario exactly matching the measurements we make here.

The measurements are then combined with the redshift-space model given in Appendix~\ref{sec:appendix}. We correct this model for the effects of the window function using a modified version of the code provided with \cite{Blake2018}, which computes a convolution matrix that multiplies the fine-binned model multipoles and returns a convolved model with the same binning as the measurements. We correct for the integral constraint using the same code, and remove the Bulk Flow before making the measurements, consistent with our findings in Section~\ref{sec:bulkflow}. We then fit the average measurements from the mocks for $k=0.02-0.20\hompc$ using a Gaussian likelihood and covariance matrix also estimated from the mocks.

From Appendix~\ref{sec:appendix}, there are five free parameters for the density and momentum power spectra: $b_{1}\sigma_{8}$, $b_{2}\sigma_{8}$, $f\sigma_{8}$, and the velocity dispersions $\sigma_{v,1}$ and  $\sigma_{v,2}$. This models is found to recover unbiased constraints when fitting either the density field or momentum field multipoles separately. However, when fitting all five multipoles simultaneously, we find that we need to use separate $\sigma_{v,1}$ parameters for the density and momentum measurements. We attribute this to the fact that we are weighting the same sample differently for the two measurements, which changes the effective depth and bias of the sample. However, as the bias and velocity dispersion terms all enter the momentum power spectrum measurements on non-linear scales, we are able to account for this with a single extra parameter (rather than also introducing a second set of bias parameters). Hence when fitting both the density and momentum power spectra simultaneously we have a total of 6 free parameters. 

\begin{figure}[h!]
\centering
\subfloat{\includegraphics[width=0.5\textwidth,trim=0mm 0mm 0mm 0mm]{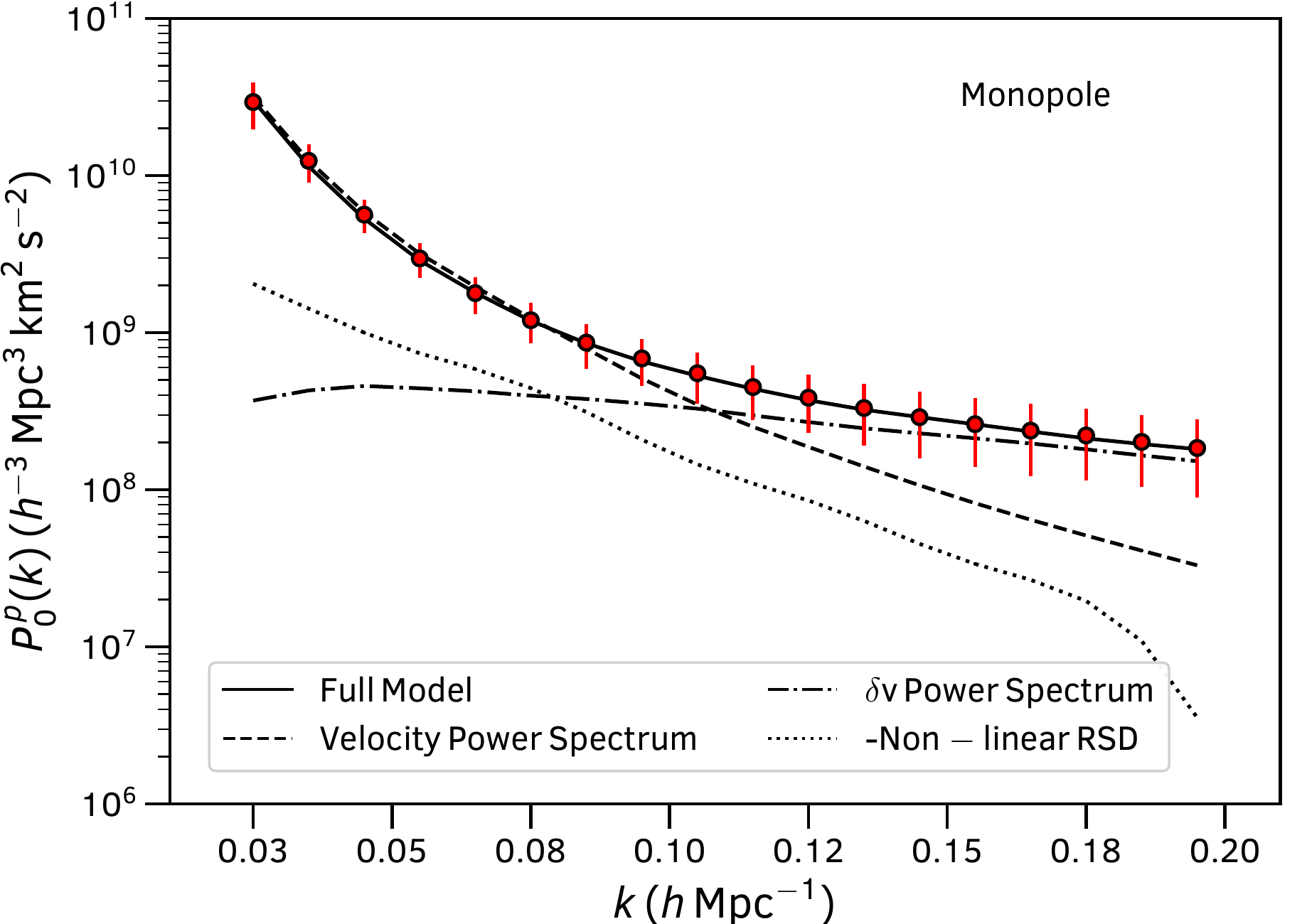}}
\subfloat{\includegraphics[width=0.5\textwidth,trim=0mm 0mm 0mm 0mm]{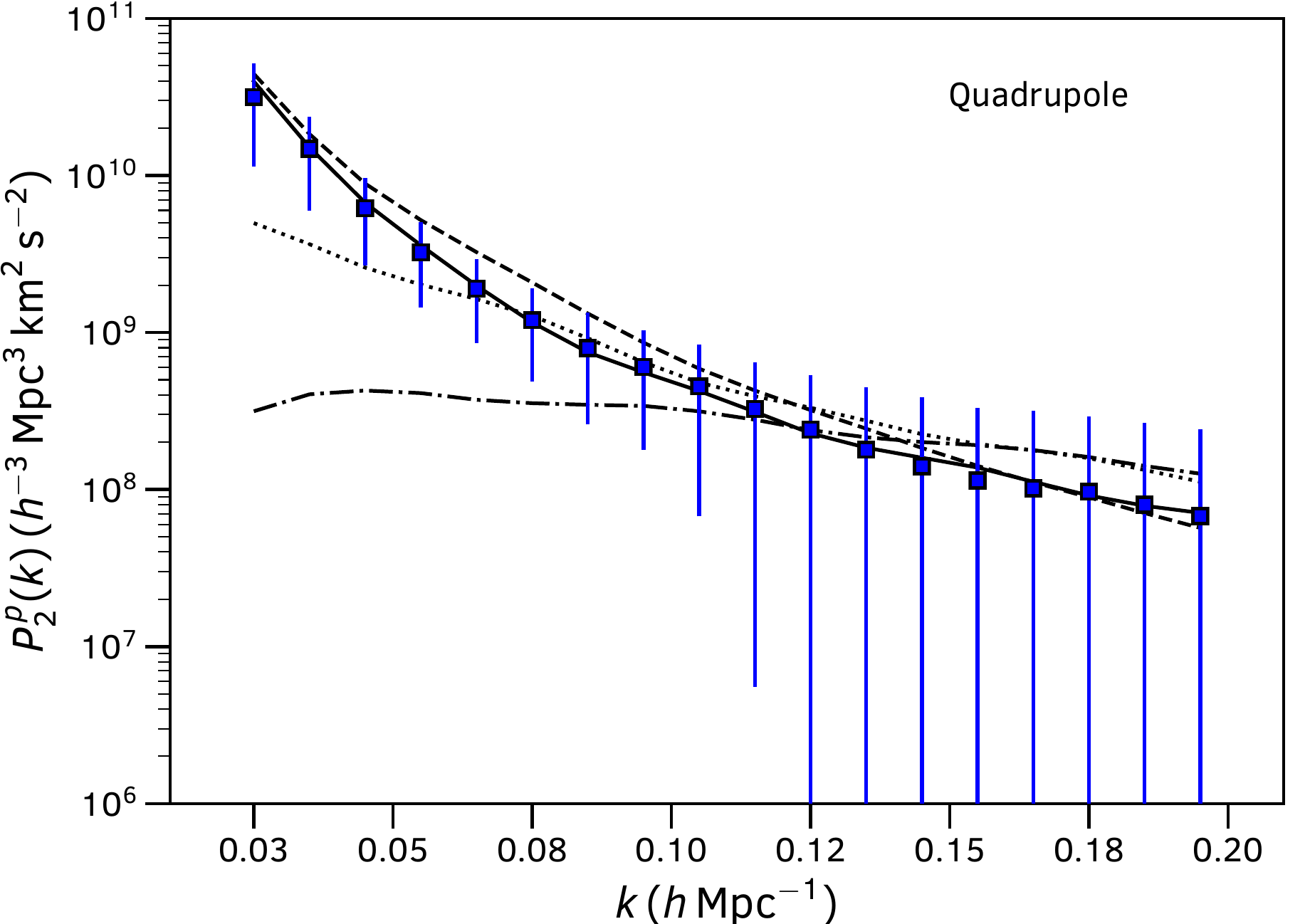}}\\
\caption{Plots of the average measurements of the monopole and quadrupole of the momentum power spectrum for our mocks alongside the best-fitting model from fitting these two measurements simultaneously whilst also correcting for the survey window function. Measurements are shown as points and errors are derived from the variance in the mocks. The solid line shows the model. We deconstruct the model into three different components; the velocity power spectrum (dashed lines), the power spectrum of the convolved density and velocity fields (dot-dashed line) and non-linear redshift-space distortions (dotted line). This last component is negative on all plotted scales, so we multiply it by $-1$ for plotting purposes. For both the monopole and quadrupole, the large scale momentum power spectrum is dominated by the velocity power spectrum component, but the other terms become important on non-linear scales.}
  \label{fig:bestfit}
\end{figure}

We start looking at how different components of the theoretical redshift-space momentum power spectrum go together to fit our data. Figure~\ref{fig:bestfit} shows best-fitting model for the monopole and quadrupole of the momentum power spectrum alongside the measurements and errors. We identify different components contributing to the best-fit model; the velocity power spectrum (the terms in $P_{11}$ in Appendix~\ref{sec:appendix} independent of galaxy bias), the power spectrum of the convolution between the density and velocity field (the $P_{11}$ terms containing galaxy bias), and non-linear redshift-space distortions ($P_{12}$, $P_{13}$ and $P_{22}$). As expected from Fig.~\ref{fig:pkmomhalos}, for the monopole the velocity power spectrum dominates on large scales, and the density-velocity convolution spectrum on non-linear scales. The non-linear RSD terms stay subdominant on all the scales we fit. For the quadrupole, the story is a little different. On the largest scales we consider, the velocity power spectrum remains dominant, but non-linear redshift-space distortions become important on relatively large scales compared to the monopole. The contributions from the convolution between the density and velocity fields and non-linear RSD are similar for $k>0.1\hompc$. Given that both the monopole and quadrupole are proportional to the growth rate on large scales, and have different dependencies on bias and non-linear RSD on small scales, we might expect there to be some improvement when adding the momentum power spectrum quadrupole. However, the effects of galaxy bias and non-linear RSD combined with the noise on non-linear scales means these are likely very degenerate and that the real benefit comes from having an independent measurement of the galaxy bias from the density power spectrum.

\begin{figure}
\centering
\includegraphics[width=0.5\textwidth,trim=0mm 0mm 0mm 0mm]{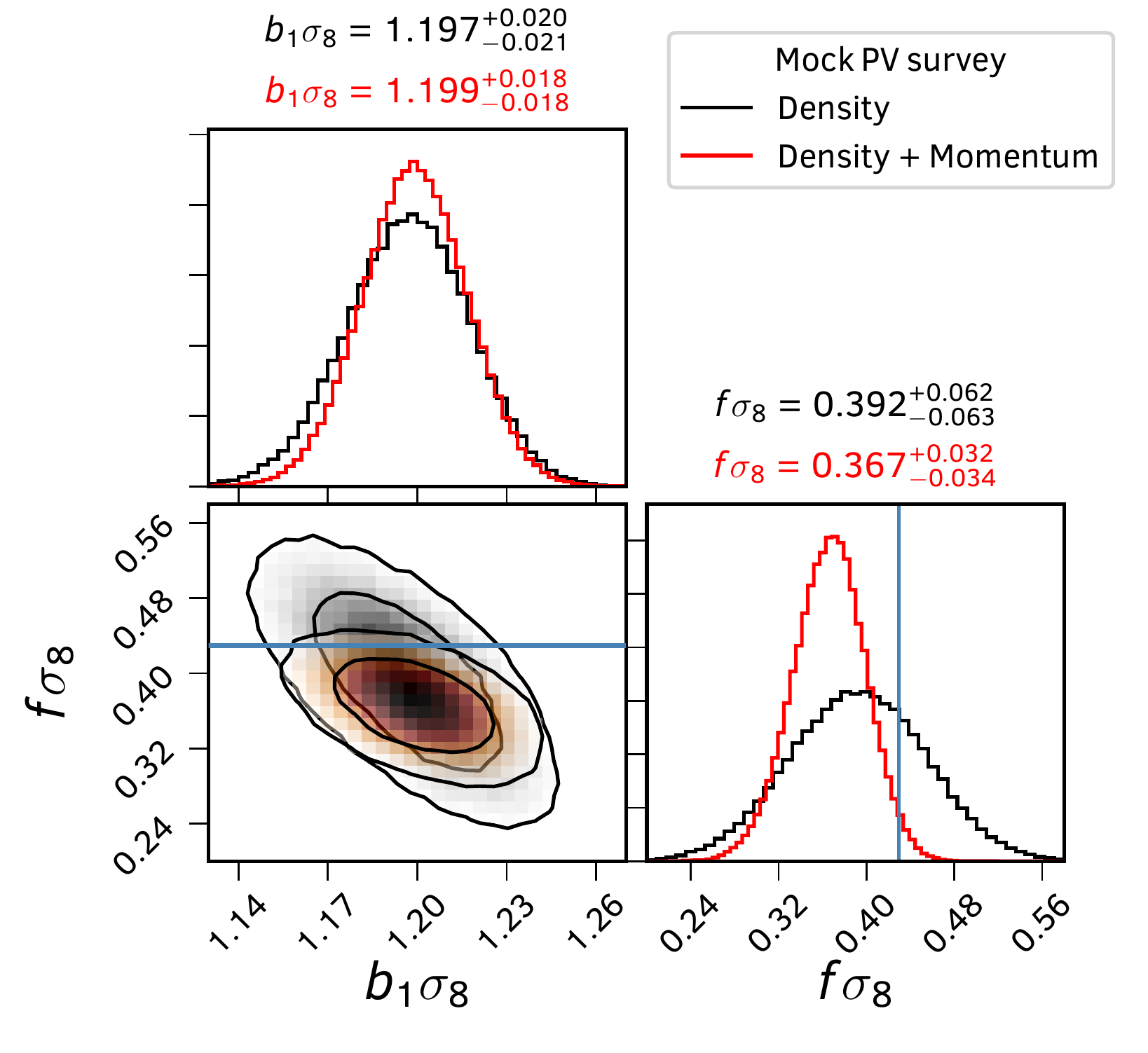}
\caption{2-D contours and marginalised histograms of $b_{1}\sigma_{8}$ and $f\sigma_{8}$ from fitting the galaxy density power spectrum monopole, quadrupole and hexadecapole and with (red) and without (black) the inclusion of the momentum power spectrum monopole and quadrupole. The numbers give the median and errors using the $16^{\mathrm{th}}$, $50^{\mathrm{th}}$ and $84^{\mathrm{th}}$ percentiles and the solid line is the expectation for the growth rate in the mocks. We see that the inclusion of the momentum power spectrum greatly improves the growth rate constraints. There is some bias in the recovered $f\sigma_{8}$ measurements due to modelling and fitting inaccuracies; these may have to remedied for fits to real data (as they are in Paper II), but do not invalidate this proof-of-concept.}
  \label{fig:bestfitcosmo}
\end{figure}

To explore this and the information contained in other combinations of measurements of the galaxy density and momentum multipoles we turn next to the growth rate constraints themselves. These are listed in Table~\ref{tab:fsigma8} for various combinations of moments of the combined density and momentum power spectra. For each case we also include forecasts based on the method in Appendix~\ref{sec:appendixB} for the growth rate measurements assuming truly optimal weights and using weights with constant $P(k)$ as given above. We also show the $b_{1}\sigma_{8}-f\sigma_{8}$ contours for the density field measurements and then when the momentum power spectrum is added in Fig.~\ref{fig:bestfitcosmo}. We found negligible information in the $\ell>4$ and $\ell>2$ moments of the density and momentum power spectra respectively, which is corroborated by our forecasting code.

\begin{table}
\setlength{\extrarowheight}{3pt}
\caption{Constraints and forecasts on the growth rate of structure from fitting various multipoles of the density and momentum power spectra from mock galaxy catalogues. We use the $16^{\mathrm{th}}$, $50^{\mathrm{th}}$ and $84^{\mathrm{th}}$ percentiles to quote the median and errors on the marginalised $f\sigma$ measurements. Given the input cosmology of the mocks, the expected value is $f\sigma_{8}\approx 0.432$. We also show forecasts using the Fisher matrix method in Appendix~\ref{sec:appendixB} which quantifies the lower error bound using the information in the velocity or density fields for different weighting schemes. We provide forecasts for the optimal weighting scheme derived in Section~\ref{sec:weights}, and where we use the common technique of using fixed values for the power spectrum in these weights. The latter case does reduce the growth rate information in both the density and momentum power spectra, but the latter is affected more greatly due to its $1/k^{2}$ dependence on linear scales.}
\centering
\begin{tabular}{lccc} \hline
\multirow{2}{*}{Measurements} & \multirow{2}{*}{$f\sigma_{8}$ (Percentiles)} & \multicolumn{2}{c}{$\sigma(f\sigma_{8})$ Predictions} \\
& & Optimum weights & Constant $P_{FKP}$ weights \\ \hline
Momentum $\ell=0$ & $0.411^{+0.056}_{-0.064}$ & $0.032$ & $0.043$ \\
Momentum $\ell=0,2$ & $0.410^{+0.053}_{-0.059}$ & $0.031$ & $0.039$ \\
Density  $\ell=0,2$ & $0.371^{+0.081}_{-0.070}$ & $0.049$ & $0.051$  \\
Density  $\ell=0,2,4$ & $0.392^{+0.062}_{-0.063}$ & $0.048$ & $0.051$ \\
Density $\ell=0$ + Momentum $\ell=0$ & $0.430^{+0.056}_{-0.054}$ & $0.031$ & $0.041$\\
Density $\ell=0,2$ + Momentum $\ell=0$ & $0.383^{+0.038}_{-0.038}$ & $0.028$ & $0.035$ \\
Density $\ell=0,2,4$ + Momentum $\ell=0$ & $0.390^{+0.039}_{-0.039}$ & $0.028$ & $0.035$ \\
Density $\ell=0$ + Momentum $\ell=0,2$ & $0.345^{+0.040}_{-0.044}$ & $0.030$ & $0.037$ \\
Density $\ell=0,2$ + Momentum $\ell=0,2$ & $0.364^{+0.034}_{-0.035}$ & $0.027$ & $0.033$ \\
Density $\ell=0,2,4$ + Momentum $\ell=0,2$ & $0.367^{+0.032}_{-0.034}$ & $0.027$ & $0.033$ \\ \hline
\end{tabular}
\label{tab:fsigma8}
\end{table}

To begin, the growth rate constraints are largely consistent with the expected value for the mocks although with a small bias likely due to non-linear modelling inaccuracies or residual non-Gaussianity in the measurements (the latter of which is solved in Paper II, but not a concern for our proof of concept). The first point of note is that the measurements using only the monopole of the momentum power spectrum are tighter than using all three moments of the galaxy density power spectrum for our sample. This is perhaps an unfair comparison as our sample contains equal numbers of redshifts and PV, whereas in reality a substantially higher number of redshifts could be obtained for the same `effort' as the 150,000 PVs we use here. Nonetheless, the fact is that we would expect the information in the density field alone to saturate relatively quickly as the cosmic variance limit is reached, and the momentum power spectrum is proportional to the growth rate on linear scales. This means we get tighter constraints on the growth rate than with the galaxy density power spectrum and it is far easier to model and extract that information from the monopole of the momentum power spectrum than the full set of galaxy density multipoles. 

The real benefit comes from combining these two sets of measurements. The combined constraints from the full set of 5 power spectrum multipoles is a factor of $\sim 1.5$-$2$ times better than the constraints from either the density or momentum field alone. We find that most of the constraints come from the first two moments of the galaxy density power spectrum and the monopole of the momentum power spectrum. Including higher order moments in the combination does provide some advantage but in practice this may not be worth the additional modelling complexity or systematic errors this could introduce. 

It is also interesting to ask how much of the total information in the velocity field is recovered using the momentum power spectrum (which is `contaminated' by bias on small scales). We can answer these by looking at the forecasts in Table~\ref{tab:fsigma8}. To provide a benchmark, we first compare the forecasts for the optimal weights and using the fixed $P_{FKP}$ weighting scheme. The use of fixed values to make the measurements reduces the information by about $10\%$ and $30\%$ for the density and momentum power spectrum respectively. We additionally remind the reader that the fixed values were chosen as the weights that minimize this difference. Hence we conclude that the use of a fixed weight impacts the constraints somewhat, and this commonly used technique could be improved upon. This is true for the galaxy density power spectrum (and so likely true for a number of published works), but is more evident in the momentum power spectrum due to the fact that this has a strong $1/k^{2}$ scale dependence. Future improvements to this work could develop a method to implement weights that vary as a function of $k$ which would benefit even measurements of the galaxy density power spectrum from redshift-only surveys.

Looking at the case for the momentum power spectrum alone, we see the measurements to be quite a bit worse than the forecasts even for non-optimum weights. Some portion of this can be attributed to the overoptimistic nature of the Fisher matrix forecasts as the measurements from the density field alone are also larger than the forecasts by $\sim25\%$, however this will also be due to the fact that the momentum power spectrum is not the velocity power spectrum and so does not contain much growth rate information on non-linear scales where the density-velocity convolution starts to dominate ($k \approx 0.1\hompc$ as seen in Fig.~\ref{fig:bestfit}). However, when combining the density and momentum power spectra the story is different. The constraints we obtain are in-line with the forecasts for our weighting scheme; simultaneously fitting the density and momentum measurements serves to break the degeneracy between galaxy bias and the growth rate in both of these. This proves that our method for measuring and modelling the combined density and velocity field information using the momentum power spectrum does an excellent job of recovering the information on the growth rate from the PV data. This serves as a proof-of-concept that the work in this paper can be used to extract accurate constraints from real data.  

A final question we can ask is how much information we may be missing by only considering the power spectra of the density and momentum field and their cross-covariance and neglecting the cross-\textit{spectrum} between these two fields. This was mentioned in Section~\ref{sec:crosscovariance}, where we argued that measuring this introduces enough complexity that we leave it for future work. Although we also leave the full details of the calculation to the next paper in this series, we can include the multipoles of the cross-spectrum, \textit{its} covariance matrix, and its cross-covariance with the multipoles of the galaxy density and momentum power spectra in the Fisher matrix method of Appendix~\ref{sec:appendixB}. Doing so, we forecast errors on $f\sigma_{8}$ of 0.026 and 0.030 for the optimum and fixed weights used herein. This is an improvement of $\sim 10\%$ over the case where we do not include this observable, which is somewhat less than the results of \cite{Adams2017}. However, as they pointed out in their comparisons with simulations, a larger improvement may be seen in data with lower number density or worse velocity errors, where the growth rate and galaxy bias are less well constrained. A larger improvement may be also be seen in the case of real data where including the cross-spectrum could help resolve unknown systematics.

\section{Conclusions} \label{sec:conclusion}
In this work we have formally derived an estimator for the redshift-space momentum (or mass-weighted velocity) power spectrum from a peculiar velocity (PV) survey. The method holds for any tracer of the velocity field, including galaxy motions from Tully-Fisher or Fundamental Plane surveys, or for Type-IA supernovae. This derivation mimics the well-known \cite{Feldman1994} (FKP) and \cite{Yamamoto2006} estimators for the galaxy density power spectrum, and has a number of useful properties. In particular:
\begin{itemize}
\item{It is an extremely efficient compression of information compared to other techniques for extracting the velocity power spectrum \citep{Johnson2014,Adams2017,Howlett2017c,Huterer2017}.}
\item{The normalisation of the galaxy and momentum power spectrum is the same.}
\item{The effect of a survey window function introduces a convolution with the momentum power spectrum identical to the galaxy density power spectrum. This can be best seen by comparing Eqs.~\ref{eq:convterms1}-\ref{eq:convterms} with the expressions for the convolution of the galaxy density power spectrum in \cite{Feldman1994}, \cite{Yamamoto2006} or \cite{Blake2018}.}
\item{The momentum power spectrum is also subject to shot-noise dependent on the galaxy number density, however there is an additional contribution from the variance in the velocity field, which will contain both an intrinsic component, and a component due to measurement error.}
\item{The covariance matrix of the momentum power spectrum, and the cross-covariance with the galaxy density power spectrum can both be calculated analytically in the Gaussian regime. We find good agreement between the Gaussian prediction and measurements from simulations on large scales, but the scale at which this breaks down is much larger than for the galaxy density power spectrum due to the convolution of density and velocity fields which enters the momentum power spectrum.}
\item{We have computed the optimal weights for the momentum power spectrum. Much like the FKP weights, these depend on the galaxy number density and power spectrum itself. We also find dependence on the intrinsic variance of the velocity field.}
\end{itemize}

We have also investigated several properties unique to PV surveys that affect the momentum power spectrum measurements. We find that the presence of errors on individual PV measurements adds as shot-noise to the momentum power spectrum, such that our estimator remains unbiased, but the variance increases significantly. Removing the bulk motion of the observer before estimating the momentum power spectrum removes power on large scales, acting as an integral constraint, but also reduces the variance accordingly. In the mocks used here, we find it beneficial to remove the bulk flow when fitting the growth rate of structure as it slightly reduces the uncertainty on the constraints but without adding bias. However we caution that this result could arise from of the relatively small bulk flow we might expect in our mocks compared to current datasets and that more theoretical work is needed to understand how this changes the measured momentum power spectrum. Finally, we show that the momentum power spectrum measured from a galaxy survey is Gaussian distributed on scales sufficiently smaller than the survey volume, allowing for extraction of cosmological constraints using only an estimate of the covariance matrix. This point is further investigated in Paper II where we validate a method to recover unbiased cosmological constrains even in the presence of greater levels of non-Gaussianity than shown here.

Finally, we have provided a complete proof of concept demonstrating that our estimator of the momentum power spectrum makes the information in the velocity field, in particular on the growth rate of structure, readily obtainable. This will be a key tool in the analysis of the next generation Taipan Galaxy survey \citep{DaCunha2017}, WALLABY survey \citep{Koribalski2012} or of Type-IA supernovae \citep{Howlett2017b,Kim2019} and will allow these surveys to overcome the limitations of cosmic variance in the low redshift universe, leading to some of the most precise test of gravity to date. However, this work is only the tip of the iceberg; in Paper II (Qin et. al., in preparation), we present the first application of all the techniques in this work to the combined 2MASS Tully-Fisher (\citealt{Hong2014}; Hong et. al, in preparation) and 6-degree Field Galaxy Redshift Survey peculiar velocity samples \citep{Campbell2014}. In subsequent papers in this series we will extend this to look at the remaining information contained within in the density-momentum cross-power spectrum and the prospects of constraining modified gravity theories.

\section*{Acknowledgements}
We thank Fei Qin, David Parkinson, Lister Staveley-Smith and Chris Blake for many useful discussions and providing comments on this manuscript. We especially thank the latter for providing their code to measure and model the multipoles of the galaxy density power spectrum.

This research was conducted by the Australian Research Council Centre of Excellence for All-sky Astrophysics (CAASTRO), through project number CE110001020. This work was performed on the Pleiades HPC cluster at the International Centre for Radio Astronomy Research at the University of Western Australia and on the OzSTAR and swinSTAR supercomputers at Swinburne University of Technology.

This research has made use of NASA's Astrophysics Data System Bibliographic Services and the \texttt{astro-ph} pre-print archive at \url{https://arxiv.org/}. All plots in this paper were made using the {\sc matplotlib} plotting library \citep{Hunter2007}.

\begin{multicols}{2}
\bibliography{/Volumes/Work/ICRAR/LaTeX/massive.bib}{}
\bibliographystyle{mnras}
\end{multicols}

\vspace{-0.5cm}
\appendix
\section{Theoretical model for the redshift-space density and momentum power spectrum} \label{sec:appendix}
In this work we model the redshift-space density and momentum power spectrum for biased tracers using the combined distribution function and Eulerian perturbation theory approach of \cite{Vlah2012,Vlah2013,Okumura2014,Saito2014}, with some small modification to higher order terms for easier computation. Full details of the model can be found therein, but requires some work piecing together various components; here we provide a more condensed overview.

We begin by defining the multipoles of the power spectra as integrals with respect to the angle, $\mu$
\begin{equation}
P^{\delta(p)}_{\ell}(k) = (2\ell+1) \int^{1}_{0}P^{\delta(p)}(k,\mu) L_{\ell}(\mu) d\mu
\end{equation}
where the quantities $P^{\delta}(k,\mu)$ and $P^{p}(k,\mu)$ are the full redshift-space anisotropic galaxy density and momentum power spectra respectively. We model the two anisotropic power spectra (dropping the explicit dependencies on $z$, $k$ and $\mu$) as
\begin{align}
P^{\delta} &= P_{00} + \mu^{2}(2P_{01} + P_{02} + P_{11}) + \mu^{4}(P_{03} + P_{04} + P_{12} + P_{13} + 1/4P_{22}), \\
P^{p} &= (aH)^{2}k^{-2}(P_{11} + \mu^{2}(2P_{12} + 3P_{13} + P_{22})),
\end{align}
where
\begin{align}
P_{00} & = b_{1}^{2}D^{2}(P_{L} + 2D^{2}(I_{00} + 3k^{2}P_{L}J_{00})) + 2b_{1}D^{4}(b_{2}K_{00} + b_{s}K^{s}_{00} + b_{3nl}\sigma^{2}_{3}P_{L}) + D^{4}(1/2b_{2}^{2}K_{01} + 1/2b^{2}_{s}K^{s}_{01} + b_{2}b_{s}K^{s}_{02}), \\
P_{01} & = fb_{1}D^{2}(P_{L} + 2D^{2}(I_{01} + b_{1}I_{10} + 3k^{2}P_{L}(J_{01}+b_{1}J_{10})) - b_{2}D^{2}K_{11} - b_{s}D^{2}K^{s}_{11}) - fD^{4}(b_{2}K_{10} + b_{s}K^{s}_{10} + b_{3nl}\sigma^{2}_{3}P_{L}), \\
P_{02} & = f^{2}b_{1}D^{4}(I_{02} + \mu^{2}I_{20} + 2k^{2}P_{L}(J_{02} + \mu^{2}J_{20})) - f^2k^{2}\sigma^{2}_{v}P_{00} - f^{2}D^{4}(b_{2}(K_{20}+\mu^{2}K_{30}) + b_{s}(K^{s}_{20}+\mu^{2}K^{s}_{30})), \\
P_{03} & = -f^{2}k^{2}\sigma^{2}_{v}P_{01}, \\ 
P_{04} & = -1/2f^{4}b_{1}k^{2}\sigma^{2}_{v}D^{4}(I_{02} + \mu^{2}I_{20} + 2k^{2}P_{L}(J_{02} + \mu^{2}J_{20})) + 1/4f^{4}b_{1}^{2}k^{4}P_{00}(\sigma_{v}^{4} + \sigma^{2}_{4}), \\
P_{11} & = f^{2}D^{2}(\mu^{2}(P_{L} + D^{2}(2I_{11} + 4b_{1}I_{22} + b_{1}^{2}I_{13} + 6k^{2}P_{L}(J_{11} + 2b_{1}J_{10}))) + b_{1}^{2}D^{2}I_{31}), \\ 
P_{12} & = f^{3}D^{4}(I_{12} + \mu^{2}I_{21} - b_{1}(I_{03} + \mu^{2}I_{30}) + 2k^{2}P_{L}(J_{02} + \mu^{2}J_{20})) - f^{2}k^{2}\sigma^{2}_{v}P_{01}, \\
P_{13} & = -f^{2}k^{2}\sigma^{2}_{v}P_{11}, \\
P_{22} & = 1/4f^{4}D^{4}(I_{23} + 2\mu^{2}I_{32} + \mu^{4}I_{33}) + f^{4}k^{4}\sigma_{v}^{4}P_{00} - f^{2}k^{2}\sigma^{2}_{v}(2P_{02} - f^{2}D^{4}(b_{2}(K_{20}+\mu^{2}K_{30}) + b_{s}(K^{s}_{20}+\mu^{2}K^{s}_{30}))).
\end{align}
In the above equations, $D(z)$ is the linear growth factor, $f(z)$ is the growth rate of structure, $H(z)$ is the Hubble parameter, $a(z)$ is the scale factor, $P_{L}(k)$ is the $z=0$ linear matter power spectrum and $\sigma_{v}$ is the velocity dispersion. We have used the galaxy bias expansion of \cite{McDonald2009a} which encapsulates local and non-local bias up to third order in the four terms $b_{1}$, $b_{2}$, $b_{s}$ and $b_{3nl}$. The remaining terms $I_{mn}$, $J_{mn}$, $K_{mn}$, $\sigma^{2}_{3}$ and $\sigma^{2}_{4}$ are integrals over the linear power spectrum. 

$I_{mn}$ and $J_{mn}$ are neatly presented in Appendix D of \cite{Vlah2012} and so we do not repeat them here. We can write the terms $K^{(s)}_{mn}$ in the same form
\begin{equation}
K^{(s)}_{mn} = \int \frac{d^{3}q}{(2\pi)^{3}} h^{(s)}_{mn}(\bk,\bq) P_{L}(q)P_{L}(|\bk-\bq|) = \frac{k^{3}}{4\pi^{2}}\int dr\,r^{2}P_{L}(kr)\int_{-1}^{1}dx\,h^{(s)}_{mn}(r,x)P_{L}(k\sqrt{1+r^{2}-2rx}),
\end{equation}
where $r=q/k$, $x=\frac{\bk\cdot\bq}{kq}$ and 
\begin{align}
h_{00} &= \frac{7x+3r-10rx^{2}}{14r(1+r^{2}-2rx)}, & h^{s}_{00} &= \frac{(7x+3r-10rx^{2})(3x^{2}-4rx+2r^{2}-1)}{42r(1+r^{2}-2rx)^{2}}, \\
h_{01} &= 1, & h^{s}_{01} &= \frac{(3x^{2}-4rx+2r^{2}-1)^{2}}{9(1+r^{2}-2rx)^{2}}, \\
& & h^{s}_{02} &= \frac{3x^{2}-4rx+2r^{2}-1}{3(1+r^{2}-2rx)}, \\
h_{10} &= \frac{7x-r-6rx^{2}}{14r(1+r^{2}-2rx)}, & h^{s}_{10} &= \frac{(7x-r-6rx^{2})(3x^{2}-4rx+2r^{2}-1)}{42r(1+r^{2}-2rx)^{2}}, \\
h_{11} &= \frac{x}{r}, & h^{s}_{11} &= \frac{x(3x^{2}-4rx+2r^{2}-1)}{3r(1+r^{2}-2rx)}, \\
h_{20} &= \frac{x^{2}-1}{2(1+r^{2}-2rx)}, & h^{s}_{20} &= \frac{(x^{2}-1)(3x^{2}-4rx+2r^{2}-1)}{6(1+r^{2}-2rx)^2}, \\
h_{30} &= \frac{2x^{2}+r-3rx^{2}}{2r(1+r^{2}-2rx)}, & h^{s}_{30} &= \frac{(2x^{2}+r-3rx^{2})(3x^{2}-4rx+2r^{2}-1)}{6r(1+r^{2}-2rx)^2},
\end{align}
We normalise $K_{01} \rightarrow K_{01}-\int \frac{d^{3}q}{(2\pi)^{3}}P^{2}_{L}(q)$, $K^{s}_{01} \rightarrow K^{s}_{01}-\frac{4}{9}\int \frac{d^{3}q}{(2\pi)^{3}}P^{2}_{L}(q)$ and $K^{s}_{02} \rightarrow K^{s}_{02}-\frac{2}{3}\int \frac{d^{3}q}{(2\pi)^{3}}P^{2}_{L}(q)$, which ensures that the higher order bias terms go to zero as $k\rightarrow 0$ \citep{Saito2014}.

The final pieces we need are 
\begin{equation}
\sigma^{2}_{3}(k) = \frac{105k^{3}}{64\pi^{2}}\int dr\,r^{2}P_{L}(kr)\biggl[\frac{4}{7}\int_{-1}^{1}dx\,h^{s}_{20}(r,x)+\frac{16}{63}\biggl] \qquad \mathrm{and} \qquad \sigma^{2}_{4} = \frac{1}{24\pi^{2}}\int \frac{dq}{q^{2}}\,D^{4}\biggl[I_{23}(q) + \frac{2I_{32}(q)}{3} + \frac{I_{33}(q)}{5}\biggl].
\end{equation}

\subsection{Choice of free parameters}
In the above equations there are a number of free parameters. In addition to the cosmological parameters that affect the shape of the linear power spectrum, these include the growth rate of structure $f$ (our main parameter of interest), galaxy bias parameters $b_{1}$, $b_{2}$, $b_{s}$ and $b_{3,nl}$, the normalisation of the linear power spectrum commonly parameterised as $\sigma_{8}$ and the velocity dispersion $\sigma_{v}$. However, measurements of the density and momentum power spectrum are too noisy to break the degeneracy between all these parameters. Instead, when fitting our mock data we fix the linear power spectrum (and hence the $I_{mn}$, $J_{mn}$ and $K_{mn}$) to the input cosmology of the mocks with $\sigma_{8}=0.815$ and adopt the parameter combinations $b_{1}\sigma_{8}$, $b_{2}\sigma_{8}$, $f\sigma_{8}$. We then write the remaining galaxy bias terms as a function of the linear bias $b_{s}=-4/7(b_{1}-1)$ and $b_{3,nl}=32/315(b_{1}-1)$ \citep{Saito2014}. A similar technique was used in \cite{Beutler2017}. This allows for constraints to be put on $f\sigma_{8}$ from the overall amplitude of the momentum power spectrum (which still allows for tests of different gravitational models; \citealt{Song2009}) whilst the systematic effects of fixing $\sigma_{8}$ can be absorbed into the nuisance terms causing negligible change in the recovered values for the growth rate.

The velocity dispersion can be evaluated on linear scales as 
\begin{equation}
\sigma_{v,L} = \frac{1}{6\pi^{2}}\int dq\,q^{2}\,D^{2}(P_{L}(q) + D^{2}(2I_{11}(q) + 6k^{2}P_{L}(q) J_{11}(q))).
\end{equation}
however as demonstrated in \cite{Vlah2012} this will not fully capture the non-linear motions of galaxies within their host halos. In applying the above model to our mock galaxy catalogues in Section~\ref{sec:proof} we found that even leaving $\sigma_{v}$ to vary leads to significant biases in the recovered values of $f\sigma_{8}$. Instead, motivated by their Table 1 we make the following redefinitions
\begin{align}
&\sigma_{v} \rightarrow f^{-2}\sigma_{v,1} & \mathrm{terms}\,P_{02}, P_{12}, P_{22}, \mathrm{vector\,part}\,P_{13}\,(P_{13}\propto b^{2}_{1}), P_{04} \notag \\
&\sigma_{v} \rightarrow f^{-2}\sigma_{v,2} & \mathrm{terms}\,P_{03}, \mathrm{scalar\,part}\,P_{13} \notag
\end{align}
We are hence swapping a single $\sigma_{v}$ (free or otherwise) for 2 free parameters $\sigma_{v,1}$ and $\sigma_{v,2}$ that enable the model to more accurately reproduce the non-linearities in the motions of galaxies. Overall we fit for 5 parameters [$b_{1}\sigma_{8}$, $b_{2}\sigma_{8}$, $f\sigma_{8}$, $\sigma_{v,1}$, $\sigma_{v,2}$].

\section{Fisher information in the combined velocity and density field multipoles} \label{sec:appendixB}
In this appendix we develop a simple Fisher matrix calculation for the cosmological information contained in measurements of the galaxy density or velocity power spectrum multipoles for arbitrary weights. This is based on a combination of \cite{Taruya2011}, \cite{Koda2014} and \cite{Howlett2017a}. Taking results from these works, the Fisher information for a set of parameters $\theta$ contained within a series of power spectrum multipoles $P_{\ell}(k)$ can be calculated as
\begin{equation}
F_{ij} = \frac{\Omega}{4\pi^{2}}\int dk\,k^{2} \sum_{\ell \ell'} \frac{\partial P_{\ell}(k)}{\partial \theta_{i}} C^{-1}_{\ell \ell'}(k,w) \frac{\partial P_{\ell'}(k)}{\partial \theta_{j}}
\label{eq:fisher}
\end{equation}
where $\Omega$ is the sky area of the survey (in steradians) and the $\mathsf{C}_{\ell \ell'}(k,w)$ is the covariance matrix between the different measured multipoles which depends on the weights $w$ assigned to the data. In this work we consider multipoles of the galaxy density and momentum power spectra (with some consideration also given to the cross-spectrum). The covariance of these various measurements has been derived in various Sections of this work and can be pieced together to create $C_{\ell \ell'}(k,w)$ depending on the exact combination of observed multipoles. Assuming that the power spectra depend only on $k$ and $\mu$ and that the weights, number density and velocity errors of galaxies vary only radially, we can write these as a modified version of Eq.~\ref{eq:covweight},
\begin{align}
C^{\delta}_{\ell\ell'}(k) &= 2(2\ell + 1)(2\ell'+1)\int_{0}^{1} d\mu \frac{\int_{0}^{r_{max}} dr \,w_{\delta}^{4}(k,\mu,r)\bar{n}_{\delta}^{4}(r) \biggl[P^{\delta}(k,\mu) + \frac{1}{\bar{n}_{\delta}(r)}\biggl]^{2}L_{\ell}(\mu)L_{\ell'}(\mu)}{\biggl[\int_{0}^{r_{max}} dr\,w_{\delta}^{2}(r)\bar{n}_{\delta}^{2}(r)\biggl]^{2}} \\
C^{p}_{\ell\ell'}(k) &= 2(2\ell + 1)(2\ell'+1)\int_{0}^{1} d\mu \frac{\int_{0}^{r_{max}} dr\,w_{p}^{4}(k,\mu,r)\bar{n}_{p}^{4}(r) \biggl[P^{p}(k,\mu) + \frac{\langle v^{2}(r) \rangle}{\bar{n}_{p}(r)}\biggl]^{2}L_{\ell}(\mu)L_{\ell'}(\mu)}{\biggl[\int_{0}^{r_{max}} dr\,w_{\delta}^{2}(r)\bar{n}_{p}^{2}(r)\biggl]^{2}} \\
C^{\delta p}_{\ell\ell'}(k) &= 2(2\ell + 1)(2\ell'+1)\int_{0}^{1} d\mu \frac{\int_{0}^{r_{max}} dr\,w_{\delta}^{2}(k,\mu,r)w_{p}^{2}(k,\mu,r)\bar{n}_{\delta}^{2}(r)\bar{n}_{p}^{2}(r) \biggl[P^{\delta p}(k,\mu)\biggl]^{2}L_{\ell}(\mu)L_{\ell'}(\mu)}{\biggl[\int_{0}^{r_{max}} dr\,w_{\delta}^{2}(r)\bar{n}_{\delta}^{2}(r)\biggl]\biggl[\int_{0}^{r_{max}} dr\,w_{p}^{2}(r)\bar{n}_{p}^{2}(r)\biggl]^{2}}
\end{align}
where we have carefully maintained the distinction between the weights and number density used in the estimation of the density and momentum power spectra (denoted via the $\delta$ and $p$ superscripts respectively). $r_{max}$ is the maximum comoving distance of the survey; in solving the above equations it is often useful to perform a change of variable from comoving distance to redshift, such that $r_{max} \rightarrow z_{max}$. 

The other piece we need to compute the Fisher matrix is the derivatives of the power spectrum multipoles as a function of cosmological parameters. For our Fisher forecasts in Table~\ref{tab:fsigma8} we are interested in the information in the velocity $u$, power spectrum (rather than the momentum $p$, power spectrum, which may contain less) and so instead of the models in Appendix~\ref{sec:appendix}, we use models for the density, velocity and cross power spectrum from \cite{Koda2014, Howlett2017a}
\begin{equation}
P^{\delta}(k,\mu) = (b_{1} + f\mu^{2})^{2}D^{2}_{g}P_{L}(k), \qquad P^{\delta u}(k,\mu) = (b_{1} + f\mu^{2})D_{g}D_{u}\frac{ia H f\mu}{k}P_{L}(k), \qquad P^{u}(k,\mu) = \biggl(\frac{a H f\mu}{k}\biggl)^{2}D^{2}_{u}P_{L}(k)
\label{eq:appmodels}
\end{equation}
where the damping terms $D_{g}(k,\mu)=[1+0.5(k\mu\sigma_{v,1})^{2}]^{-1/2}$ and $D_{u}(k)=\mathrm{sinc}(k\sigma_{v,2})$ account for non-linear redshift-space distortions (although not to the same extent as the model in Appendix~\ref{sec:appendix}). The models in Eq.~\ref{eq:appmodels} are then integrated with respect to the Legendre polynomials to obtain the multipoles. The derivatives with respect to the parameters (in this case galaxy bias, growth rate and two velocity dispersion parameters for 4 in total) can be performed before the integration. The above equations for the covariances still hold when we swap out the momentum power spectrum for the velocity power spectrum. Given these derivatives and the above equations we can compute Eq.~\ref{eq:fisher} relatively simply for arbitrary combinations of multipoles. Finally, the Fisher matrix can be inverted to get the covariance matrix for the parameters, where the diagonal element corresponding to the growth rate is the predicted variance after marginalising over the other parameters in the matrix. In applying this to our mock catalogues we use the number density as a function of redshift averaged over all our mocks and marginalise over $b_{1}\sigma_{8}$, $\sigma_{v,1}$ and $\sigma_{v,2}$ when computing the growth rate errors. 

The above method also allows us to numerically minimise the predicted error on the growth rate as a function of the two numbers $P^{p}_{FKP}$ and $P_{FKP}$ used as inputs to Eq.~\ref{eq:weights} and the standard FKP weight. When doing this we are interested in actually modelling the momentum power spectrum (not just the velocity component, as the optimal constant values should take into account the fact that the density-velocity convolution term enters on non-linear scales) and so we \textit{do} use the models of Appendix~\ref{sec:appendix}. We compute the derivatives using these models and then substitute the weights for each $P^{p}_{FKP}$ and $P_{FKP}$ into the above covariance matrices. Performing this optimisation for the mock catalogues in Section~\ref{sec:proof} returns rounded values $P_{FKP}=6000\mpcohV$ and $P^{p}_{FKP}=2.4\times 10^{9}\kmsmpcohV$. Alternatively, the Fisher matrix could be numerically optimised for a particular choice of $k$ and $\mu$, such that the model anisotropic density and momentum power spectra are first calculated for this $k$ and $\mu$, and then these two values are used to compute the fixed weights. 

\end{document}